\documentclass[11pt]{article}
\usepackage[top=1in,bottom=1in,right=1in,left=1in]{geometry}

\usepackage{amsmath}
\usepackage{amsfonts}
\usepackage{amssymb}
\usepackage{graphicx}
\usepackage{subfigure}
\usepackage{multirow}
\usepackage{colortbl}
\usepackage{url}
\usepackage{float}

\newcommand{\vect}[1]{\mathbf{#1}}
\newcommand{\vx}{\vect{x}}
\newcommand{\ep}{\varepsilon}

\begin{document}


\title{An RBF-FD polynomial method based on polyharmonic splines for the Navier-Stokes equations:
Comparisons on different node layouts}

\author{Gregory A. Barnett \\
University of Colorado \\
Department of Applied Mathematics \\
Boulder, CO 80309 USA\\
gregory.barnett@colorado.edu \\
\\
Natasha Flyer \\
National Center for Atmospheric Research \\
Institute for Mathematics Applied to Geosciences\\
Boulder, CO 80305 USA \\
flyer@ucar.edu\footnote{Corresponding author}\\
\\
Louis J. Wicker\\
NOAA National Severe Storms Laboratory\\
Norman, OK 73072 USA\\
louis.wicker@noaa.gov
}

\maketitle

\begin{abstract}
Polyharmonic spline (PHS) radial basis functions (RBFs) are used together with polynomials to create local RBF-finite-difference (RBF-FD) weights on different node layouts for spatial discretization of the compressible Navier-Stokes equations at low Mach number, relevant to atmospheric flows. Test cases are taken from the numerical weather prediction community and solved on bounded domains. Thus, attention is given on how to handle boundaries with the RBF-FD method, as well as a novel implementation for the presented approach. Comparisons are done on Cartesian, hexagonal, and quasi-uniformly scattered node layouts. Since RBFs are independent of a coordinate system (and only depend on the distance between nodes), changing the node layout amounts to changing one line of code. In addition, consideration and guidelines are given on PHS order, polynomial degree and stencil size. The main advantages of the present method are: 1) capturing the basic physics of the problem surprisingly well, even at very coarse resolutions, 2) high-order accuracy without the need of tuning a shape parameter, and 3) the inclusion of polynomials eliminates stagnation (saturation) errors.
\end{abstract}


\section{Introduction}

In applications of radial basis functions (RBFs) for fluid modeling (both incompressible and compressible), infinitely smooth RBFs have traditionally been used due to their spectral convergence properties, with multiquadrics and Gaussians being the most popular \cite{SDY02,CS,CS05,CDNT,EKLRS08,FF2011,FLBWSC12,SSL}. However, fluid flows in nature can exhibit complex rapidly developing features with such steep gradients that spectral accuracy can not be realized on resolutions that are observable or practical. This study offers a different perspective on RBF-based fluid modeling with the following aspects: 1) using odd-ordered polyharmonic spline (PHS) RBF, $r^{2m+1}$ for $m\in\Bbb{N}$ ($r$ is the Euclidean distance between where the RBF is centered and where is evaluated), and 2) in conjunction with higher-order polynomials (degree $\geq 4$). From a historical perspective, using this combination for RBF-FD has not been considered, most likely for the following reasons:

\begin{enumerate}

\item Before the development of RBF-FD or other flavors of local RBFs \cite{SSL,SSQ09,SPLM}, applications of RBFs were global. Then, \textit{\textit{if}} piecewise smooth RBFs were used, they were used in conjunction with low order polynomials, e.g $1,x,y$ in 2-D. The only role of the polynomial was to guarantee non-singularity of the RBF interpolation matrix, which needs to be inverted to derive the differentiation matrices \cite{HW2005,FassBook}. The role of capturing the physics of complicated fluid flows was the left to the RBFs.

\item Even when used in a global sense, these RBFs were not as popular as infinitely smooth RBFs. For example, $r^3$, results in an RBF that jumps in the third derivative, giving at best fourth-order accuracy in 1-D (with the order of convergence increasing as the dimension increases (c.f.~\cite{Pow92}) assuming smooth data). The curse lies in the fact that as $m$ increases, leading to a smoother RBF, the condition number of the interpolation matrix gravely increases. Thus, in the past, one was limited to keeping $m$ small and having low algebraic accuracy.

\item Lastly, using polynomials on a global scale can be dangerous, since it can lead to Runge phenomena near the boundaries. In contrast, on a local scale as in the RBF-FD method, one is only interested in the approximation at the center of the stencil and not at the edges.

\end{enumerate}

As a result, a new way to use PHS RBFs combined with polynomials in the context of RBF-FD is introduced, such that high-order accuracy is gained with excellent conditioning of RBF-FD interpolation matrix and no saturation error is encountered. Furthermore, there is no need to bother with selecting an optimal shape parameter, which plays an important role in the accuracy of the solution when using infinitely smooth RBFs \cite{FassZhang2007,BaMoKi11,DaOa11,Sch11,Ch12}). We will demonstrate the performance of the modified RBF-FD method for 1) the advection of a scalar in a strong shear flow (a hyperbolic PDE introduced by \cite{LeVeque} and popularized by \cite{BlosseyDurran}) and 2) the 2D nonhydrostatic compressible Navier-Stokes on bounded domains applied to test cases common in the numerical weather prediction community. Although already broad in scope, the authors further wish to classify the differences, if any, that occur in applying this methodology on different node layouts: 1) Cartesian, 2) hexagonal, and 3) scattered. The rational being that classical finite difference methods, based on polynomials, are usually implemented on Cartesian lattices; hexagonal layouts are optimal in terms of node packing in 2D, supplying information along 3 distinct directions in contrast to Cartesian layouts where information is aligned only along 2 directions; and scattered nodes allow for geometric flexibility of the domain and the ability of node refinement.

The paper is organized as follows: Section \ref{sec:RBF} very briefly introduces RBFs. Section \ref{calc_wtgs} discusses the calculation of RBF-FD weights using polynomials. Section \ref{sec:saturation} demonstrates how the inclusion of polynomials with PHS eliminates stagnation (saturation) errors. Section \ref{nodeSets} discusses the node sets that are used and how boundaries and hyperviscosity are handled. Section \ref{TestCases} applies the methodology on the various node layouts, giving detailed results from test cases that are standard in the numerical weather prediction community. Lastly, Section \ref{summary} summarizes the observations of this paper.


\section{A brief introduction to Radial Basis Functions} \label{sec:RBF}

An RBF is a $d$-dimensional radially symmetric function $\phi:\mathbb{R}^d\rightarrow\mathbb{R}$ that depends only on the Euclidean distance between where the RBF is centered, $\vx_c$ and where it is evaluated, $\vx$. That is, regardless of dimension, its argument is always a scalar defined by $r = \|\vx - \vx_c\|_2$, where $\|\cdot\|_2$ denotes the Euclidean distance. RBFs come in two flavors: 1) Piecewise smooth and 2) infinitely smooth. The former features a jump in some derivative and thus can only lead to algebraic convergence. The latter is a $C^\infty$ function and can lead to spectral convergence when the data is sufficiently smooth. Only PHS RBFs do not depend on a parameter $\ep$ that controls the shape of the RBF (which influences both the conditioning of the matrices and the accuracy of the results \cite{Tarwater,Sch95}). This last comment is a strong motivation of this paper since, by using PHS RBF, one avoids the difficulty of dealing with a shape parameter and yet can achieve high-order accuracy. For theoretical aspects of PHS (a class of conditionally positive definite radial functions), see \cite{FassBook, HW2005}. Common RBFs of both categories are given in Table \ref{tbl:RBFs}, where $\vx_c$ represents where the RBF is centered.
\begin{table}[h]
\centering
\caption{Some common choices for radial functions $\phi(r)$ \label{tbl:RBFs}}
\vspace{0.05in}
\begin{tabular}{ll}
\hline
\textbf{Piecewise smooth RBFs}  &   $\phi(r = \|\vx - \vx_c\|_2)$ \\
\hline
Polyharmonic Splines (PHS) \cite{Du77}     & $r^{2m}\log r,\; m\in \Bbb{N}$\\
                                           & $r^{2m+1},\; m \in \Bbb{N}^0$ \\
Matern \cite{Matern60}                     & $\frac{2^{1-m}}{\Gamma(m)} r^{m}K_{m}(\varepsilon r),m>0$,\,\, (Bessel $K$-function)\\
Compact support (`Wendland' \cite{Wend95}) & $(1-\varepsilon r)_{+}^{m}p(\varepsilon r),$ $p$ certain polynomials, $m\in N$\\
\hline
\textbf{Infinitely smooth RBFs}  & \\
\hline
Gaussian (GA)                    & $e^{-(\varepsilon r)^{2}}$\\
Multiquadric (MQ)                & $\sqrt{1+(\varepsilon r)^{2}}$\\
Inverse Multiquadric (IMQ)       & $1\left/\sqrt{1+(\varepsilon r)^{2}}\right.$\\
Inverse Quadratic (IQ)           & $1\left/\left(1+(\varepsilon r)^{2}\right)\right.$\\
\hline
\end{tabular}
\end{table}

\section{Calculation of the PHS RBF-FD differentiation weights with polynomials} \label{calc_wtgs}
The differentiation weights are derived so that the resulting linear system becomes exact for all RBF interpolants $s(\mathbf{x})$ of the form $s(\mathbf{x}) = \sum_{i=1}^n \lambda_i \phi(\|\vx - \vx_i\|_2) + \{ p_l(\mathbf{x})\}$ with the constraints $\sum_{i=1}^n \lambda_i p_l(\mathbf{x_i}) = 0$, where $ p_l(\mathbf{x})$ are all polynomials up to degree $l$ in the dimension of the problem. These constraints enforce that the RBF basis reproduces polynomials up to degree $l$ as well as ensure that the far-field RBF expansion is regularized (i.e. does not blow up) \cite{FDWC}. They also are known as the vanishing moment conditions \cite{ISKE2003}.

It can then be shown (see Section 5.1.4 in \cite{FFBook}) that the above leads to the following linear system for the differentiation weights,

\begin{align}
\left[\begin{array}{ccc|ccc}
\left\|\mathbf{x}_1-\mathbf{x}_1\right\|_2^{2m+1} & \cdots & \left\|\mathbf{x}_1-\mathbf{x}_n\right\|_2^{2m+1} & 1      & x_1    & y_1    \\
\vdots                                       & \ddots & \vdots                                       & \vdots & \vdots & \vdots \\
\left\|\mathbf{x}_n-\mathbf{x}_1\right\|_2^{2m+1} & \cdots & \left\|\mathbf{x}_n-\mathbf{x}_n\right\|_2^{2m+1} & 1      & x_n    & y_n    \\
\hline
1                                            & \cdots & 1                                            & 0      & 0      & 0      \\
x_1                                          & \cdots & x_n                                          & 0      & 0      & 0      \\
y_1                                          & \cdots & y_n                                          & 0      & 0      & 0
\end{array}\right]
\left[\begin{array}{c}
w_1     \\
\vdots  \\
w_n     \\
\hline
w_{n+1} \\
w_{n+2} \\
w_{n+3}
\end{array}\right]
=&
\left[\begin{array}{c}
\left.L\left\|\mathbf{x}-\mathbf{x}_1\right\|_2^{2m+1}\right|_{\mathbf{x}=\mathbf{x}_c}    \\
\vdots                                                                                \\
\left.L\left\|\mathbf{x}-\mathbf{x}_n\right\|_2^{2m+1}\right|_{\mathbf{x}=\mathbf{x}_c}    \\
\hline
\left.L1\right|_{\mathbf{x}=\mathbf{x}_c}                                             \\
\left.Lx\right|_{\mathbf{x}=\mathbf{x}_c}                                             \\
\left.Ly\right|_{\mathbf{x}=\mathbf{x}_c}
\end{array}\right]
\label{weights}
\end{align}
where for simple illustration purposes, we have included only up linear polynomials. The weights $w_{n+1}$ to $w_{n+3}$ are ignored after the matrix is inverted. Solving (\ref{weights}) will give one row of the differentiation matrix (DM) that contains the weights for approximating $L$ at $x_c$. Thus this process is repeated $N$ times over all nodes in the domain, giving a preprocessing cost of $O(n^3 N)$. Since $n<<N$, it should be noted that the DM usually becomes over $99\%$ empty. As a result, we do not actually store the DM but only its nonzero entries.


\section{Stagnation error, PHS order, and polynomial degree} \label{sec:saturation}

Stagnation (saturation) error is defined as the convergence either stagnating or increasing as resolution increases. For infinitely smooth RBFs, as the resolution increases (i.e. $r$ decreases), the shape parameter must increase to maintain the condition number of the matrix in (\ref{weights}), resulting in stagnation error since the more peaked the infinitely smooth RBFs become as $\ep$ increases, the less accurate the approximation. In contrast, PHS RBF \textit{with} polynomials can achieve high-order algebraic convergence without encountering saturation error or the difficulty of finding an optimal value of the shape parameter $\ep$ for good accuracy, which has been a central focus of quite a few studies \cite{Foley94,HLC07,FassZhang2007,BaMoKi11,DaOa11}. It should be noted that if polynomials are not included with the PHS RBF, stagnation error is encountered since boundary errors at the edge of the RBF-FD stencil can be quite large and penetrate toward the center of the stencil where the interpolant or any derivative is being approximated. Further investigation of the effects of adding polynomials to RBF-FD approximations for both infinitely smooth and PHS RBFs is given in \cite{FFBB}.

\begin{figure}[H]
\begin{align*}
\begin{array}{cc}
\includegraphics[width=.5\textwidth]
{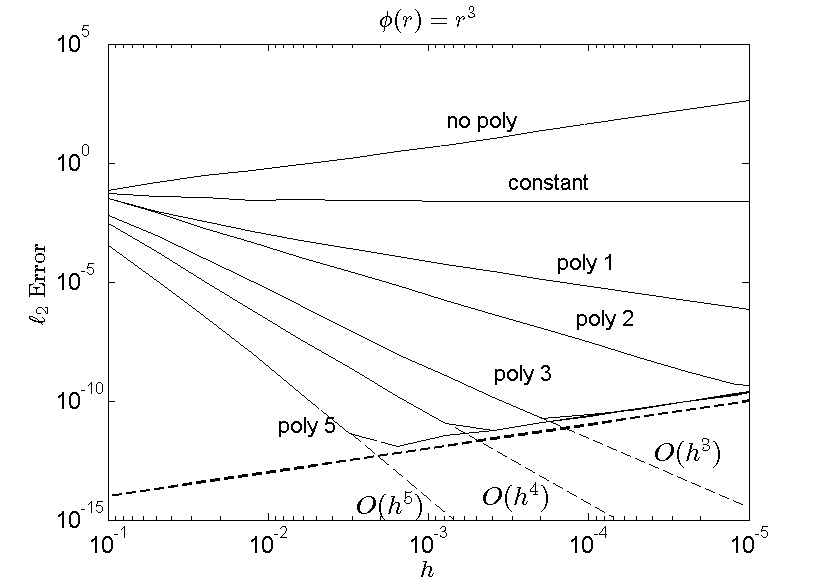} \quad & \quad
\includegraphics[width=.5\textwidth]
{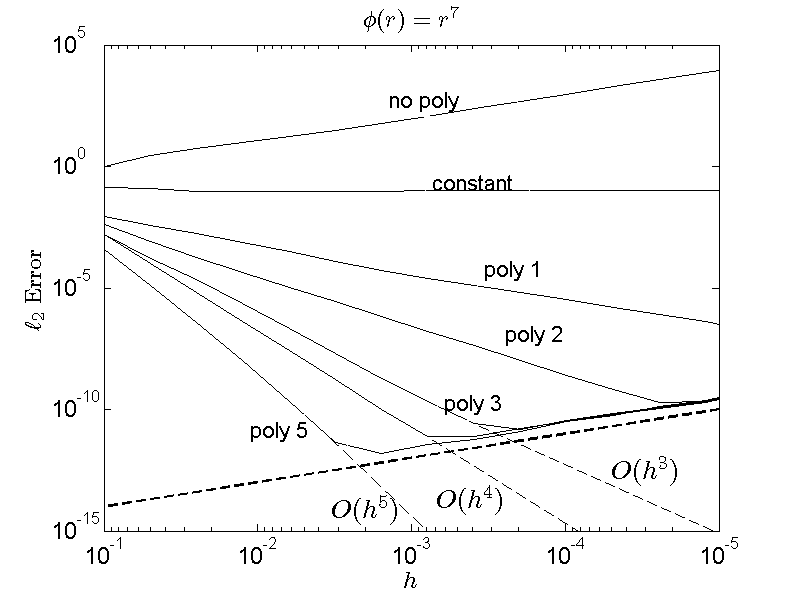}
\end{array}
\end{align*}
\caption{The convergence rate in approximating $d/dx$ of the function $f(x, y) = 1 + \sin(4x) + \cos(3x) + \sin(2y)$ on a 37 node RBF-FD stencil based on $r^3$ and $r^7$, augmented with polynomials as described in the text. The dashed line marks machine round-off errors in standard double precision of $10^{-15}/h$ for approximating the first derivative.}
\label{r3_vs_r7_poly}
\end{figure}

Since locally all smooth functions are well represented by Taylor expansions, then under refinement, the RBF-FD approximation must reproduce polynomial behavior. In the following numerical studies, it indeed was found that the convergence rate is dictated not by the order of the PHS, $m$, but by the highest degree of polynomials, $l$, used. In addition, for PDEs with only first-order spatial derivatives, as in all the test cases, the convergence rate can be expected to be $O(h^{l})$. The reason being is that the error for polynomial interpolation is $O(h^{l+1})$, but one order in $h$ is lost in approximating a first derivative. These observations are in excellent agreement with Figure \ref{r3_vs_r7_poly}, where $d/dx$ of the smooth function $f(x, y) = 1 + \sin(4x) + \cos(3x) + \sin(2y)$ is approximated with two different PHS RBFs, $r^3$ and $r^7$, augmented with polynomials (e.g. poly 3 is augmentation of the matrix in (\ref{weights}) with the 10 polynomials up to degree 3 in 2D). The approximation is at the center of 37 node hexagonal stencil with the evaluation nodes of the derivative in its vicinity. The slopes in the two panels of Figure \ref{r3_vs_r7_poly} are identical, the only difference being that the constant that multiplies the order of convergence is slightly smaller for $r^7$, thus giving a marginally higher accuracy for a given resolution.  It is still important to note that the RBFs play a crucial role in safety against singularities due to particular node layouts.

\section{Node sets, Ghost nodes, and Hyperviscosity}\label{nodeSets}

Section \ref{sub:nodesets} overviews the various node sets considered in this study. Section \ref{sub:ghostnodes} discusses how to increase accuracy near boundaries with the use of ghost nodes and Section \ref{sub:hyper} discusses the need for and type of hyperviscosity used, introducing a novel way of implementing hyperviscosity with PHS and polynomials.

\subsection{Node-sets } \label{sub:nodesets}
\begin{figure}[H]
\centering
\includegraphics[width=0.75\textwidth]{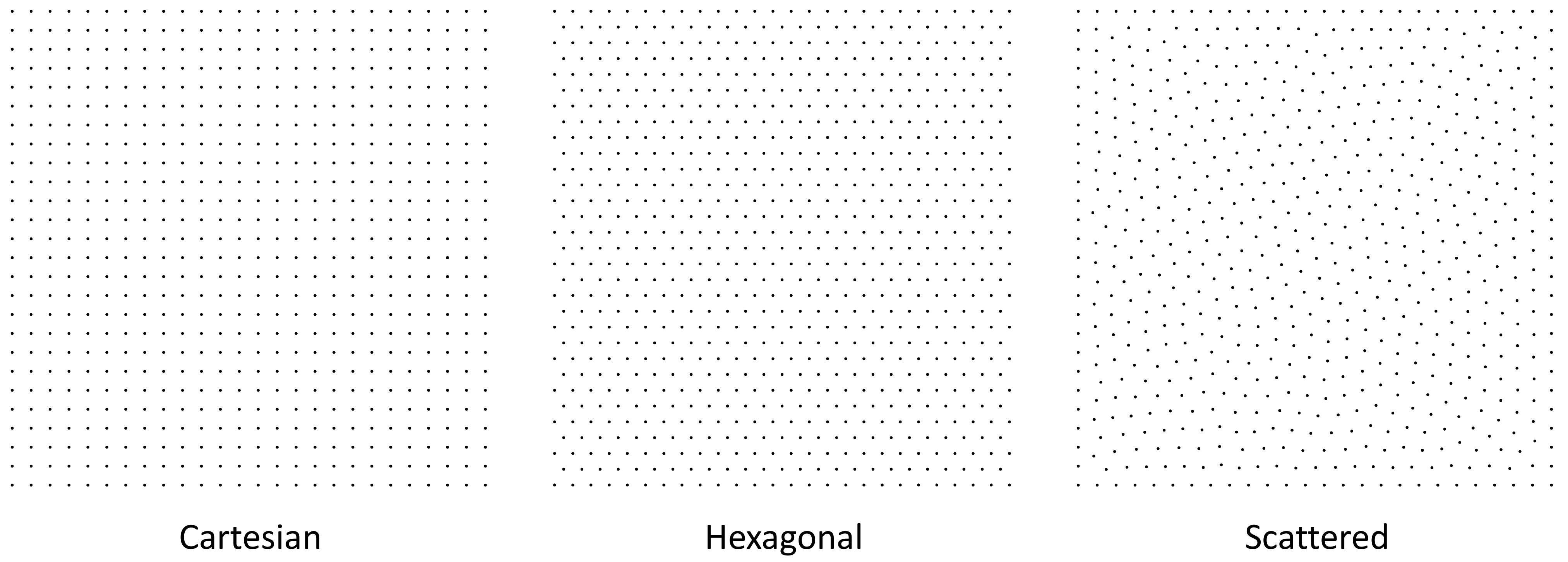}
\caption{Three different types of node-distributions that are used to solve the 2D test problems.}
\label{nodes_cartHexScat}
\end{figure}

Unlike traditional methods, RBF-FD has the advantage of being equally simple to apply on any set of nodes.  Figure \ref{nodes_cartHexScat} shows the three types of node-layouts that will be considered in the present study. First, we consider Cartesian since they are the lattices classical finite differences (FD) are usually implemented on. Next, it is well known that hexagonal node sets are the optimal packing strategy in 2D (i.e. for a fixed area, one can fit the most number of nodes). They have also been considered an optimal layout for differentiation stencils, since information is aligned along three different directions as opposed to only two with Cartesian layouts. Although FD have been sporadically implemented over the decades on hexagonal node layouts, they have never caught favor do to the complexity of the implementation. However, now with RBF-FD, implementation is simple. Lastly, scattered node layouts are considered as they have the great advantage of geometric flexibility that will be needed in future applications of RBF-FD on irregular domains and/or with local node refinement.

\subsection{Ghost nodes} \label{sub:ghostnodes}

Near boundaries, stencils become one-sided, leading to a deterioration of the approximation due to Runge phenomenon.  In order to ameliorate this effect, one layer of ghost nodes can be placed just outside the domain.  Once the ghost nodes are placed, function values at these locations can be solved for by enforcing additional constraints at the boundary nodes.  For example, if the upper boundary of a rectangular domain is free-slip, then $\partial u/\partial z=0$, and this condition can be used to solve for values of $u$ at the ghost nodes. Similarly, the PDE itself can be enforced on the boundary, giving an extra constraint.

\begin{figure}[H]
\centering
\includegraphics[width=0.65\textwidth]{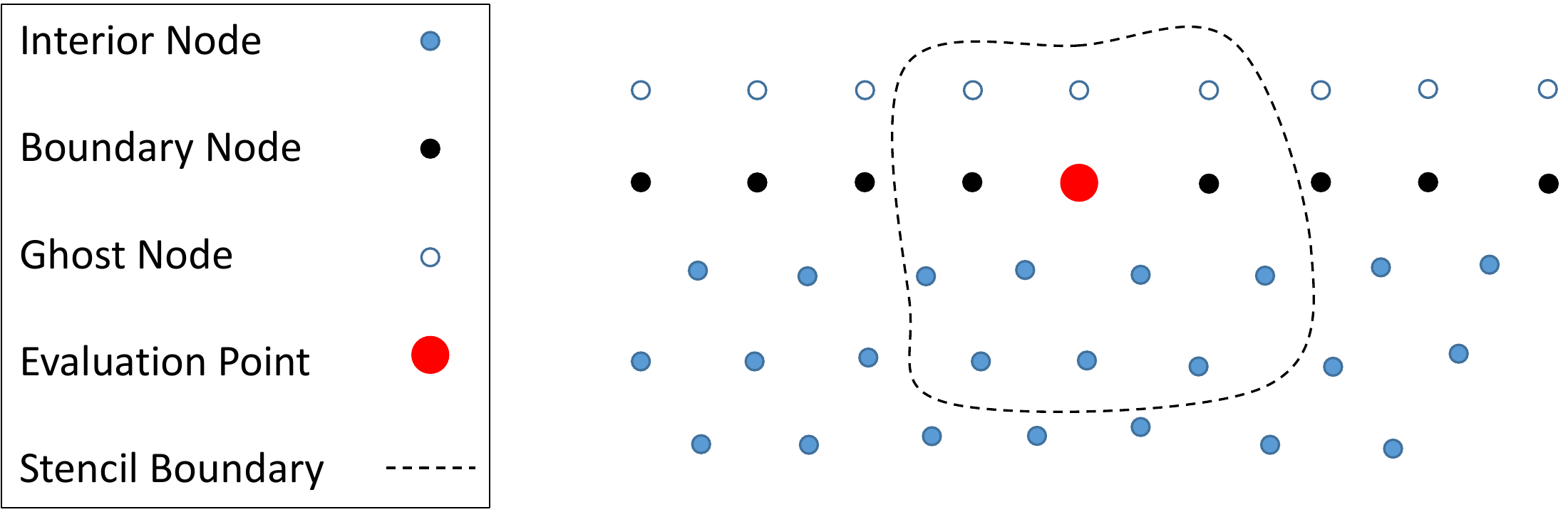}
\caption{An example of an RBF-FD stencil that might be used for enforcing $\partial u/\partial z=0$ on the top boundary.}
\label{stencilDiagram}
\end{figure}

For each node on the top boundary, one ghost node is placed just outside the boundary, as shown in Figure \ref{stencilDiagram}. The function values for $u$ at the ghost nodes are obtained by enforcing $\partial u/\partial z=0$ at each of the top boundary nodes simultaneously. This will lead to a coupled system with as many equations as there are ghost nodes. In the case illustrated in Figure \ref{stencilDiagram}, the system will be tridiagonal, as there are three unknown values at the ghost points for each evaluation node on the boundary.

The following is a more detailed discussion on how the ghost node values are calculated. Suppose there are $N_I$ interior nodes, $N_B$ boundary nodes, and $N_G$ ghost nodes, and let the total number of nodes be $N=N_I+N_B+N_G$.  For each top boundary node, approximate the differentiation weights for $\partial /\partial z$  as given in Section \ref{calc_wtgs}. This will result in a sparse $N_B\times N$ DM (here, called $W$). Thus,
\begin{align*}
\frac{\partial u}{\partial z}=&\,0 \quad \text{is approximated by}\\
W\underline{u}=&\,\underline{0}.
\end{align*}
The function values $\underline{u}$ are organized according to where they are located:
\begin{align*}
\underline{u}_I:&\text{ function values at interior nodes}\\
\underline{u}_B:&\text{ function values at boundary nodes}\\
\underline{u}_G:&\text{ function values at ghost nodes}
\end{align*}
Then, the condition $W\underline{u}\approx\underline{0}$ can be written as
\begin{align}
W_I\underline{u}_I+W_B\underline{u}_B+W_G\underline{u}_G=&\,\underline{0},\label{bcEqn}
\end{align}
where the matrix $W$ has similarly been split into pieces according to the three different types of nodes:
\begin{align*}
W_I\,\left(N_B\times N_I\right):&\text{ weights applied on interior nodes}\\
W_B\,\left(N_B\times N_B\right):&\text{ weights applied on boundary nodes}\\
W_G\,\left(N_B\times N_G\right):&\text{ weights applied on ghost nodes}
\end{align*}
Finally, \eqref{bcEqn} is used to solve for $\underline{u}_G$, the function values at the ghost nodes:
\begin{align*}
\underline{u}_G=-W_G^{-1}\left(W_I\underline{u}_I+W_B\underline{u}_B\right).
\end{align*}
Once the function values at the ghost node are known, they can be used for the approximation of other derivatives that appear in the governing equations.

\subsection{Hyperviscosity with PHS and polynomials} \label{sub:hyper}

When the viscosity of the fluid $\mu$ is small (such as the case with air $\approx 10^{-5} m^2/s$), there is essentially no natural diffusion in the governing equations, and high-frequency errors will grow to dominate a numerical solution.  To achieve time stability with the RBF-FD method, it has been shown that adding a relatively small amount of hyperviscosity to the right-hand-side of the governing equations eliminates the contaminating high-frequency noise while keeping the numerically relevant portion of the solution intact \cite{FoL11,FLBWSC12,Bollig12}.

The hyperviscosity operator takes the form $\gamma\Delta^k$, where $k$ is the power the Laplacian and $\gamma$ is a scaling parameter. The integer $k$ controls which frequencies are most affected, with larger values of $k$ giving stronger damping of high frequencies and weaker damping of low frequencies. As has been shown in \cite{FoL11,FLBWSC12,Bollig12}, for good stability and accuracy, the parameter $\gamma$ is directly proportional to the number of nodes in the domain $N$ or conversely the resolution $h$, as well as $k$. Thus, for a square-type domain in 2D $h\sim 1/\sqrt{N}$, and $\gamma = c h^{-2k}$, where $c$ is a constant that is generally set for the problem at hand and independent of the resolution $h$, (e.g. for the NS test cases $c = 2^{-6}$ regardless of the resolution or node layout used).

The hyperviscosity operator $\Delta^k$ is particularly simple to apply to an odd-powered PHS RBF, regardless of the spatial dimension.  The Laplace operator in $d$ dimensions for a radially symmetric function is given by $\Delta = \partial{}^2/\partial{r^2} + ((d-1)/r)\partial{}/\partial{r}$. Apply this to $\phi(r) = r^m$ results in
\begin{equation}
\Delta\left(\left\|\mathbf{x}\right\|^m\right)=m\left[m+\left(d-2\right)\right]\left\|\mathbf{x}\right\|^{m-2}.
\label{eqn:laplacian}
\end{equation}
In other words, applying the Laplace operator to a PHS RBF of degree $m$ gives a new PHS RBF of degree $m-2$. Using the above relationship, one can evaluate $\Delta^k\left(\left\|\mathbf{x}\right\|^m\right)$, and the new RBF will be continuous provided that $\left(m\ge2k+1\right)$. Higher-order Laplacians can be implemented by applying \eqref{eqn:laplacian} repeatedly. It should noted that the order of the PHS used for spatial discretization need not be that used for hyperviscosity. However, it was found experimentally that the simplest approach for the needed inclusion of polynomials was to use up to the same degree as that for discretization.


\section{Numerical Studies}\label{TestCases}

The first test case, inviscid transport of a scalar variable in a strongly sheared vortex flow, is a case of pure advection with a known analytical solution, so that the convergence properties of the method can be tested. It was originally proposed in \cite{LeVeque} and then considered in the context of applying limiters in \cite{BlosseyDurran, Skamarock06}. The second set of tests is based on the work presented in \cite{Straka}, where a cold descending bubble in a neutrally-stratified atmosphere develops into a traveling density current with the formation of Kelvin-Helmholtz rotors. It is now considered a classic test case in nonhydrostatic atmospheric modeling. The third test \cite{Robert93} (with similar studies in \cite{Grabowski}) simulates a rising thermal air bubble and nicely illustrates how instability patterns at the leading edge of the thermal are dependent on the node layout when the the dynamic viscosity is that of air.

In all cases, $n=37$ node stencils are used for spatial discretization, since both Cartesian and hexagonal layouts have perfectly symmetric stencils at that number, as seen in Appendix A. Although not essential, stencil symmetry is beneficial in that it it provides information evenly for approximating an operator at the center of the stencil. In all cases, time stepping is done with a 4th-order Runge-Kutta scheme (RK4).


\subsection{Advective transport of a scalar variable}

In this test, a circular scalar field is stretched and deformed into a crescent by vortex-like velocity field that then reverses and returns it back to its original position and shape. The governing equation is defined on $[0,1]\times[0,1]$ in $x$ and $y$ and given by
\begin{align}
\frac{\partial\psi}{\partial t}
=&
-\frac{\partial}{\partial x}\left(u\psi\right)-\frac{\partial}{\partial y}\left(v\psi\right).\nonumber
\end{align}
The scalar $\psi$ is advected by the following divergence-free velocity field
\begin{align*}
u\left(x,y,t\right)=u_{\theta}\left(r,t\right)\sin\theta,\quad\quad\quad v\left(x,y,t\right)=-u_{\theta}\left(r,t\right)\cos\theta.
\end{align*}
with period $T$, where
\begin{align*}
u_{\theta}\left(r,t\right)=\frac{4\pi r}{T}\left[1-\cos\left(\frac{2\pi t}{T}\right)\frac{1-(4r)^6}{1+(4r)^6}\right],
\end{align*}
and
\begin{align*}
r=\sqrt{\left(x-0.5\right)^2+\left(y-0.5\right)^2},\quad\quad\quad\theta=\tan^{-1}\left(\frac{y-0.5}{x-0.5}\right).
\end{align*}

\begin{figure}
\centering
\includegraphics[width=0.5\textwidth]{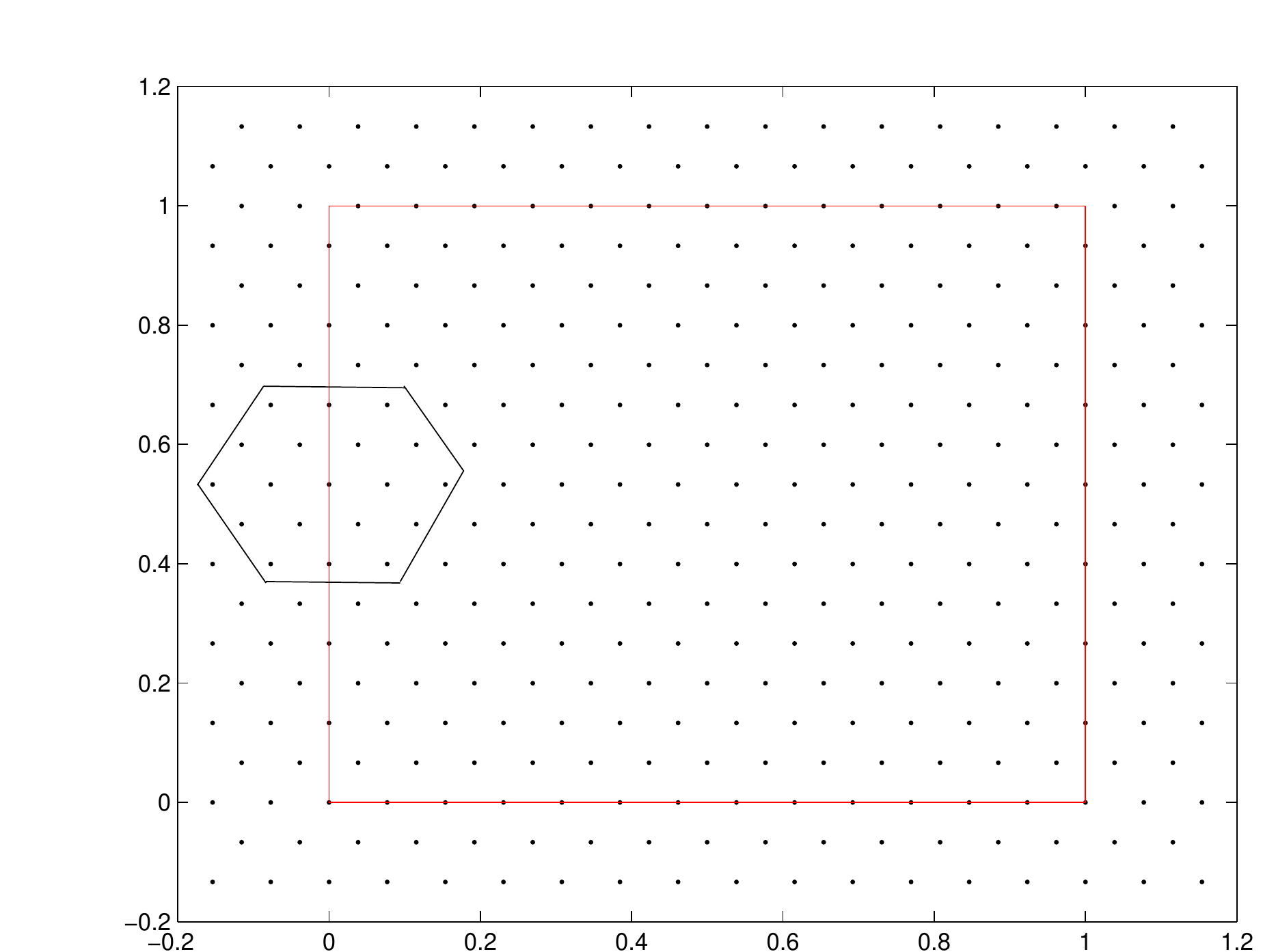}
\caption{An illustration of a hexagonal 19 node boundary stencil in which the nodes and there associated function values are reflected outside the domain. This setup is used for the advection of a scalar transport test case. }
\label{hex_stencil}
\end{figure}

In order to create a test problem with no boundary effects, the nodes on the interior of the domain near the boundary, as well as the function values associated with them, are simply reflected over the boundary (thus no boundary conditions are needed), forming perfectly symmetric boundary stencils - half of which are then ghost nodes and half interior nodes. The boundary is then time stepped with the rest of the interior of the domain. An example of a symmetric 19 node boundary stencil is given in Figure \ref{hex_stencil}.

The initial condition for $\psi$ is a cosine bell
\begin{align}
\left.\psi\right|_{t=0}=
\begin{cases}
\frac{1+\cos\left(\pi\tilde{r}\right)}{2} & \quad \tilde{r}\le1 \\
0 & \quad\tilde{r}>1,\\
\end{cases}
\end{align}
where $\tilde{r}=5\sqrt{\left(x-0.3\right)^2+\left(y-0.5\right)^2}$. Using a classic Runge-Kutta 4th-order (RK4), $\psi$ is advanced in time from $t=0$ until $t=T$ (one period), at which point it should have ideally returned to its original height and position.

\subsubsection{Accuracy of solution, convergence, and effect of different node layouts}

\begin{figure}
\centering
\includegraphics[width=0.8\textwidth]{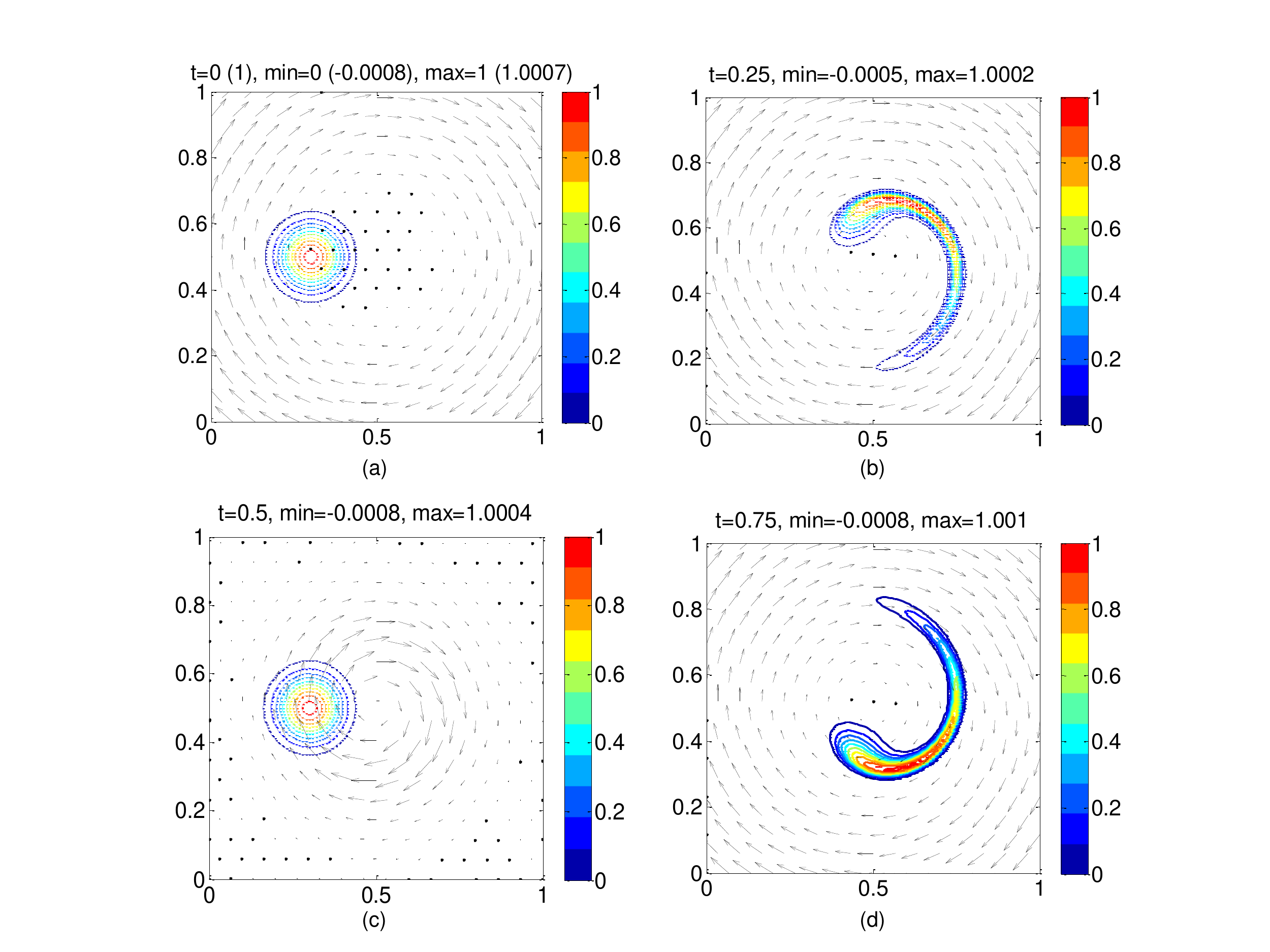}
\caption{Time series of the solution for $\psi$ at: (a) $t=0$(1), (b) $t=0.25$, (c) $t=0.5$, (d) $t=0.75$, with the corresponding minimum and maximum values at each time. Contour lines are in intervals of 0.05. An $n=37$ node stencil with $r^9$ PHS and up to $4^{\text{th}}$-order polynomials on a hexagonal node layout of 201 by 201 is used. The time-step is h/15, where here $h=1/200=0.005$. A hyperviscosity of $-2^{-14}\Delta^4$ is also implemented.}
\label{time_series_BD}
\end{figure}

Figure \ref{time_series_BD} shows the time evolution for solution (assuming a period of $T=1$), with corresponding velocity field, on a hexagonal node layout. The specifications of the resolution, time-step, basis functions and hyperviscosity used are given in the caption of the figure. Although, this is a high resolution case with a total of $N=40401$ nodes in the domain or a node spacing of $h=0.005$, it should be noted that the maximum amplitude of $\psi$ has only increased by $0.07\%$ and has only gone below zero by $-0.08\%$, as can be seen in Figure \ref{time_series_BD}a. In fact, when the scalar field is in its state of highest deformation at $t=0.25,0.75$, the error in the maximum amplitude is not more than 0.001. Furthermore, the CFL criterion for an RK4 stability domain dictates a time step of
$\Delta t < \Delta x \cdot 2\sqrt{2}/(\max{\mathbf{u}})$, which for this case translates into $\Delta t < 7.0(10)^{-4}$, a factor of only 2.12 larger than the time step taken of $\Delta t = h/15 = 0.005/15 = 3.3(10)^{-4}$.

\begin{figure}[H]
\begin{align*}
\begin{array}{cc}
\includegraphics[width=.4\textwidth]
{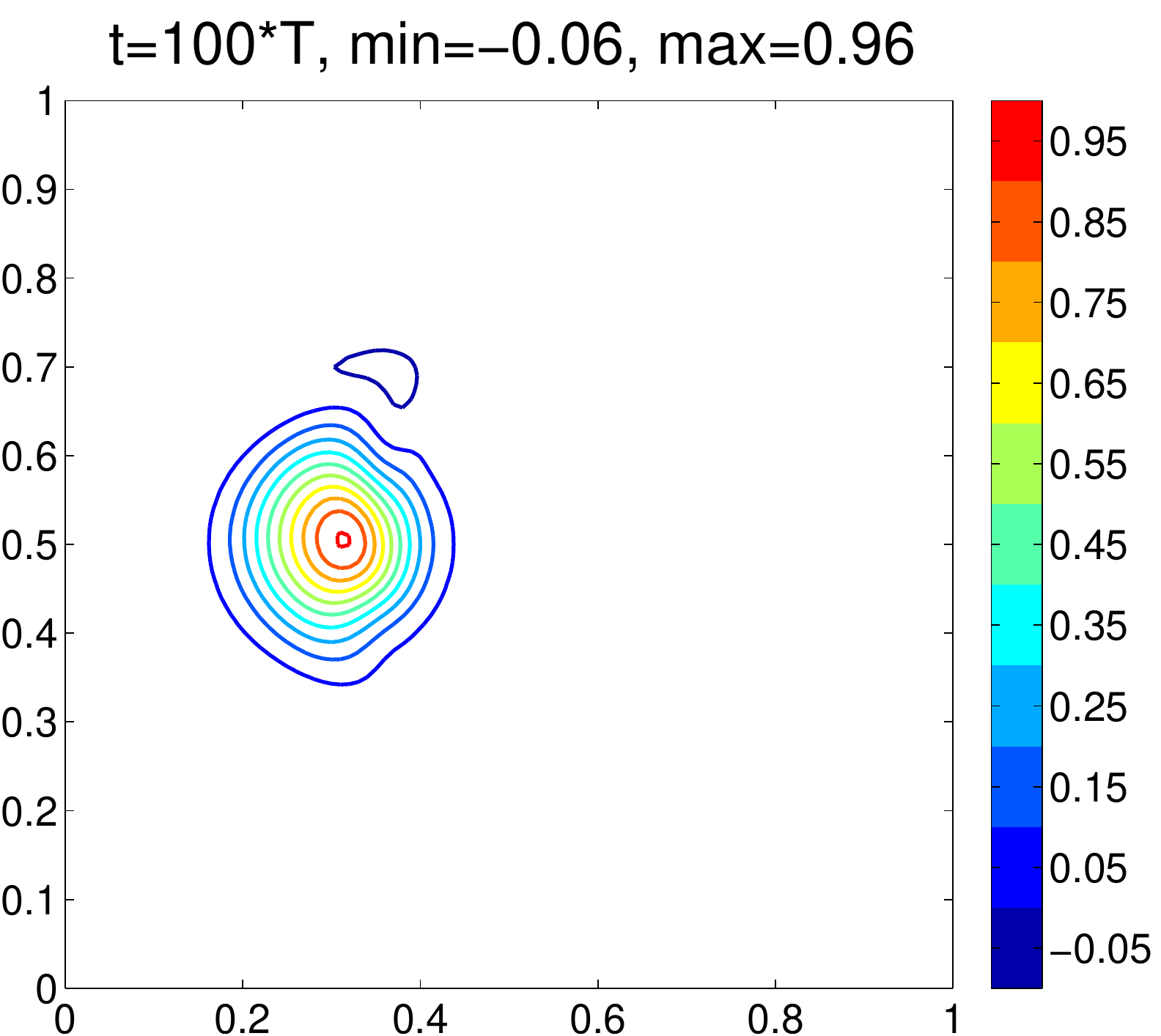} &
\includegraphics[width=.4\textwidth]
{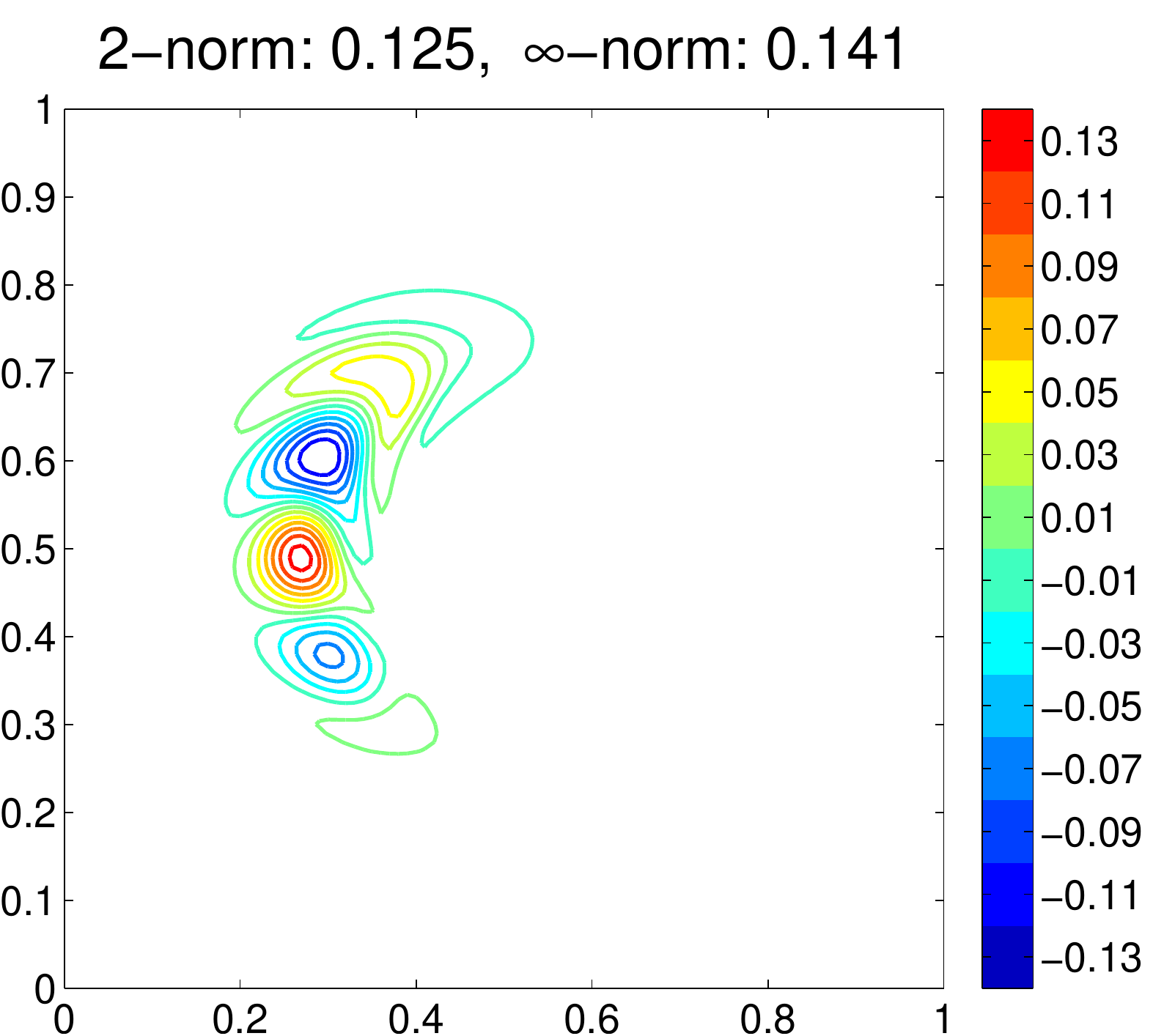}
\end{array}
\end{align*}
\caption{Solution (left panel) and error (right panel) at $t=100T$ on 40,401 $\left(h=0.005\right)$ hexagonal nodes using $\phi(r)=r^9$ with up to $4^{\text{th}}$ degree polynomials on a $37$-node stencil and $\Delta^3$-type hyperviscosity.}
\label{100T}
\end{figure}

In order to observe the long-time errors in the method, the solution is advanced for 100 periods as shown in the left panel of Figure \ref{100T}. Even after so many revolutions, the height of the scalar field has decreased only by $4\%$, with a slight distortion from its circular shape. The right panel of Figure \ref{100T} shows a dispersive error pattern with a maximum value of 0.13. The $\ell_2$ and $\ell_{\infty}$ errors are 0.125 and 0.141, respectively.

\begin{figure}[H]
$$
\begin{array}{cccc}
\text{} & \text{Cartesian} & \text{Hexagonal} & \text{Scattered} \\
\rotatebox{90}{~~~~~~~~~~\,$h = 0.02$} &
\includegraphics[width=.3\textwidth]{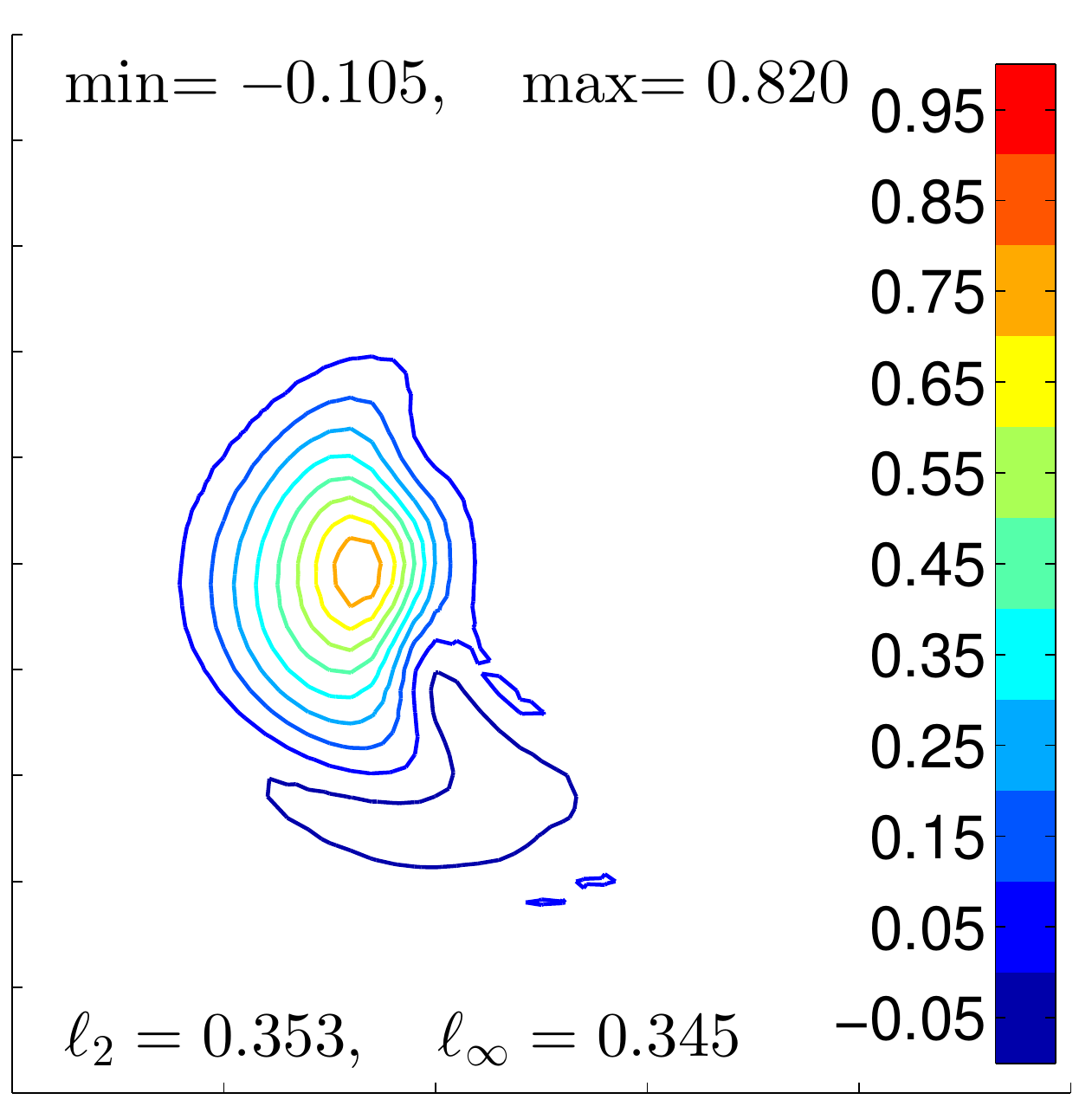} &
\includegraphics[width=.3\textwidth]{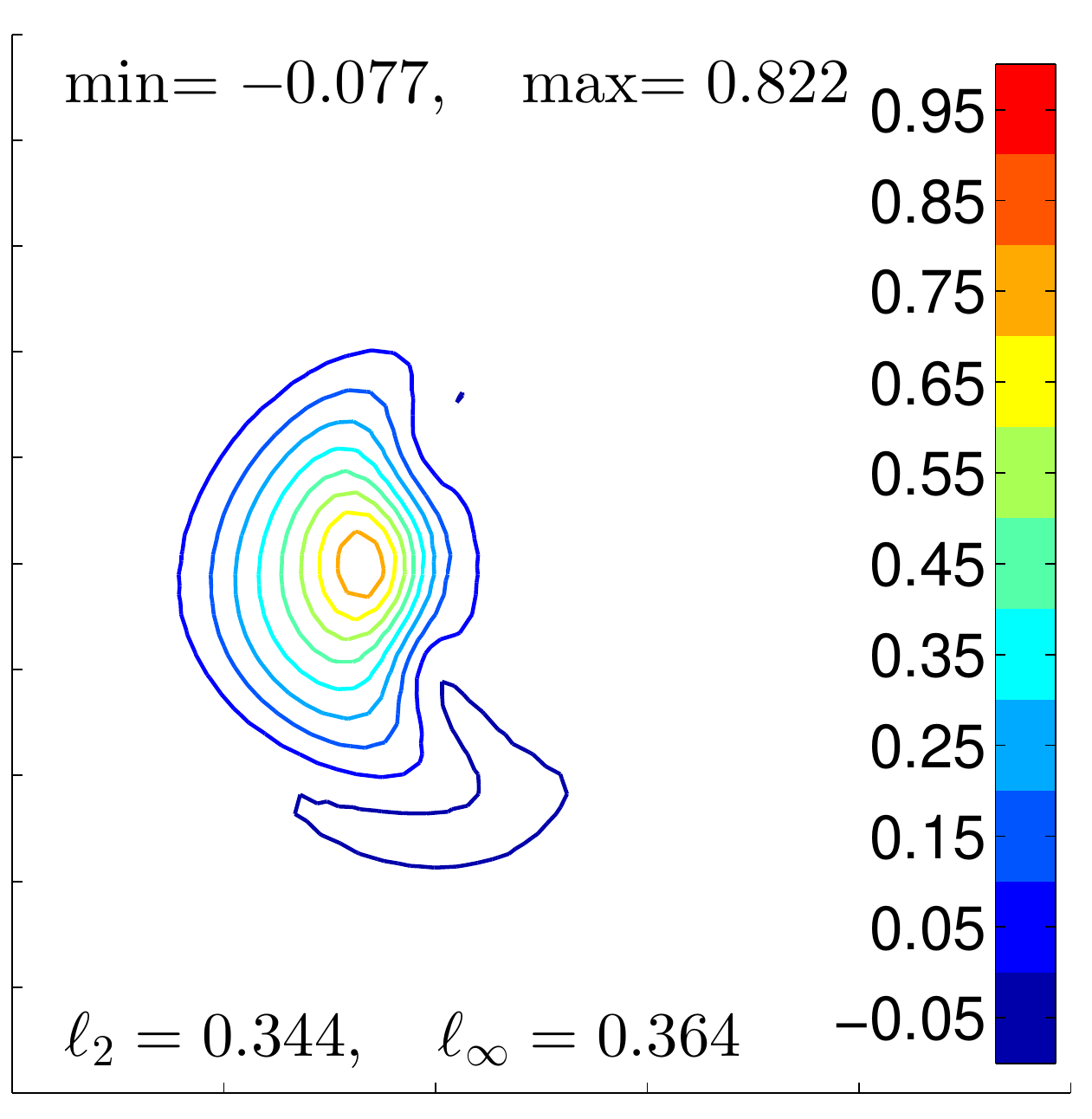}  &
\includegraphics[width=.3\textwidth]{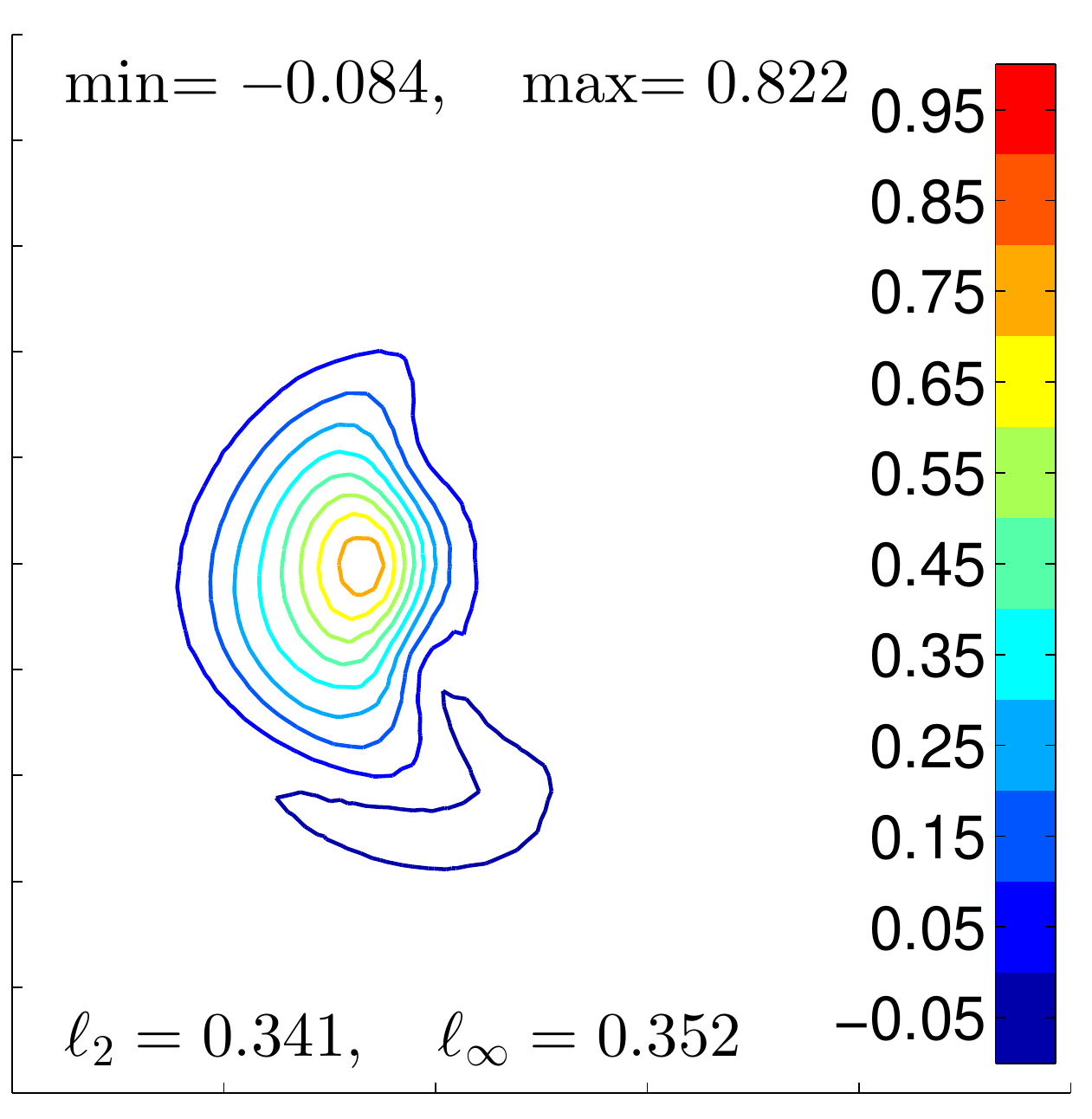} \\
\rotatebox{90}{~~~~~~~~~~\,$h = 0.01$} &
\includegraphics[width=.3\textwidth]{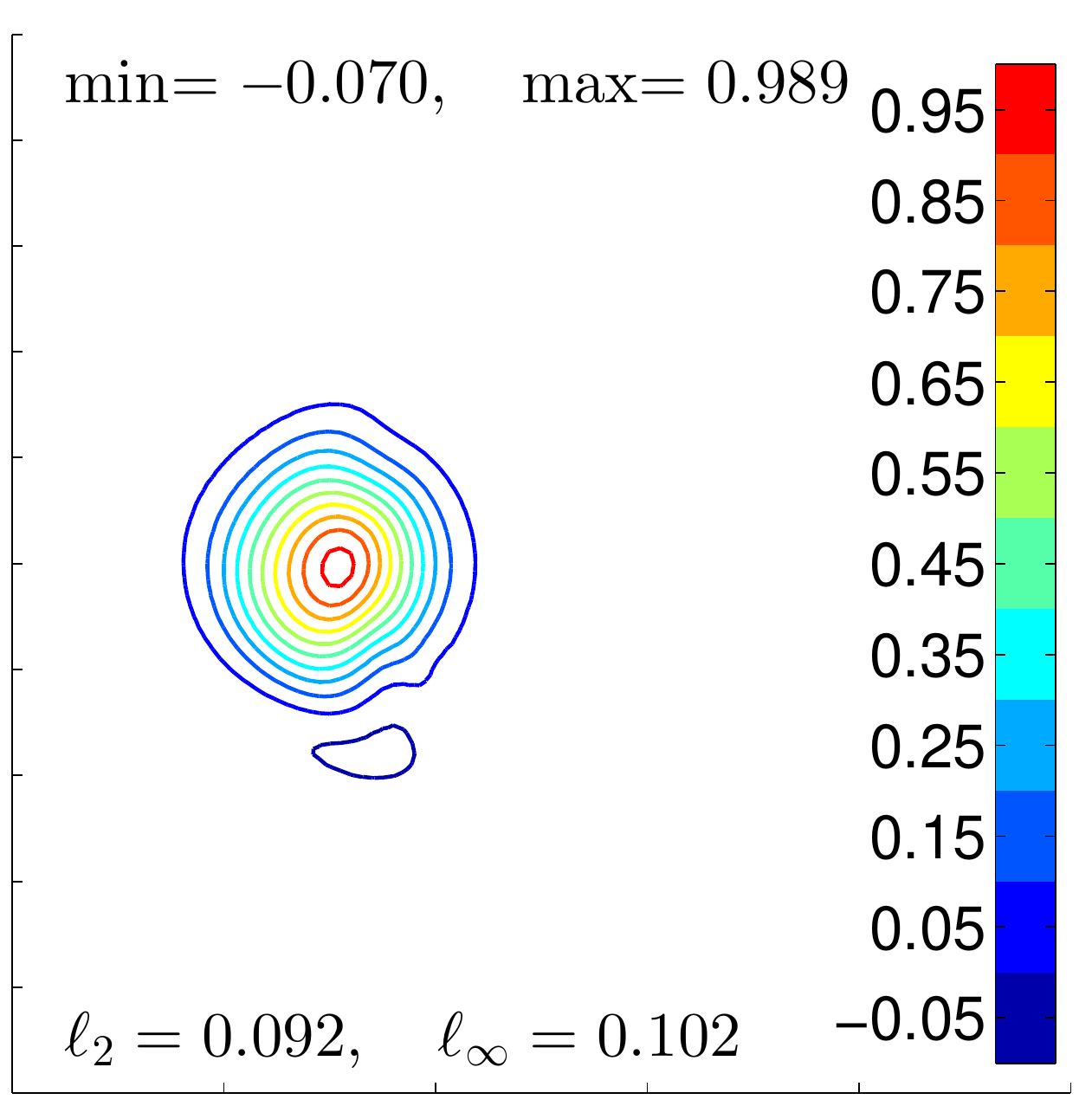} &
\includegraphics[width=.3\textwidth]{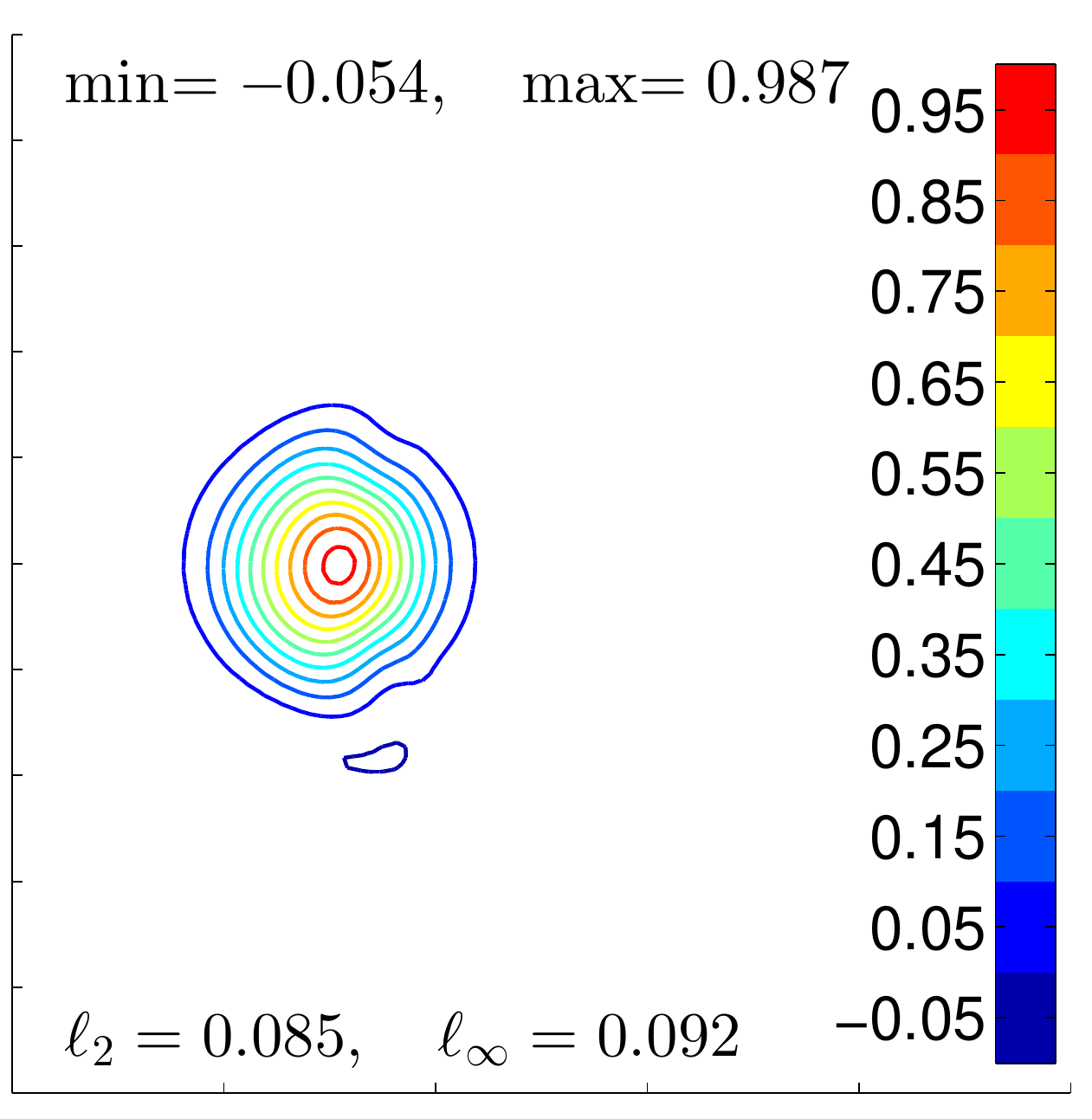}  &
\includegraphics[width=.3\textwidth]{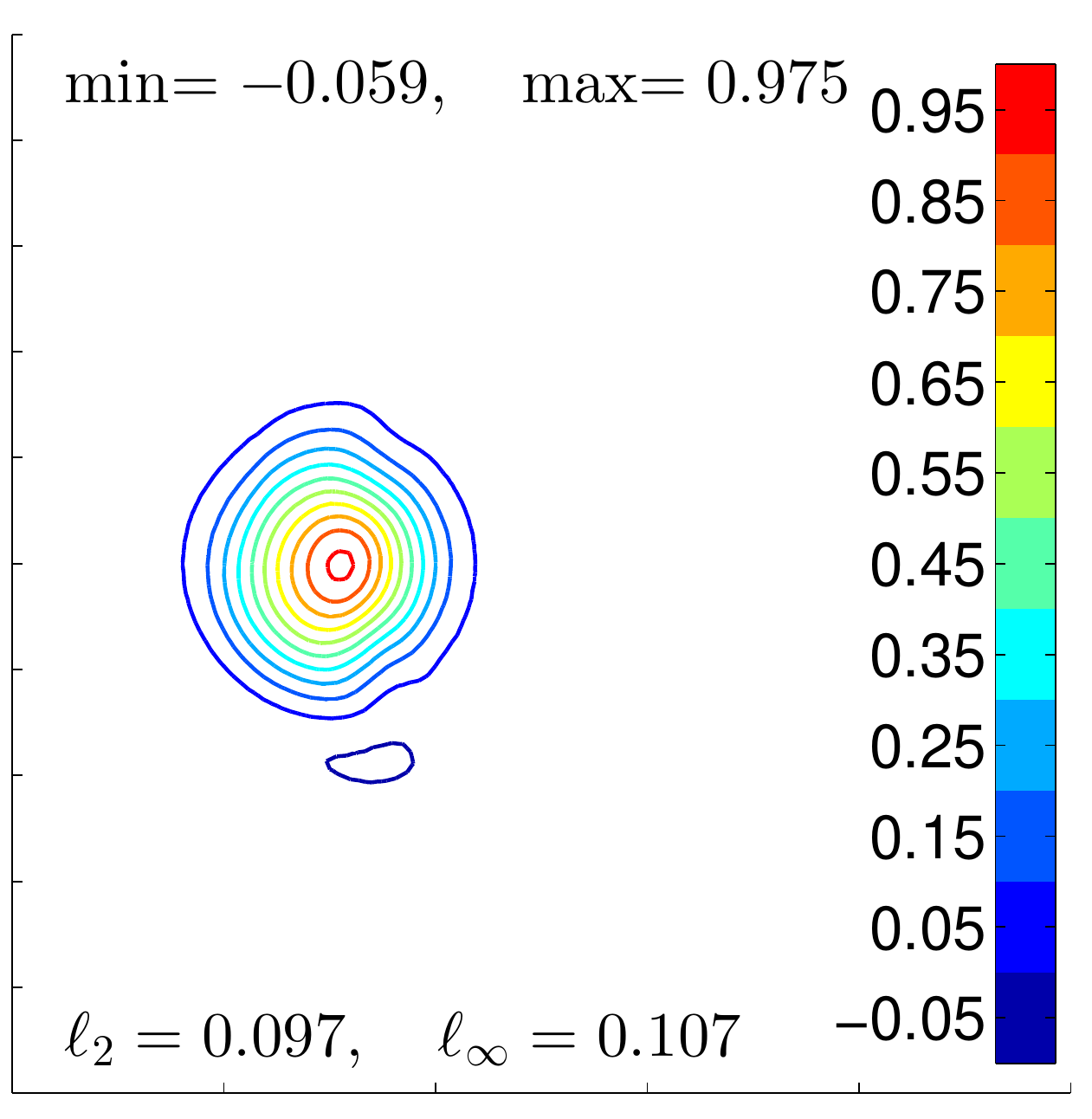} \\
\rotatebox{90}{~~~~~~~~~~\,$h = 0.005$} &
\includegraphics[width=.3\textwidth]{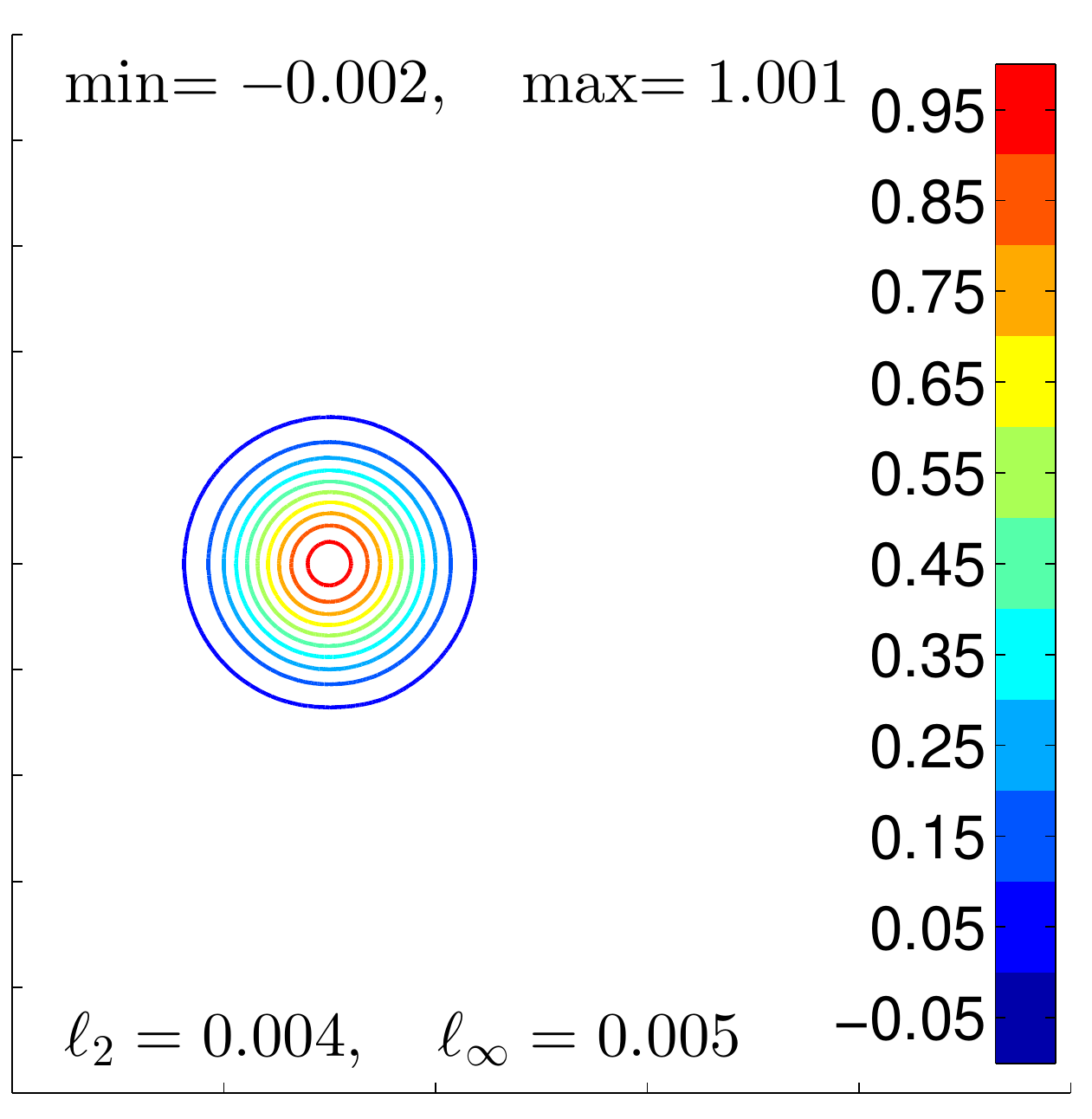} &
\includegraphics[width=.3\textwidth]{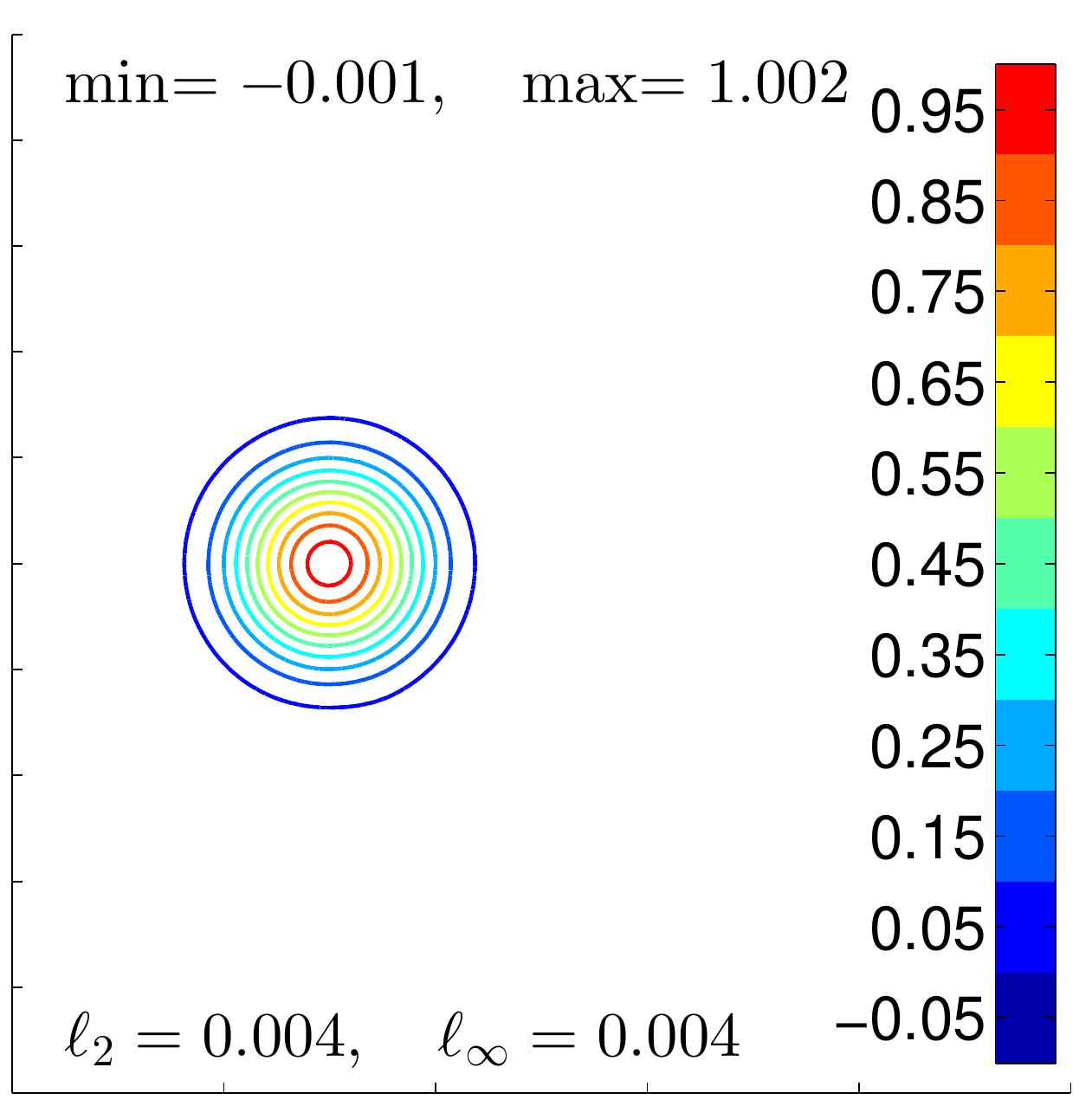}  &
\includegraphics[width=.3\textwidth]{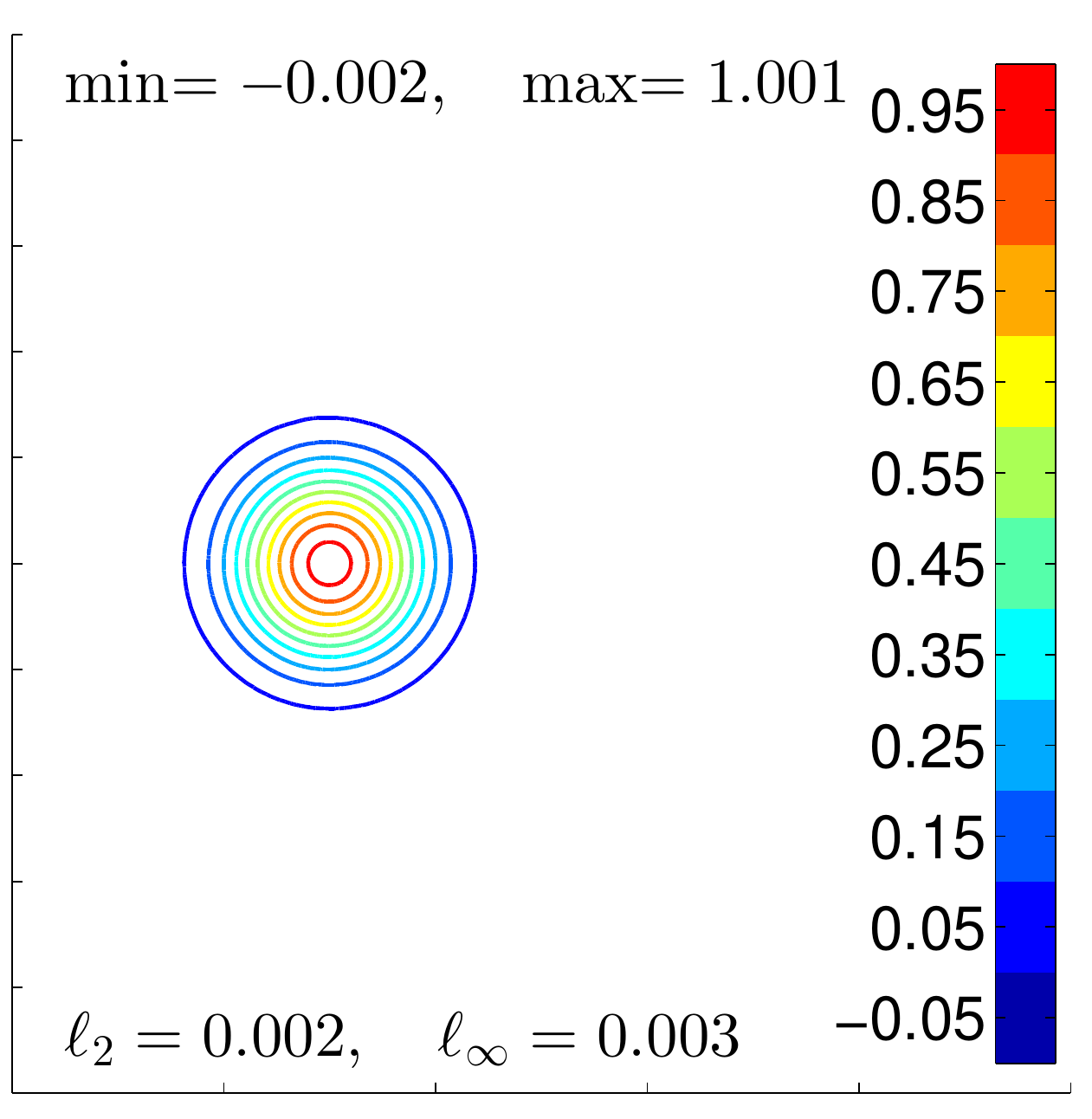} \\
\end{array}
$$
\caption{Plots of solutions at $t=T=1$ using the three different types of node-sets.  $\phi(r)=r^3$, polynomials up to degree $5$ on a $37$-node stencil and $\Delta^3$-type hyperviscosity were used.  The amount of hyperviscosity, $\gamma$, varies between node-sets, but is on the $O(10)^{-12}$ to $O(10)^{-14}$.}
\label{comp_BD_layouts}
\end{figure}

\begin{figure}[H]
\centering
\includegraphics[width=.5\textwidth]{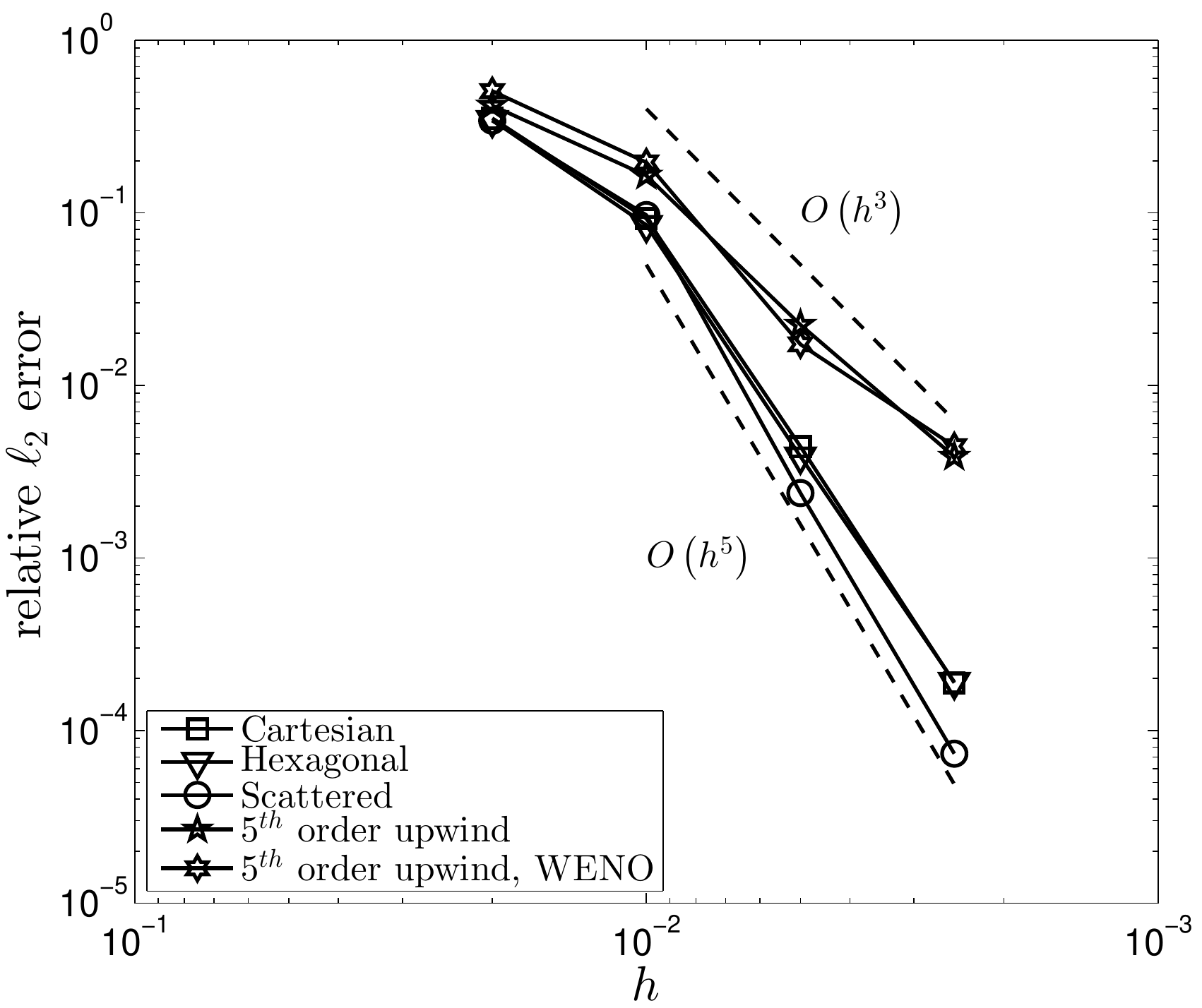}
\caption{Convergence plots of relative error $\left(\frac{\left\|\psi_{\text{approx}}-\psi_{\text{exact}}\right\|_2}{\left\|\psi_{\text{exact}}\right\|_2}\right)$ vs. $\left(h = 0.02, 0.01, 0.005, 0.0025\right)$.  In all cases, $\phi(r)=r^3$ with polynomials up to degree $5$ on a $37$-node stencil and $\Delta^3$-type hyperviscosity were used.  The error decreases $O\left(h^5\right)$, which is expected since up to $5^{\text{th}}$ degree polynomials were included and only first derivatives need to be approximated in the PDE.}
\label{fig:convergence_BD}
\end{figure}

For a 37 node stencil, the highest degree polynomials that can be used on all three node sets is fifth degree. Both hexagonal and scattered node sets can handle sixth degree polynomials but not Cartesian layouts. This is because on such a lattice the nodes approach non-unisolvency, resulting in the column vectors of the polynomial portion of the matrix in \eqref{weights} becoming linearly dependent. Also, in order to demonstrate that the polynomial degree controls the convergence and not the PHS order, $r^3$ PHS are now used instead of $r^9$ with up to fifth-degree polynomials on Cartesian, hexagonal and scattered nodes. Figure \ref{comp_BD_layouts} illustrates this for three resolutions, $h = 0.02, 0.01, 0.005$. For any given resolution, all nodes sets perform roughly the same, both with regard to the minimum and maximum function values and the errors in the $\ell_2$ and $\ell_\infty$ norm. Comparing the the maximum and minimum of the solution for hexagonal nodes on $h=0.005$ between Figures \ref{comp_BD_layouts} and \ref{time_series_BD}, it can been seen that using $r^9$ gives slightly better accuracy. This phenomena was noted in Section \ref{sec:saturation}.  Figure \ref{fig:convergence_BD} shows the convergence rate corresponding to the cases given in Figure \ref{comp_BD_layouts}. From the discussion in Section \ref{sec:saturation}, the convergence rate should be $O(h^l)$, where $l$ is the highest degree of polynomials used, in this case 5. Fifth-order convergence is indeed seen for all node sets in Figure \ref{fig:convergence_BD}. Also in this figure, the RBF-FD method is compared to a 5th-order upwind scheme with and without a WENO limiter, both of the latter exhibiting a third-order convergence rate. The reason for the comparison is that this order FD upwinding scheme is the type used in the Weather Research and Forecasting (WRF) Model (http://www.wrf-model.org/). The time steps for both methods are comparable, with less than a $1\%$ difference.

It should be remembered that this test case was set up to investigate what the numerical results for the proposed RBF-FD method would be under no boundary effects. So, if there is the unusual circumstance that the solution does not interact with any boundaries in a bounded domain \textit{and} no refinement will be needed, then solving the problem on a Cartesian lattice will give just as good results as hexagonal. Furthermore, the fact that scattered nodes performed just as well as the other two layouts is of great benefit since it paves the way for the ability to implement local node refinement.

\subsection{Governing Equations for Navier-Stokes test cases}

In all test cases below, the set of governing equations is the 2D nonhydrostatic compressible Navier-Stokes equations at low Mach number, M$\approx 0.1$, in a rectangular or square domain. The equations are given by

\begin{align*}
\frac{\partial u}{\partial t}=&-u\frac{\partial u}{\partial x}-w\frac{\partial u}{\partial z}-c_p\theta\frac{\partial\pi}{\partial x}+\mu\Delta u,\\
\frac{\partial w}{\partial t}=&-u\frac{\partial w}{\partial x}-w\frac{\partial w}{\partial z}-c_p\theta\frac{\partial\pi}{\partial z}-g+\mu\Delta w,\\
\frac{\partial\theta}{\partial t}=&-u\frac{\partial\theta}{\partial x}-w\frac{\partial\theta}{\partial z}+\mu\Delta \theta,\\
\frac{\partial\pi}{\partial t}=&-u\frac{\partial\pi}{\partial x}-w\frac{\partial\pi}{\partial z}-\frac{R_d}{c_v}\pi\left(\frac{\partial u}{\partial x}+\frac{\partial w}{\partial z}\right),
\end{align*}
where $u$ and $w$ are the velocities in the horizontal and vertical directions, respectively, $\pi=\left(\frac{P}{P_0}\right)^{R_d/c_p}$ is the non-dimensional Exner pressure $\left(P_0=1\times10^5\text{ Pa}\right)$, and $\theta=\frac{T}{\pi}$ is the potential temperature.  The constants $c_p=1004$ and $c_v=717$ are the specific heat at constant pressure and the specific heat at constant volume, respectively, with the gas constant for dry air being $R_d=c_p-c_v=287$.  Additional parameters are $g=9.81 m/s^2$ , the gravitational constant, and $\mu$, the dynamic viscosity. Furthermore, it is assumed that all quantities to be solved for, $\left[u,w,\theta,\pi\right]^{T}$, are perturbations $(')$ to a background state $(\,\bar{}\,)$ that is in hydrostatic balance, i.e. the fluid is initially at rest, $\overline{u}=\overline{w}=0$, and the background Exner pressure is a linear function of height z, ${\displaystyle \frac{d\overline{\pi}}{dz}=-\frac{g}{c_{p}\overline{\theta}}}$. Substituting this latter relation into the equations above and writing $\theta = \overline{\theta} + \theta'$ and $\pi = \overline{\pi} + \pi'$ (where the $(')$ symbol has been dropped for reading clarity) yields the governing equations to be used for computation:
\begin{eqnarray}
\frac{\partial u}{\partial t} &=& -u\frac{\partial u}{\partial x} - w\frac{\partial u}{\partial z} - c_p\left(\overline{\theta}+\theta\right)\frac{\partial\pi}{\partial x} + \mu\Delta u, \label{uEq}\\[1ex]
\frac{\partial w}{\partial t} &=& -u\frac{\partial w}{\partial x} - w\frac{\partial w}{\partial z} - c_p\left(\overline{\theta}+\theta\right)\frac{\partial\pi}{\partial z} + \frac{g\theta}{\overline{\theta}} + \mu\Delta w, \label{wEq}\\[1ex]
\frac{\partial\theta}{\partial t} &=& -u\frac{\partial\theta}{\partial x} - w\frac{\partial\theta}{\partial z} + \mu\Delta\theta,\label{thetaEq}\\[1ex]
\frac{\partial\pi}{\partial t} &=& -u\frac{\partial\pi}{\partial x} - w\left(\frac{d\overline{\pi}}{dz}+\frac{\partial\pi}{\partial z}\right) - \frac{R_d}{c_v}\left(\overline{\pi}+\pi\right)\left(\frac{\partial u}{\partial x}+\frac{\partial w}{\partial z}\right).
\label{piEq}
\end{eqnarray}
In the following studies, the perturbation notation $(')$ is generally included when reporting results to keep in mind these are perturbation quantities.

\subsection{Numerical Set-up for the NS cases}

The governing equations \eqref{uEq}-\eqref{piEq} are solved numerically using a method-of-lines (MOL) approach. PHS RBFs, $\phi(r)=r^7$, with polynomials up to fourth degree are used on a stencil-size of $n=37$ to approximate all spatial derivatives locally. The remaining system of first order ODEs in time is solved with RK4. A $\Delta^3$-type hyperviscosity is applied in all cases to damp high-frequency modes.
The time-step for all test cases as a function of resolution is
\begin{table}[H]
\centering
\begin{tabular}{r|llllll}
node-spacing ($h$):   &   800m   &   400m   &   200m             &   100m             &   50m              &   25m                 \\
\hline
$\Delta t$:           &   2s     &   1s     &   $\frac{1}{2}$s   &   $\frac{1}{4}$s   &   $\frac{1}{8}$s   &   $\frac{1}{16}$s      \\
\end{tabular}
\label{tbl:timestep}
\end{table}
Table \ref{tbl:nonlin} below gives the domain size, and number of nodes used as a function of resolution for the numerical studies of the NS test cases.
\begin{table}[H]
\caption{Information regarding the computational domain for each test case.  The number of nodes ($N$) is for hexagonal nodes.}
\centering
\begin{tabular}{l l r}
\hline\hline
Test (domain size ($x \times z$ in km))                              & Resolution (m) & $\approx N$     \\ [0.5ex]
\hline
Straka Density Current \cite{Straka} ($[-25.6,25.6] \times [0,6.4]$) & 800            & 720     \\
                                                                     & 400            & 2,700   \\
                                                                     & 200            & 10,000  \\
                                                                     & 100            & 38,500  \\
                                                                     & 50             & 152,650 \\ [0.5ex]
\hline
Translating Density Current\cite{Wicker02} ($[0,36] \times [0,6.4]$)   & 800            & 500     \\
                                                                     & 400            & 1,900   \\
                                                                     & 200            & 7,040   \\
                                                                     & 100            & 27,040  \\
                                                                     & 50             & 107,350 \\ [0.5ex]
\hline
Rising Thermal Bubble ($[0,10] \times [0,10]$)                       & 200            & 2,980   \\
                                                                     & 100            & 11,760  \\
                                                                     & 50             & 46,720  \\
                                                                     & 25             & 185,430 \\ [0.5ex]
\hline
\end{tabular}
\label{tbl:nonlin}
\end{table}

\subsection{Density Current}\label{densityCurrent}

\begin{figure}[H]
$$
\begin{array}{c}
\includegraphics[width=0.8\textwidth]{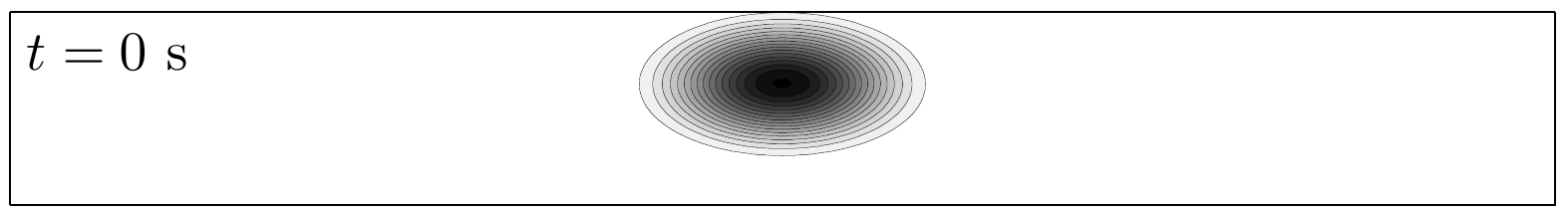}   \\
\includegraphics[width=0.8\textwidth]{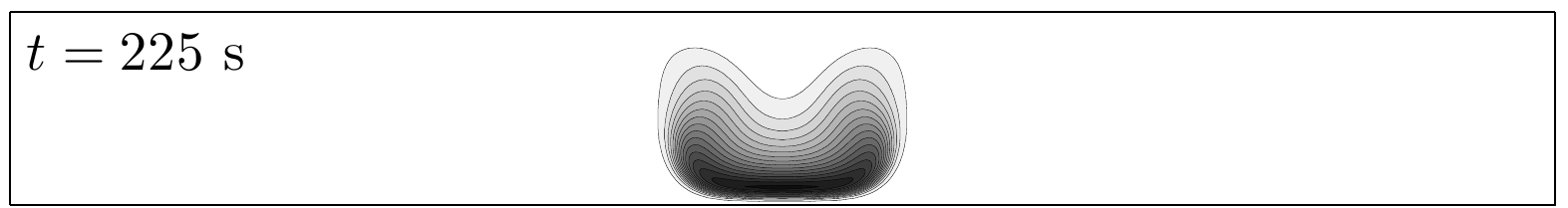} \\
\includegraphics[width=0.8\textwidth]{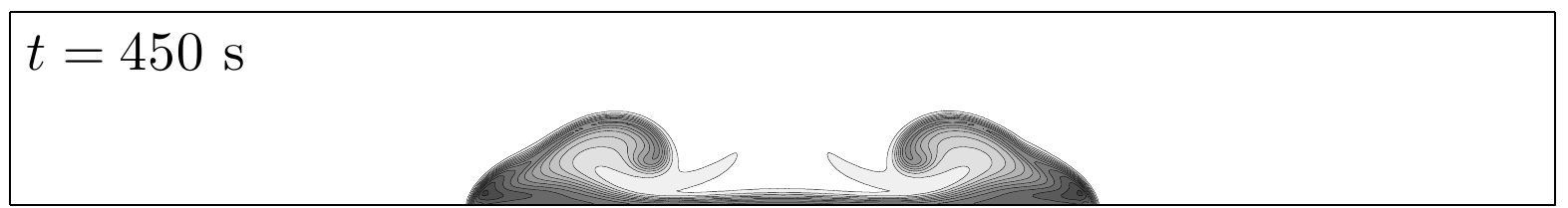} \\
\includegraphics[width=0.8\textwidth]{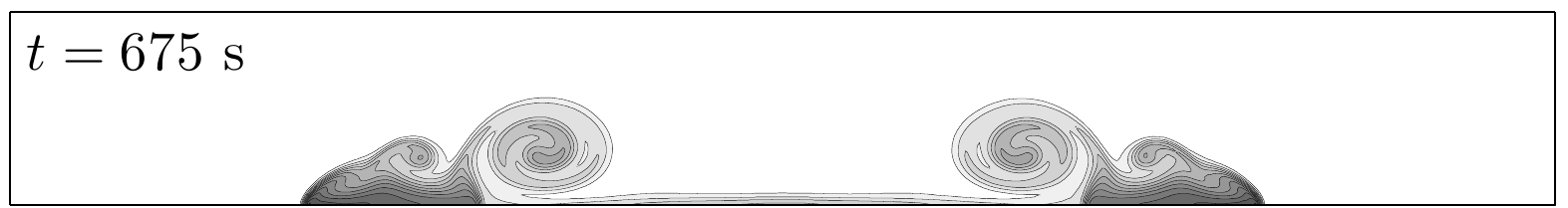} \\
\includegraphics[width=0.8\textwidth]{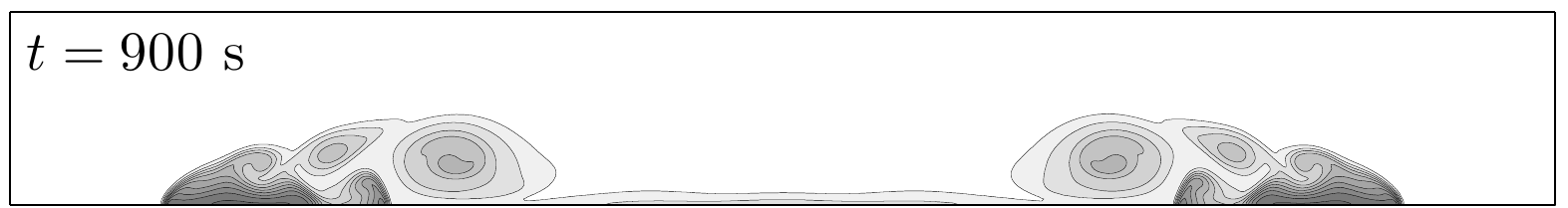}
\end{array}
$$
\caption{Time evolution of the potential temperature perturbation $\theta'$ for the density current test case with $\mu = 75 m^2/s$ at a 100m resolution on hexagonal nodes.}
\label{fig:strakaTime}
\end{figure}

In the density current test case \cite{Straka}, a hydrostatic neutral atmosphere is perturbed by a $C^1$ bubble in the potential temperature. A mass of cold air falls to the ground and develops three smooth and distinct Kelvin-Helmoltz rotors as it spreads to the sideways. This test has become widely used in weather modeling community for assessing the ability in new numerical schemes to capture the physics in nonhydrostatic fluid flows \cite{Ooyama,Giraldo08,Skamarock08,NairFV}. Figure \ref{fig:strakaTime} shows the behavior of the numerical solution in time from $t=0$s until the final time, $t=900$s.

The computational domain is $\left[-25.6,25.6\right]\times\left[0,6.4\right]\text{ km}^2$, and the governing equations \eqref{uEq}-\eqref{piEq} are solved with a viscosity of $\mu=75\text{ m}^2/\text{s}$.

\vspace{0.1in}
\textbf{Define $\overline{\theta}$ and $\overline{\pi}$}
\vspace{0.1in}

Let $\overline{T}(z)=T_s-\frac{g}{c_p}z$ be the background state for temperature, where $T_s=300$ is the temperature at the ground surface in Kelvin. Then, the background states for potential temperature and Exner pressure are given by
\begin{align*}
\overline{\theta}=T_s, \quad\quad \overline{\pi}(z)=\frac{\overline{T}(z)}{\overline{\theta}}=1-\frac{g}{c_pT_s}z.
\end{align*}

\vspace{0.1in}
\textbf{Define initial conditions}
\vspace{0.1in}

The vector of unknowns is initially zero except for the potential temperature.
\begin{align*}
\left.u\right|_{t=0}=0, \quad \left.w\right|_{t=0}=0, \quad \left.\pi'\right|_{t=0}=0.
\end{align*}
The $\left(C^1\right)$ initial condition for $\theta'$ is derived via a cool cosine bubble in the temperature $T$ defined by
\begin{eqnarray}
\left.T'\right|_{t=0}=&\left\{\begin{array}{rl}-\frac{15}{2}\left\{1+\cos\left[\pi_cr(x,z)\right]\right\}, & \quad r(x,z)\le 1, \\ 0, & \quad r(x,z)>1,\end{array}\right.\label{Tprime}
\end{eqnarray}
where $\pi_c=3.14159\ldots$ is the standard trigonometric constant and
\begin{align*}
r(x,z)=\sqrt{\left(\frac{x-x_c}{x_r}\right)^2+\left(\frac{z-z_c}{z_r}\right)^2}, \quad\quad\quad \begin{array}{c}\left(x_c,z_c\right)=\left(0\text{ km},3\text{ km}\right),\\\left(x_r,z_r\right)=\left(4\text{ km},2\text{ km}\right).\end{array}
\end{align*}
Then, the initial condition for $\theta'$ can be found by dividing by $\overline{\pi}$:
\begin{align*}
\left.\theta'\right|_{t=0} = \left[\theta-\overline{\theta}\right]_{t=0} = \left.\frac{T}{\pi}\right|_{t=0}-T_s = \left.\frac{\overline{T}+T'}{\overline{\pi}}\right|_{t=0}-T_s = \left[\overline{\theta}+\frac{T'}{\overline{\pi}}\right]_{t=0}-T_s = \frac{\left.T'\right|_{t=0}}{\overline{\pi}}.
\end{align*}

\vspace{0.1in}
\textbf{Define boundary conditions}
\vspace{0.1in}

The problem is periodic in the $x$ direction with the following conditions on the top and bottom boundaries in $z$:
\begin{align*}
w'=0~, \quad\quad \frac{\partial u'}{\partial z}=0, \quad\quad \frac{\partial\theta'}{\partial z}=0.
\end{align*}
These are the only boundary conditions necessary to solve the governing equations.  However, enforcing the vertical momentum equation \eqref{wEq} on the top and bottom boundaries and assuming that perturbation in the pressure gradient balances the perturbation in the potential temperature leads to the following condition for $\pi'$,
\begin{align*}
\frac{\partial\pi'}{\partial z}=\frac{g\theta'}{c_p\overline{\theta}\left(\overline{\theta}+\theta'\right)}.
\end{align*}
Furthermore, since the dynamic viscosity $\mu$ for air is $\approx 10^{-5}$, a good approximation on the top and bottom boundaries is $\Delta w'=0$ or $\partial^2w'/\partial z^2=0$ since $w=0$ on these boundaries. While these two extra boundary conditions on $\pi'$ and $w'$ conditions are not required, they allow for the use of ghost nodes in all four variables. In summary, the lateral boundaries are periodic, and the complete set of boundary conditions enforced on the top and bottom boundaries is:
\begin{align*}
w=\frac{\partial^2w}{\partial z^2}=\frac{\partial u}{\partial z}=\frac{\partial\theta'}{\partial z}=0, \quad\quad \frac{\partial\pi'}{\partial z}=\frac{g\theta'}{c_p\overline{\theta}\left(\overline{\theta}+\theta'\right)}.
\end{align*}


\begin{figure}[H]
\centering
\includegraphics[width=.5\textwidth]{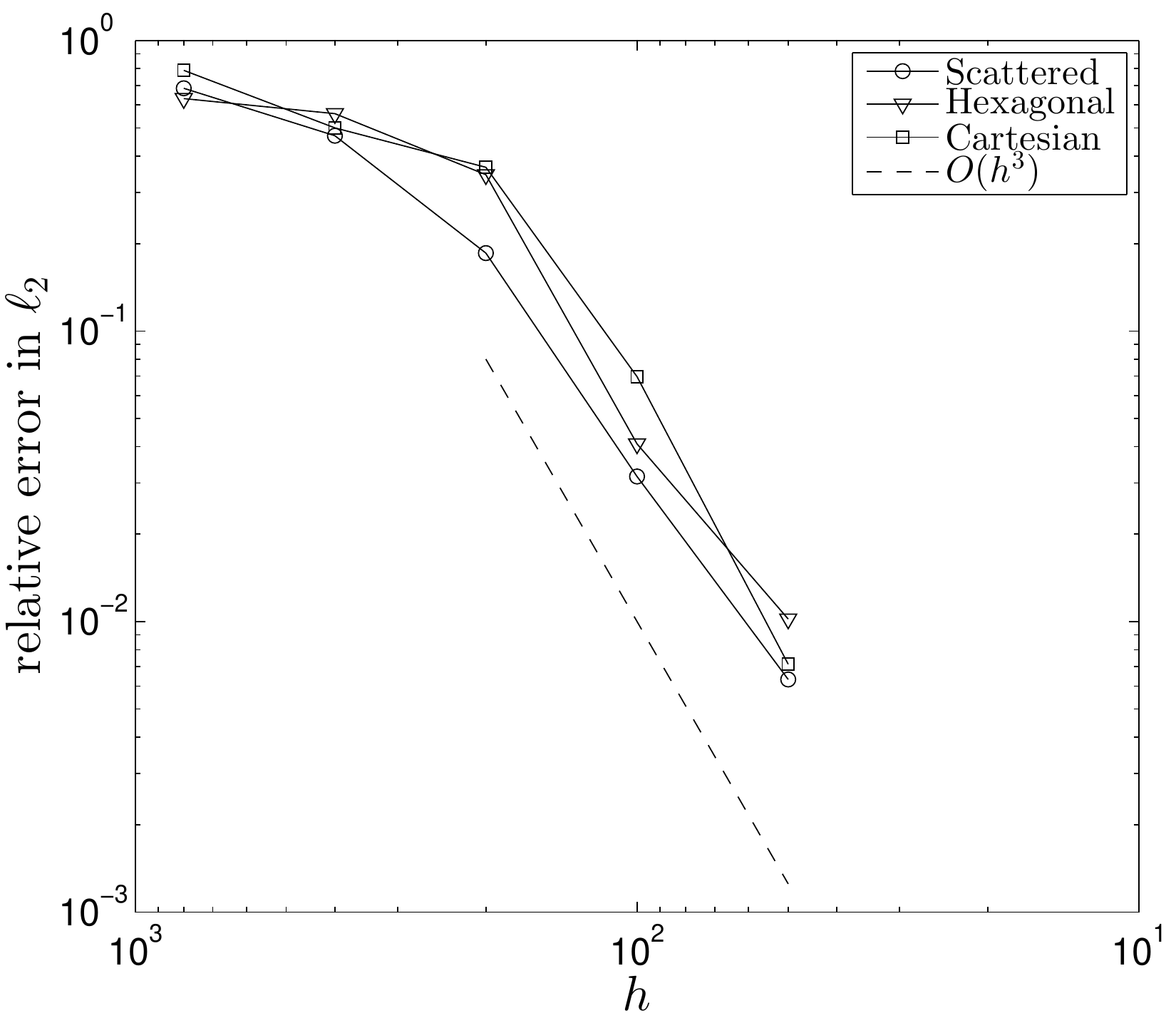}
\caption{Convergence behavior for $\theta'$ in the Straka density current test case with $\mu = 75\, m^2/s$.   The $h=$800m, 400m, 200m, 100m, and 50m errors were calculated using the 25m RBF-FD reference solution.}
\label{convergence_straka}
\end{figure}

\begin{figure}[H]
$$
\begin{array}{cccc}
\text{} & \text{Cartesian} & \text{Hexagonal} & \text{Scattered} \\
\rotatebox{90}{~~~\,800~m} &
\includegraphics[width=.3\textwidth]{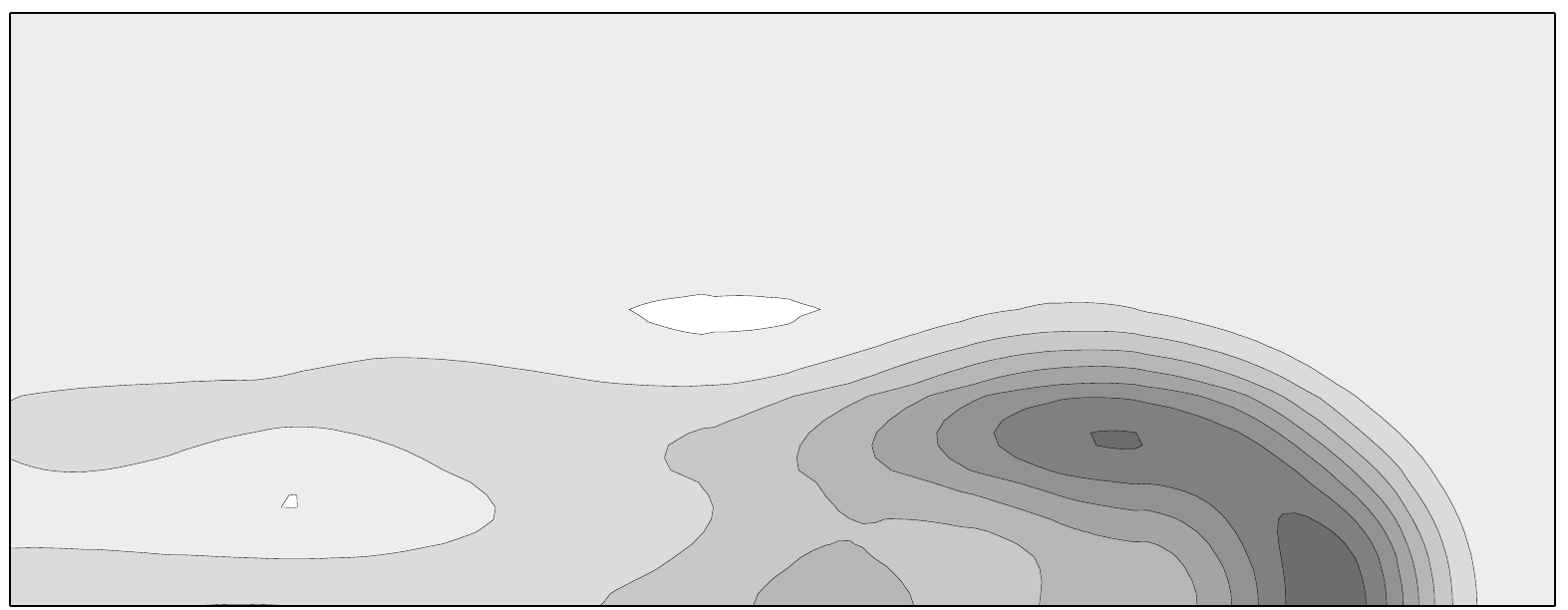} &
\includegraphics[width=.3\textwidth]{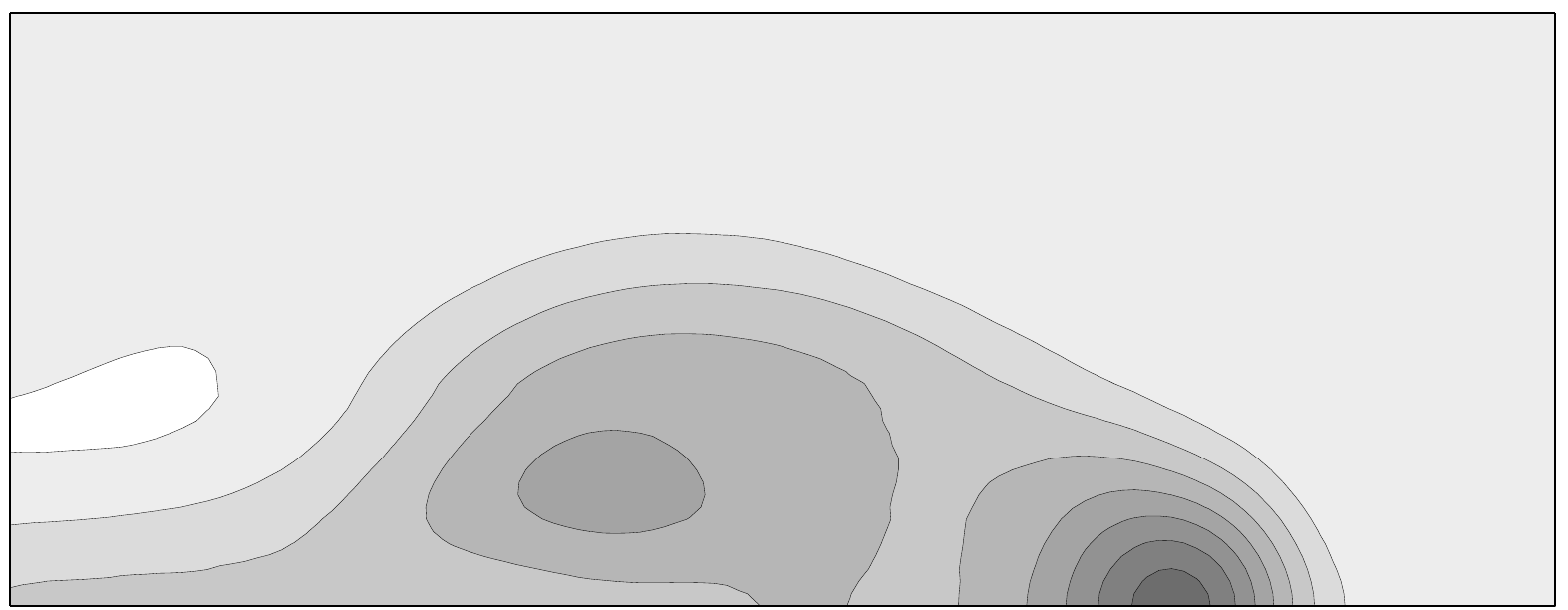} &
\includegraphics[width=.3\textwidth]{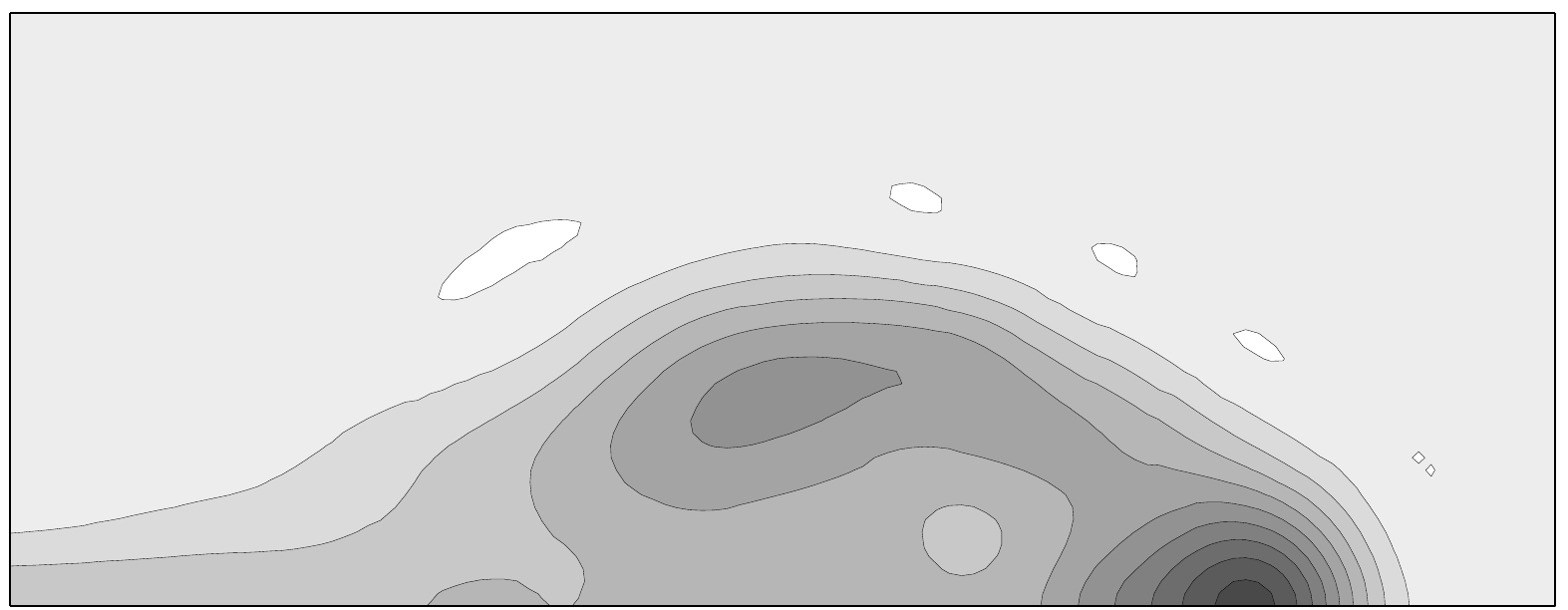} \\
\rotatebox{90}{~~~\,400~m} &
\includegraphics[width=.3\textwidth]{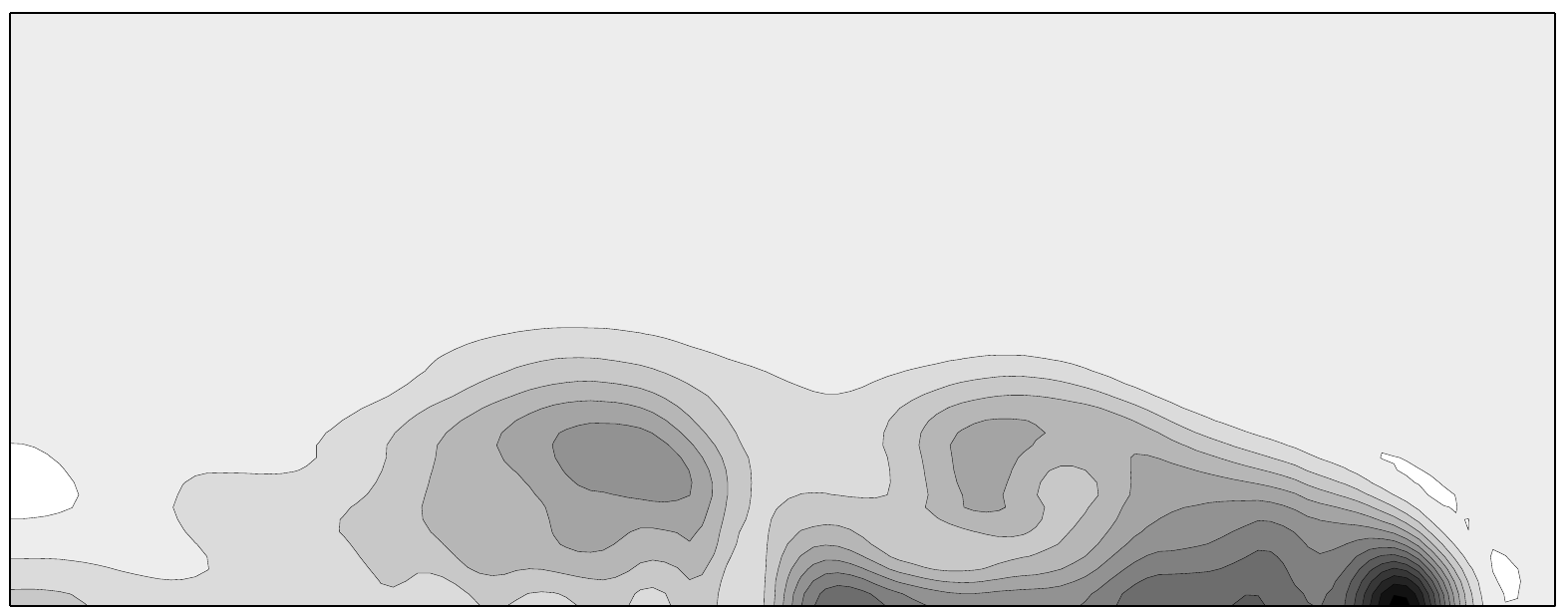} &
\includegraphics[width=.3\textwidth]{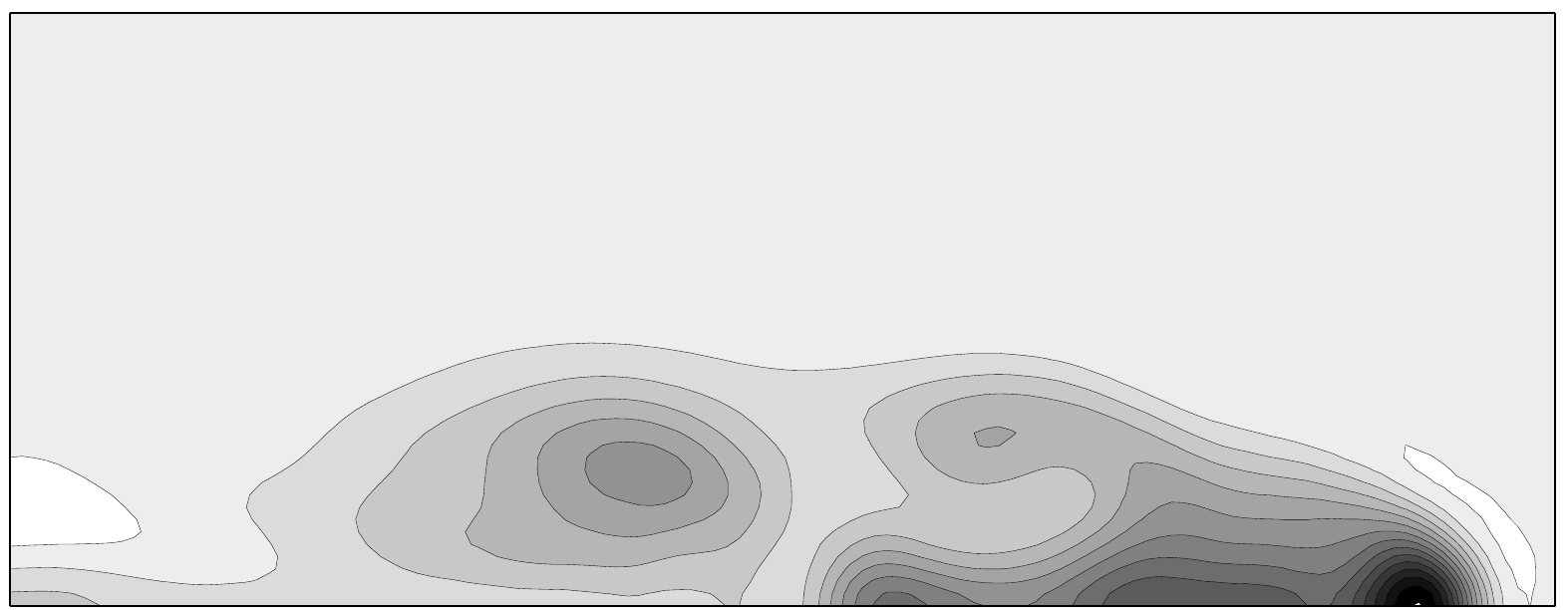} &
\includegraphics[width=.3\textwidth]{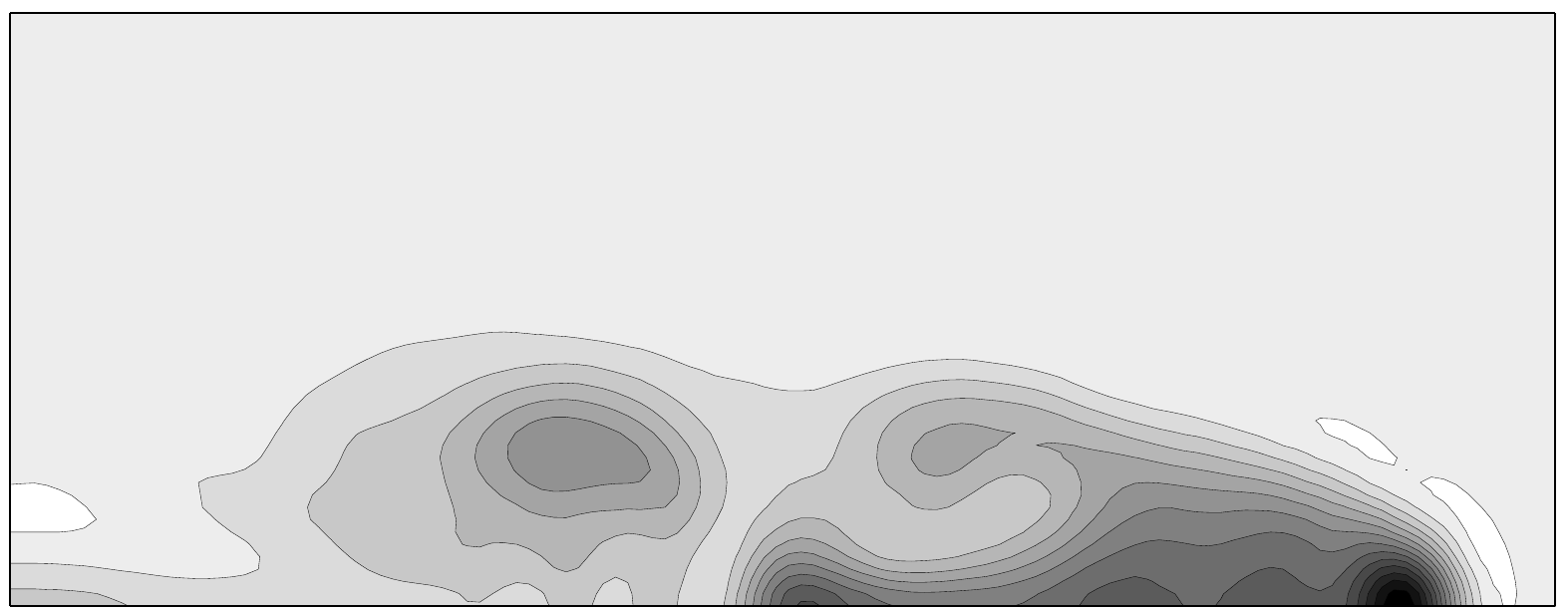} \\
\rotatebox{90}{~~~\,200~m} &
\includegraphics[width=.3\textwidth]{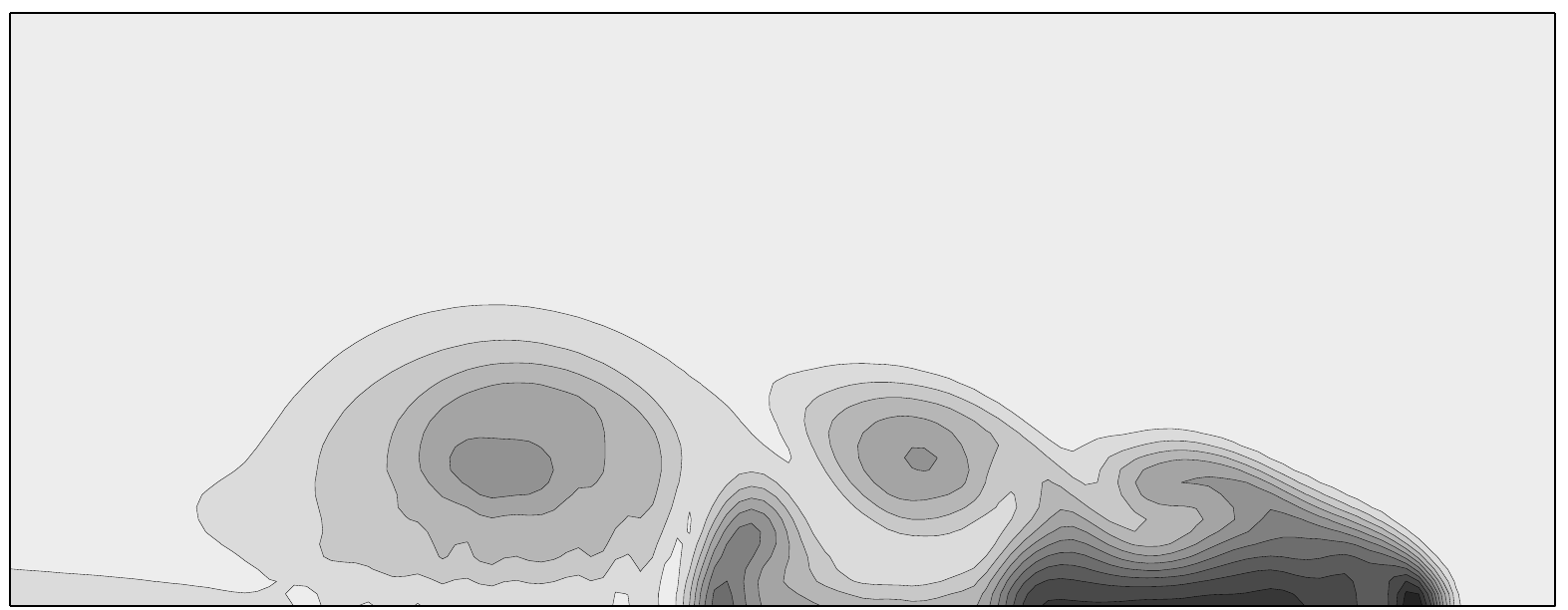} &
\includegraphics[width=.3\textwidth]{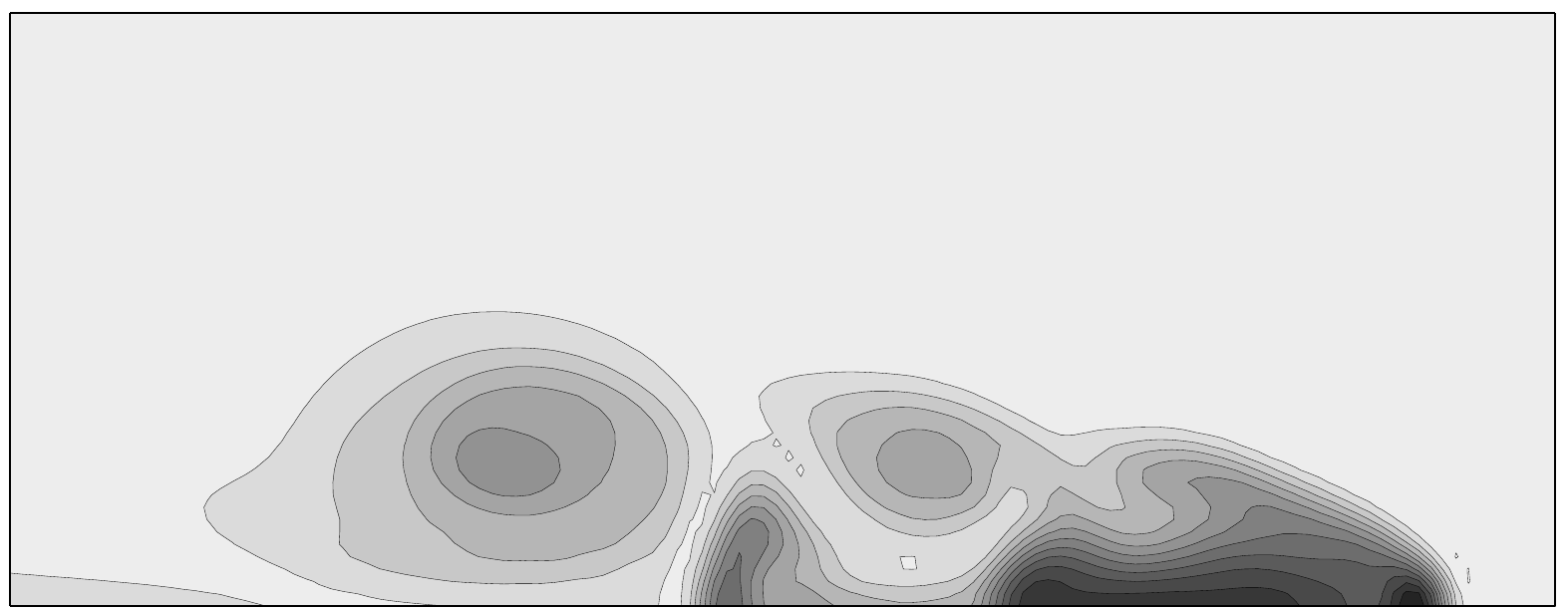} &
\includegraphics[width=.3\textwidth]{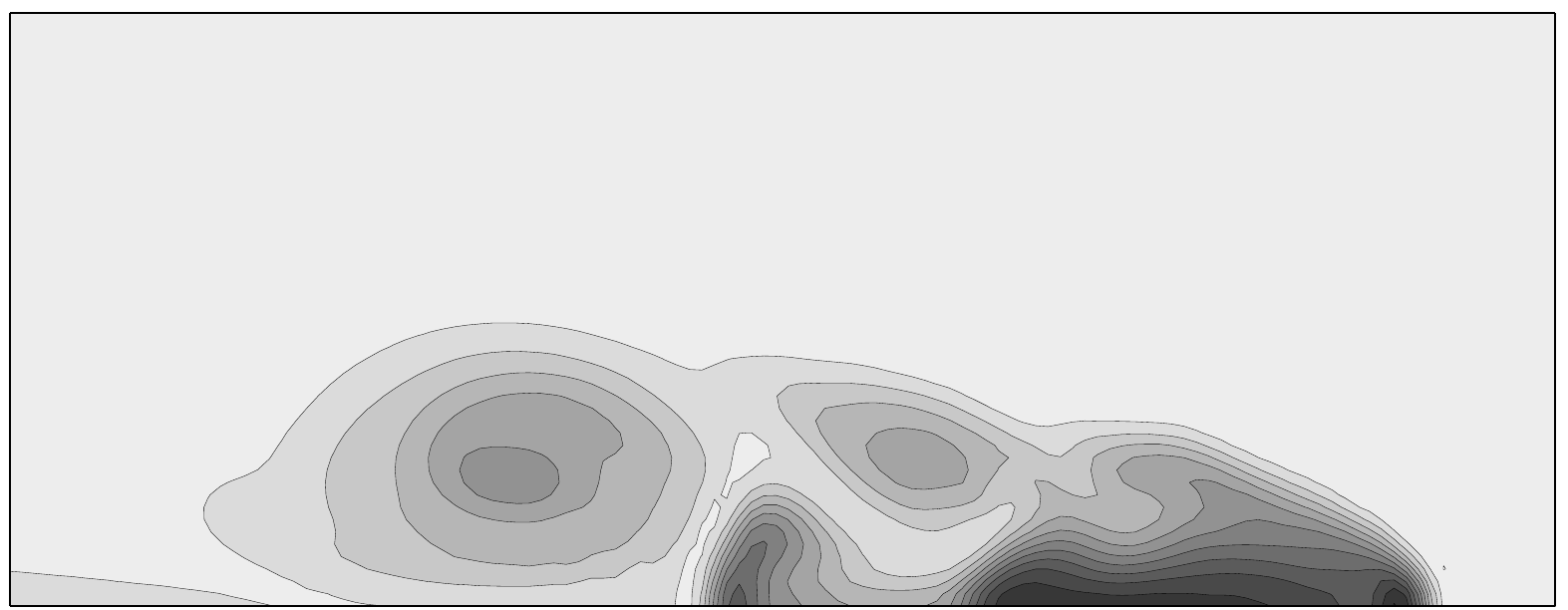} \\
\rotatebox{90}{~~~~~~~~100~m} &
\includegraphics[width=.3\textwidth]{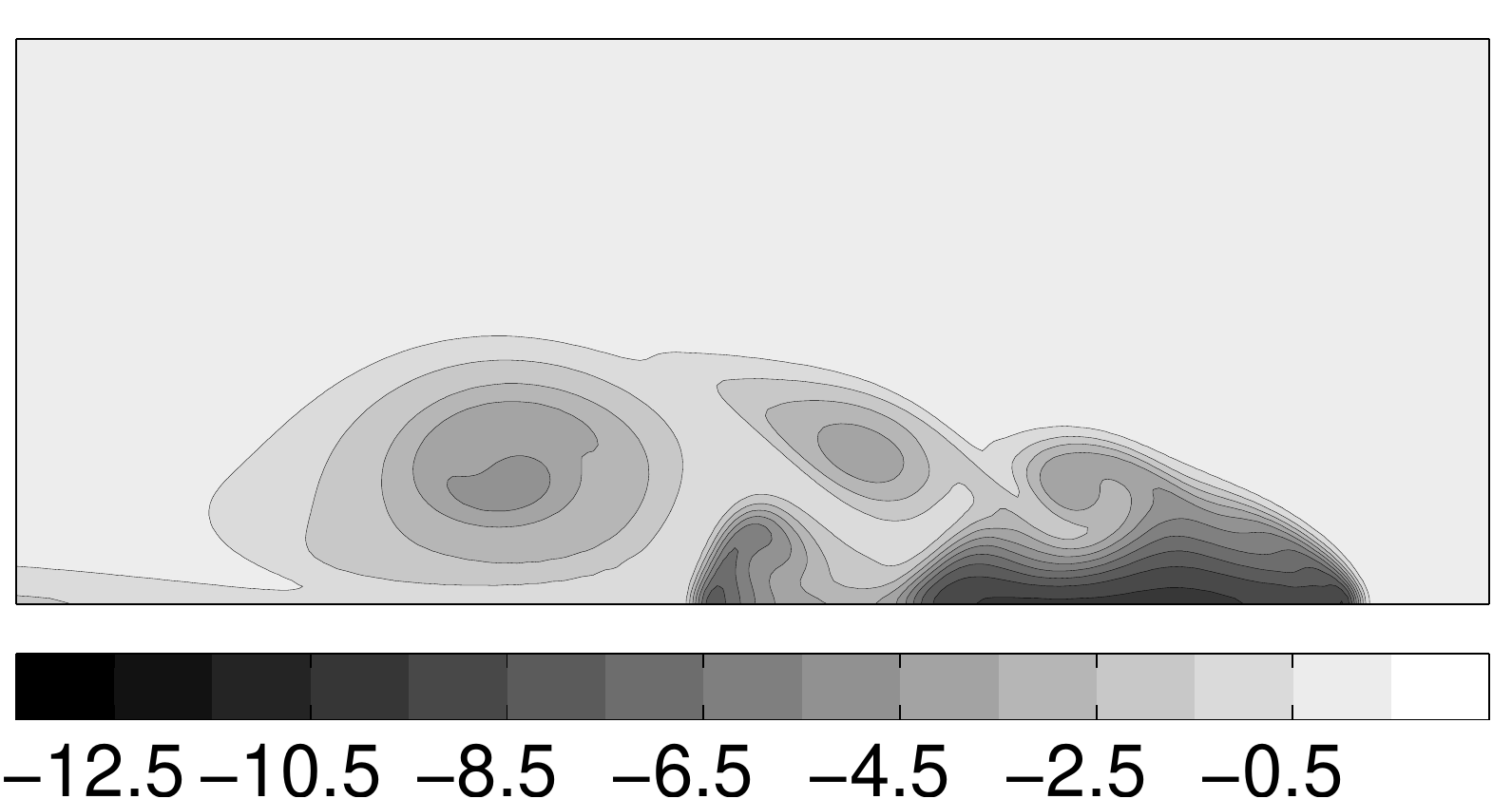} &
\includegraphics[width=.3\textwidth]{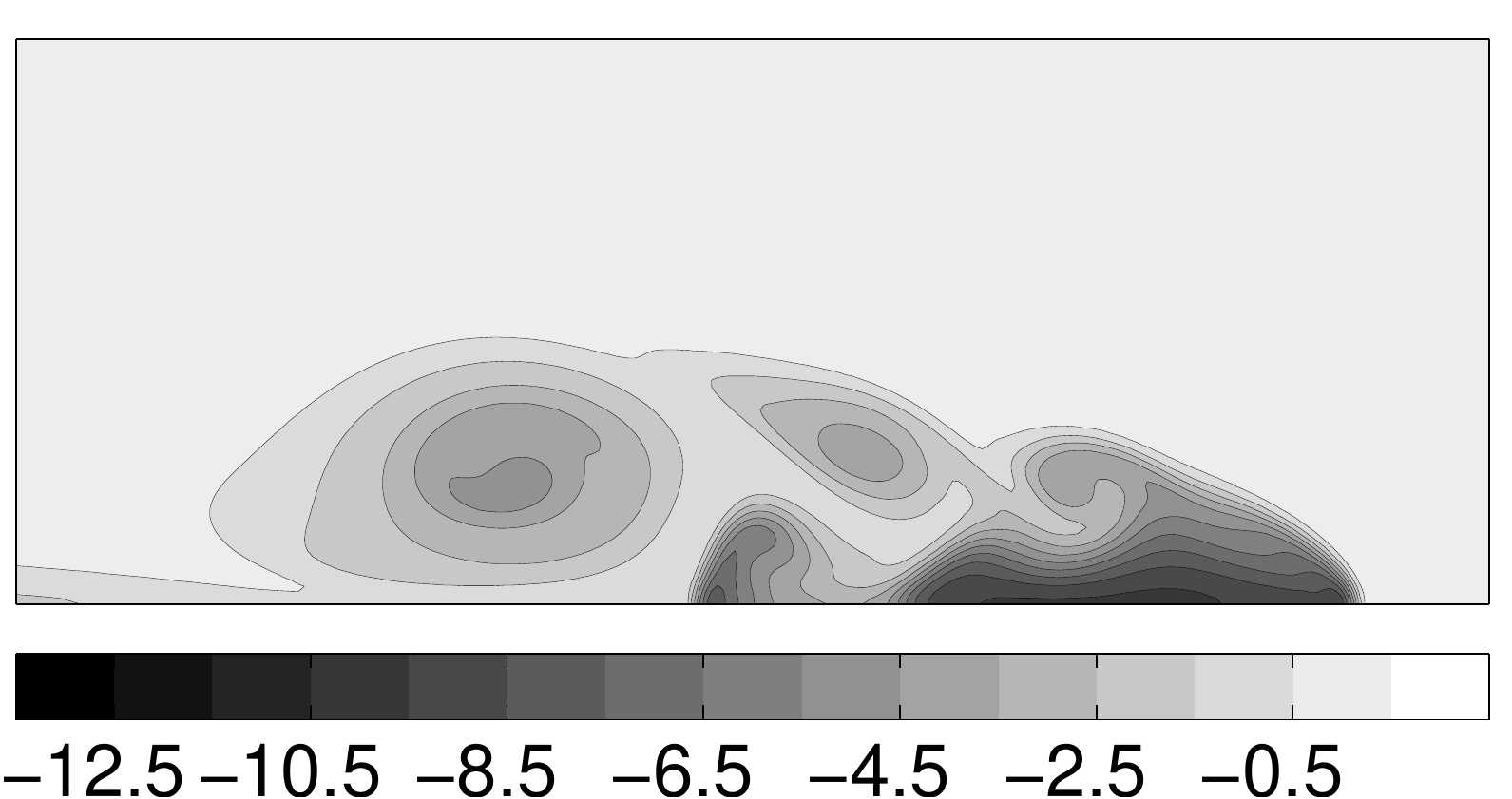} &
\includegraphics[width=.3\textwidth]{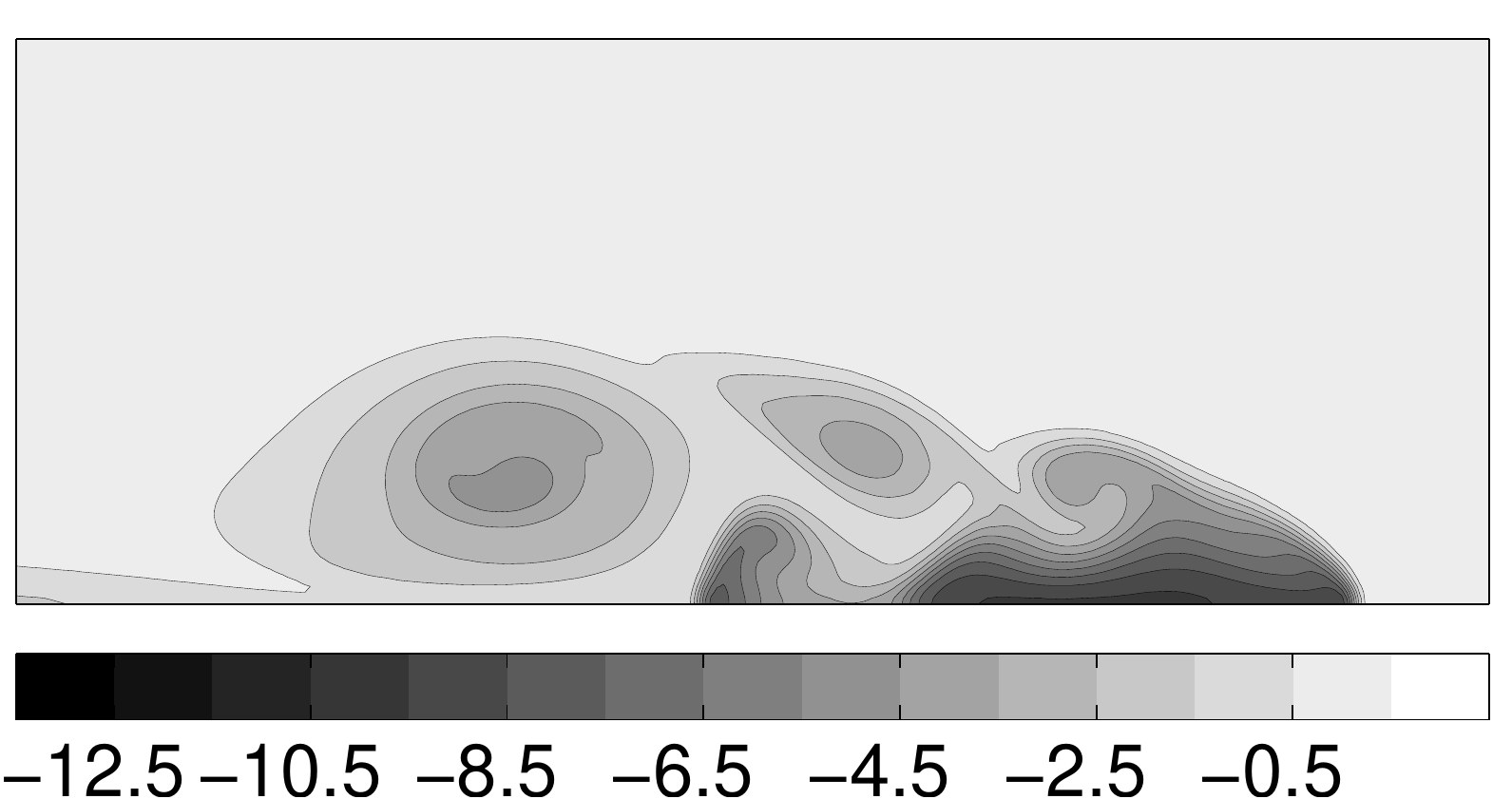} \\
\end{array}
$$
\caption{The solution at the final time for the density current test case solved on different node layouts for resolutions 800m, 400m, 200m, 100m. Only half the solution is shown to enlarge details. Contours for the density current begin at -0.5K and are in intervals of 1K. The white areas are enclosed by a contour of 0.5K.}
\label{fig:straka_nodes}
\end{figure}


As can be seen in Figure \ref{convergence_straka}, the error in the $\ell_2$ norm is approximately the same regardless of the node layout with a third-order convergence rate for node resolutions 200m and less. However, convergence rates do not illustrate how the physics is being resolved with regard to where the data is sampled (i.e. in terms of the node layout). As a result, in Figure \ref{fig:straka_nodes}, the solution of the density current test case is given on the three different node layouts discussed in Section \ref{nodeSets} for four different resolutions, varying from 800m to 100m. In the highest resolution displays (100m), although all node layouts seemed to have converged to the same solution, differences can be noted in Table \ref{tbl:straka}, where a 25m test run on hexagonal nodes is used as a reference solution. Results from a 50m test run on the 3 different node sets are also given in the table. On hexagonal and scattered nodes the minimum $\theta'$ has indeed converged by 100m, while the maximum $\theta'$ is still a fifth of a degree off. Similar error percentages can be found in $w'$  and in the front location at these fine resolutions of 100m and 50m, noting that Cartesian nodes perform slightly worse. Notice that the front for the 100m Cartesian is at the same location as achieved by a 200m hexagonal layout.

At coarser resolutions, such as 400m and 800m, values in the table will be far off the converged 25m solution. Instead, noting physical features of the solution in Figure \ref{fig:straka_nodes}, such as 1) at what resolution do the rotors begin to form, their shape and where, and 2) how much cold air has been entrenched in each rotor, will give a better idea of the capability of the node layout to capture the physics. The following observations can be made:

\begin{enumerate}
\item At 800m - approximately 720 nodes in the domain: The hexagonal and scattered node calculations give more clear evidence of the first (largest) rotor being formed. The -3.5K contour in the hexagonal case (circular inner most contour in the first rotor) is even close to its final position if compared to the 100m case. Although the first rotor for the scattered case is not quite as nicely formed as in the hexagonal case, it has entrenched more cold air, having a -4.5K contour (teardrop shape). Notice that at 100m the -4.5K contour is the coldest that appears. In comparison, the 800m Cartesian has barely any rotor formation and is much more wildly oscillatory, which is also noted by the fact that the maximum $\theta'$ is 2.43K, at least 1.3K larger than for the other node layouts. See Table \ref{tbl:straka}. Also note in the table that the error in the front location decreases from $4\%$ to $2\%$, when hexagonal nodes are used opposed to Cartesian, with scattered given an intermediate error of $3\%$. Both hexagonal and scattered nodes undershoot the correct position while Cartesian overshoot it.
 \item At 400m - approximately 2700 nodes in the domain: At this resolution, oscillations due to boundary error effects especially in the first rotor are very pronounced on the Cartesian layout; this carries over even to the 200m resolution for this node case. For scattered nodes there are minor oscillations in the solution. For the hexagonal case, barely any are evident. Formation of the second rotor has the nicest intact shape with the least amount of oscillation in both the hexagonal and scattered, with the latter having entrenched a slightly larger amount of cold air (notice the size of the -3.5K contour teardrop-shaped area in the second rotor of the scattered case).
\end{enumerate}

The differences between the columns of subplots reflect only the intrinsic resolution capabilities of the different node layouts for capturing the physics. The traditional Cartesian choice is the least effective one. At every resolution level, the hexagonal and scattered choices give better accuracy than the Cartesian one. The advantage of generalizing from hexagonal to quasi-uniformly scattered nodes, is that it then becomes easy to implement spatially variable node densities, i.e. to do local refinement in select critical areas. It is very important to note that this major increase in geometric flexibility (from hexagonal to quasi-uniformly scattered) hardly has any negative effect at all on the accuracy that is achieved, nor on the algorithmic complexity of the code.

\begin{table}[H]
\centering
\caption{Resolution $\left(h\right)$, minimum and maximum values for $\theta'$ and $w$, and front location at various resolutions for the density current test case with $\mu = 75 m^2/s$. The front location was determined by the $-0.5\text{K}$ contour line.}
\begin{tabular}{c|c|c|c|c|c|c}
            &   h (m)   &   $\min\left\{\theta'\right\}$   &   $\max\left\{\theta'\right\}$   &   $\min\{w'\}$  &   $\max\{w'\}$   &   front (m)   \\
\hline\hline
Cartesian   &   800     &   -7.74                          &   2.43                           &   -9.19         &   11.00         &   16,079      \\
            &   400     &   -13.45                         &   1.10                           &   -15.21        &   16.36         &   16,013      \\
            &   200     &   -12.15                         &   0.57                           &   -16.59        &   17.49         &   15,799      \\
            &   100     &   -9.84                          &   0.27                           &   -16.14        &   13.45         &   15,500      \\
            &   50      &   -9.71                          &   0.04                           &   -15.96        &   12.86         &   15,424      \\
\hline
Scattered   &   800     &   -8.60                          &   1.11                           &   -10.11        &   10.05         &   15,477      \\
            &   400     &   -12.03                         &   1.13                           &   -13.26        &   12.79         &   15,747      \\
            &   200     &   -10.40                         &   0.42                           &   -15.90        &   13.60         &   15,597      \\
            &   100     &   -9.70                          &   0.21                           &   -16.00        &   13.12         &   15,447      \\
            &   50      &   -9.70                          &   0.02                           &   -15.95        &   12.87         &   15,422      \\
\hline
Hexagonal   &   800     &   -6.90                          &   1.00                           &   -11.53        &   9.61          &   15,101      \\
            &   400     &   -13.38                         &   0.98                           &   -12.93        &   10.11         &   15,721      \\
            &   200     &   -11.42                         &   0.44                           &   -15.90        &   14.34         &   15,501      \\
            &   100     &   -9.70                          &   0.20                           &   -15.90        &   12.96         &   15,444      \\
            &   50      &   -9.70                          &   0.01                           &   -15.93        &   12.90         &   15,420      \\
\hline
Reference   &   25      &   -9.70                          &   0.00                           &   -15.93        &   12.90         &   15,418
\label{tbl:straka}
\end{tabular}
\end{table}

\subsubsection{Low-Viscosity Density Current $\mu = 2\times10^{-5}\text{ m}^2/\text{s}$ } \label{eulerStraka}

Here, the density current test case is repeated, except with the dynamic viscosity $\mu$ set to that of air. The purpose of this test case is to show that one can stably time step the RBF-FD method in a completely turbulent regime.  The same amount of hyperviscosity as well as the same time step are used in this test case as in the one with $\mu = 75\, m^2/s$. Time stability is governed solely by the fact that the time step could not exceed the speed of sound in air.

\begin{figure}[H]
$$
\begin{array}{cccc}
\text{} & \text{Cartesian} & \text{Hexagonal} & \text{Scattered}  \\
\rotatebox{90}{~~~100~m}                                          &
\includegraphics[width=.31\textwidth]{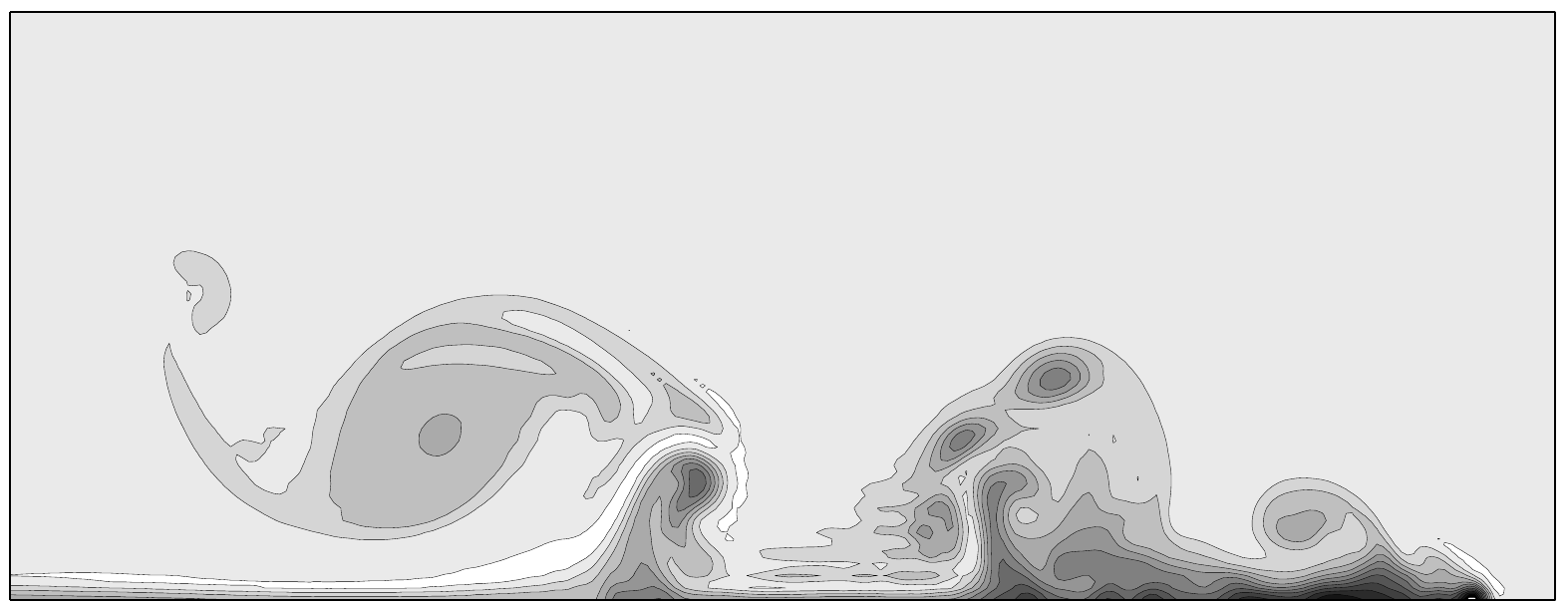} &
\includegraphics[width=.31\textwidth]{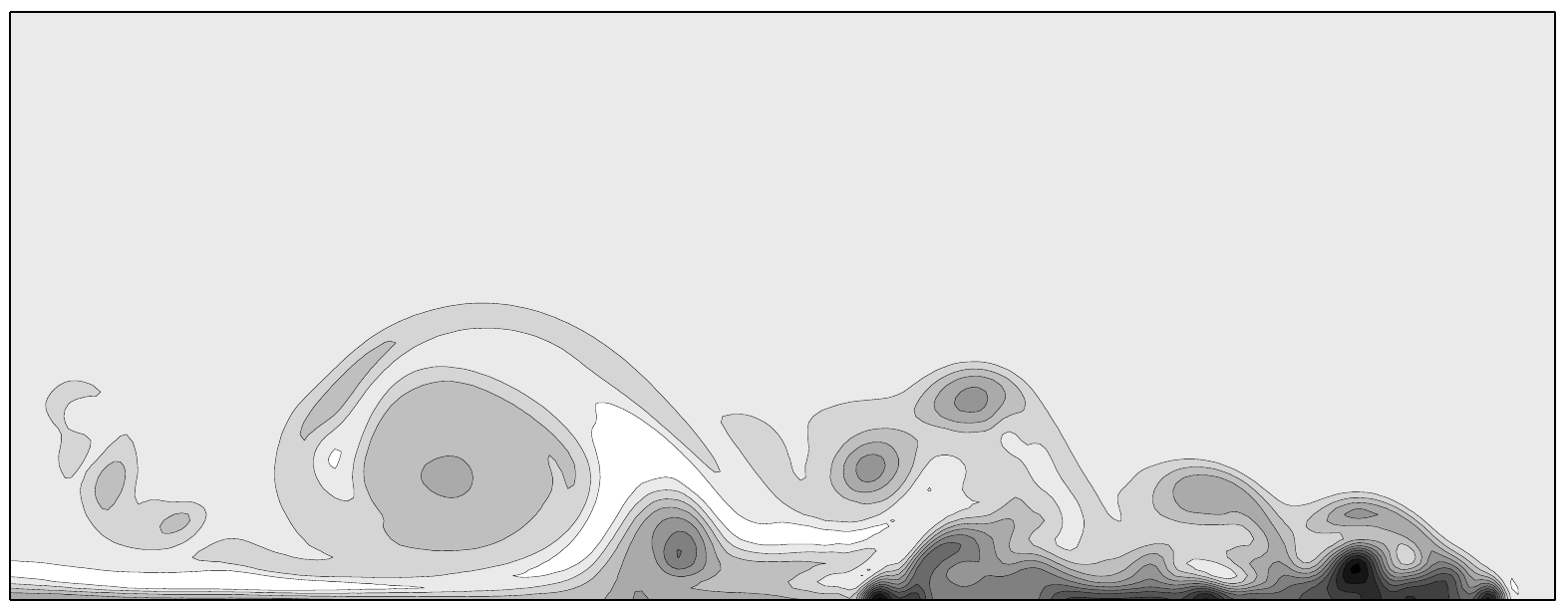} &
\includegraphics[width=.31\textwidth]{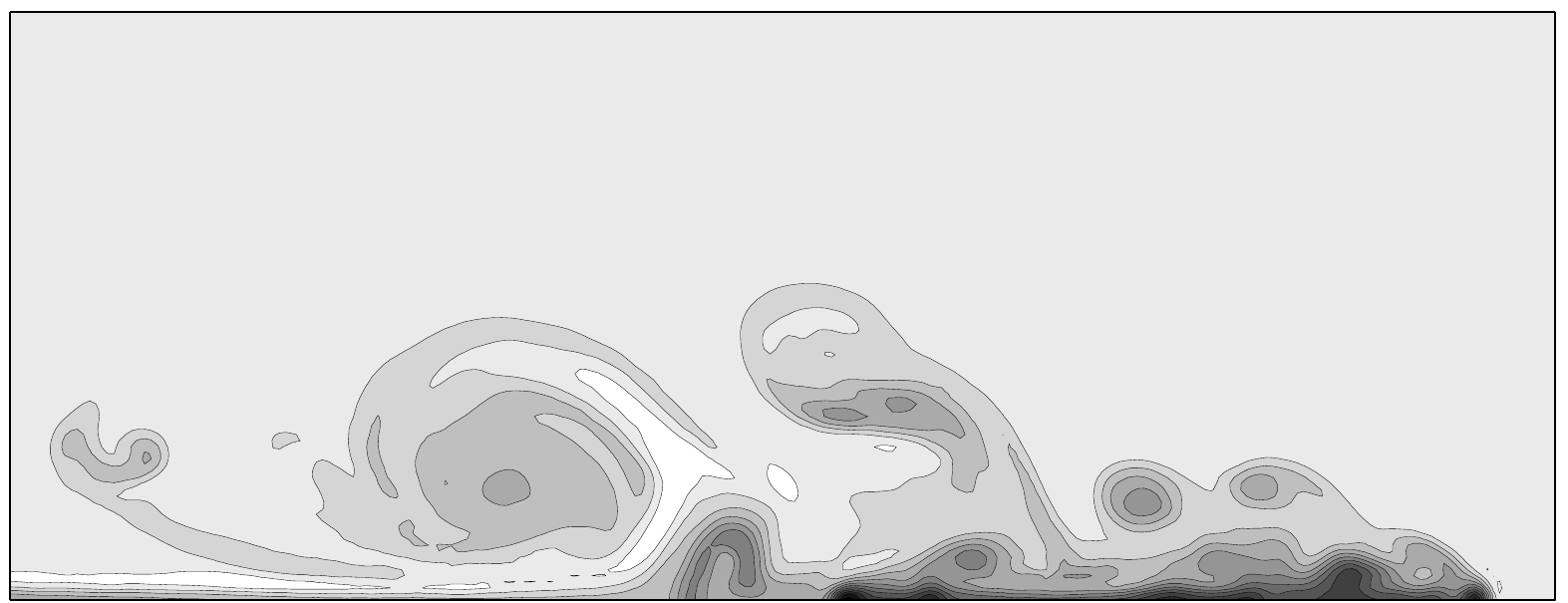}       \\
\rotatebox{90}{~~~~50~m}                                          &
\includegraphics[width=.31\textwidth]{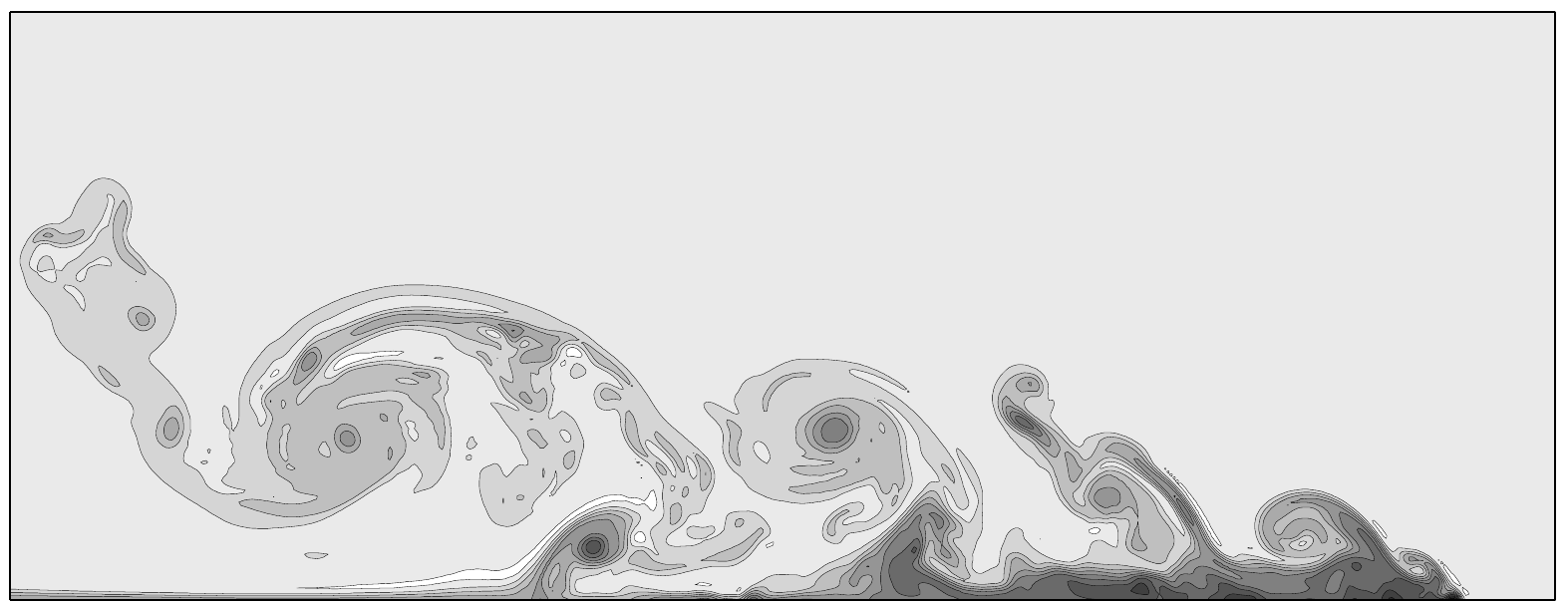}  &
\includegraphics[width=.31\textwidth]{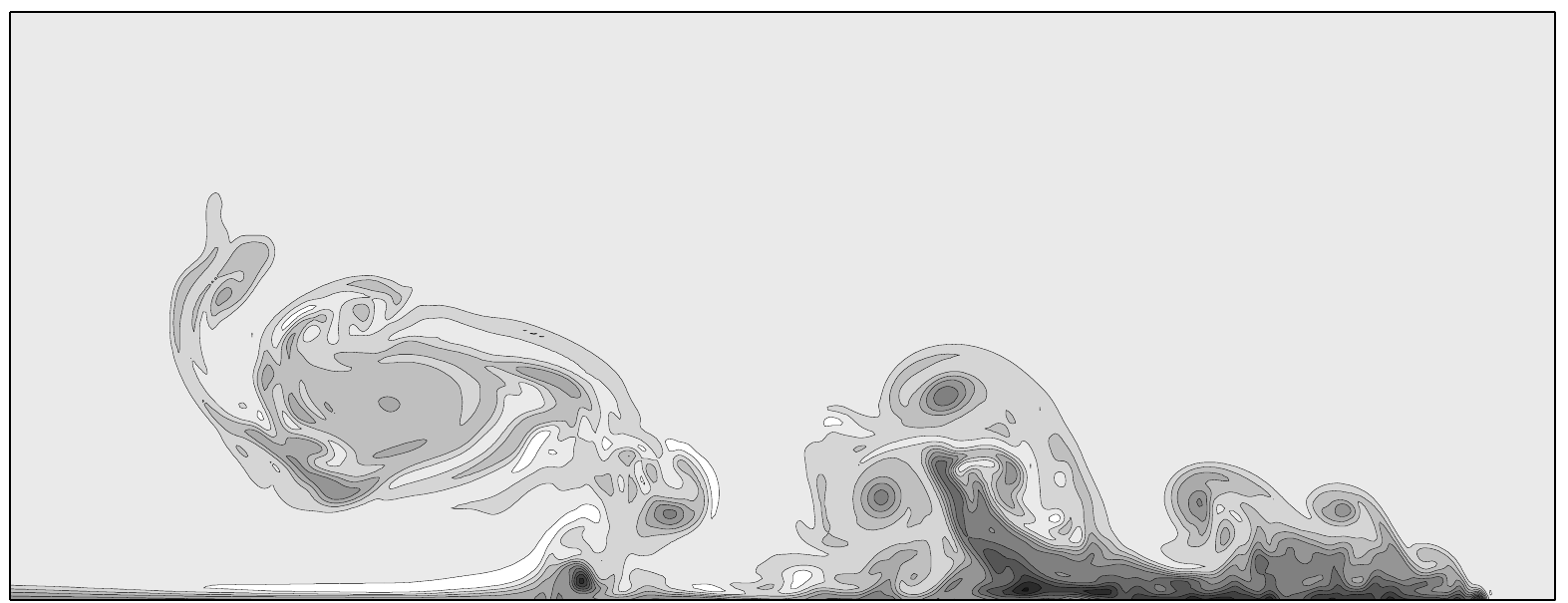}  &
\includegraphics[width=.31\textwidth]{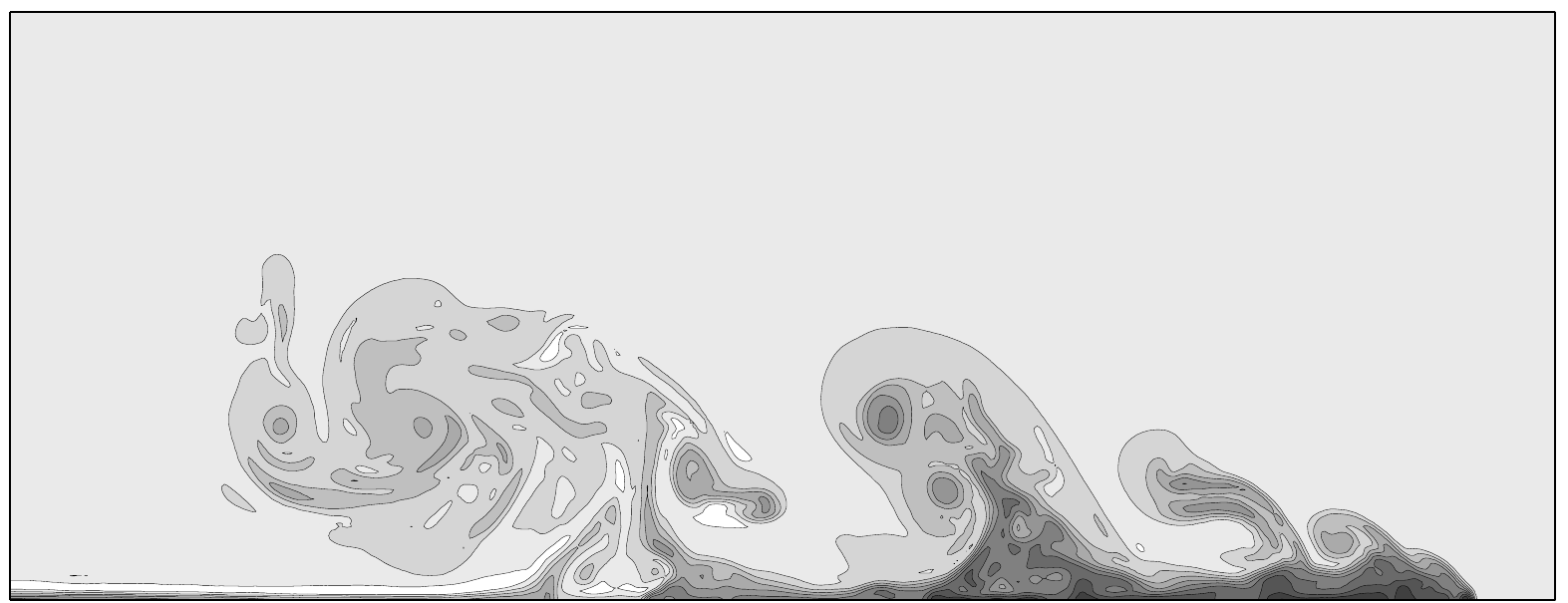}        \\
\rotatebox{90}{~~~~~~~~~25~m}                                     &
\includegraphics[width=.31\textwidth]{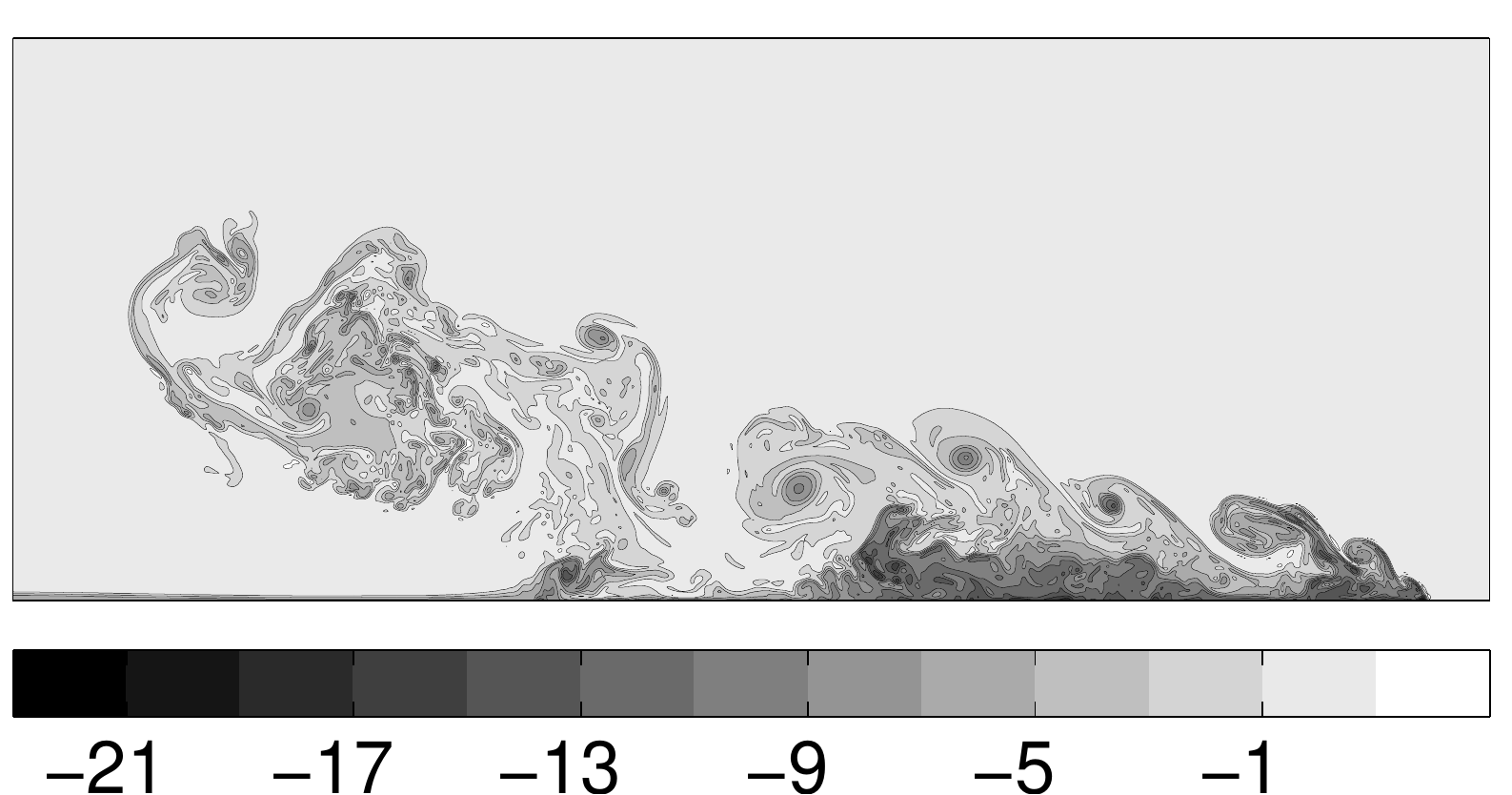}  &
\includegraphics[width=.31\textwidth]{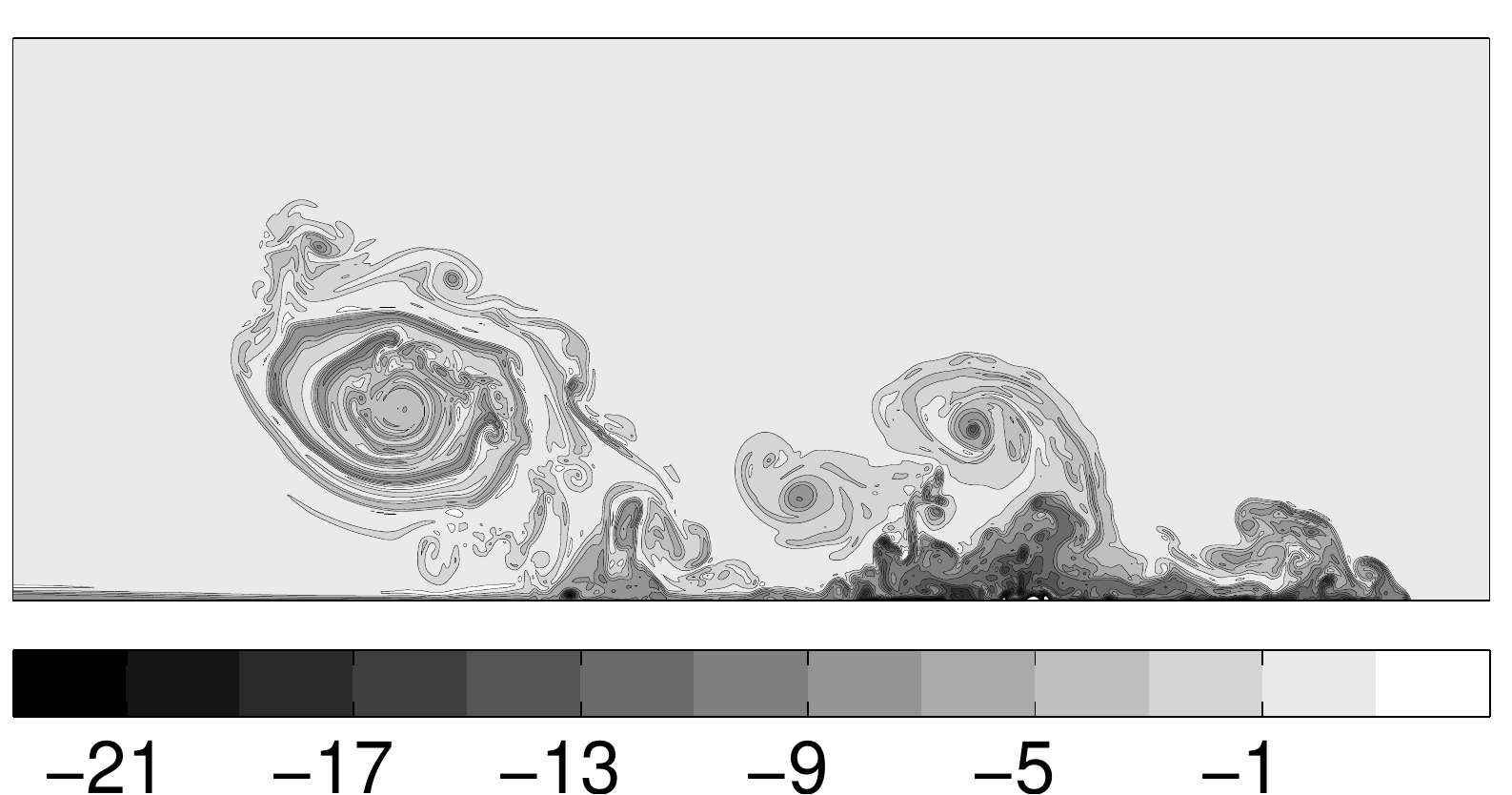}  &
\includegraphics[width=.31\textwidth]{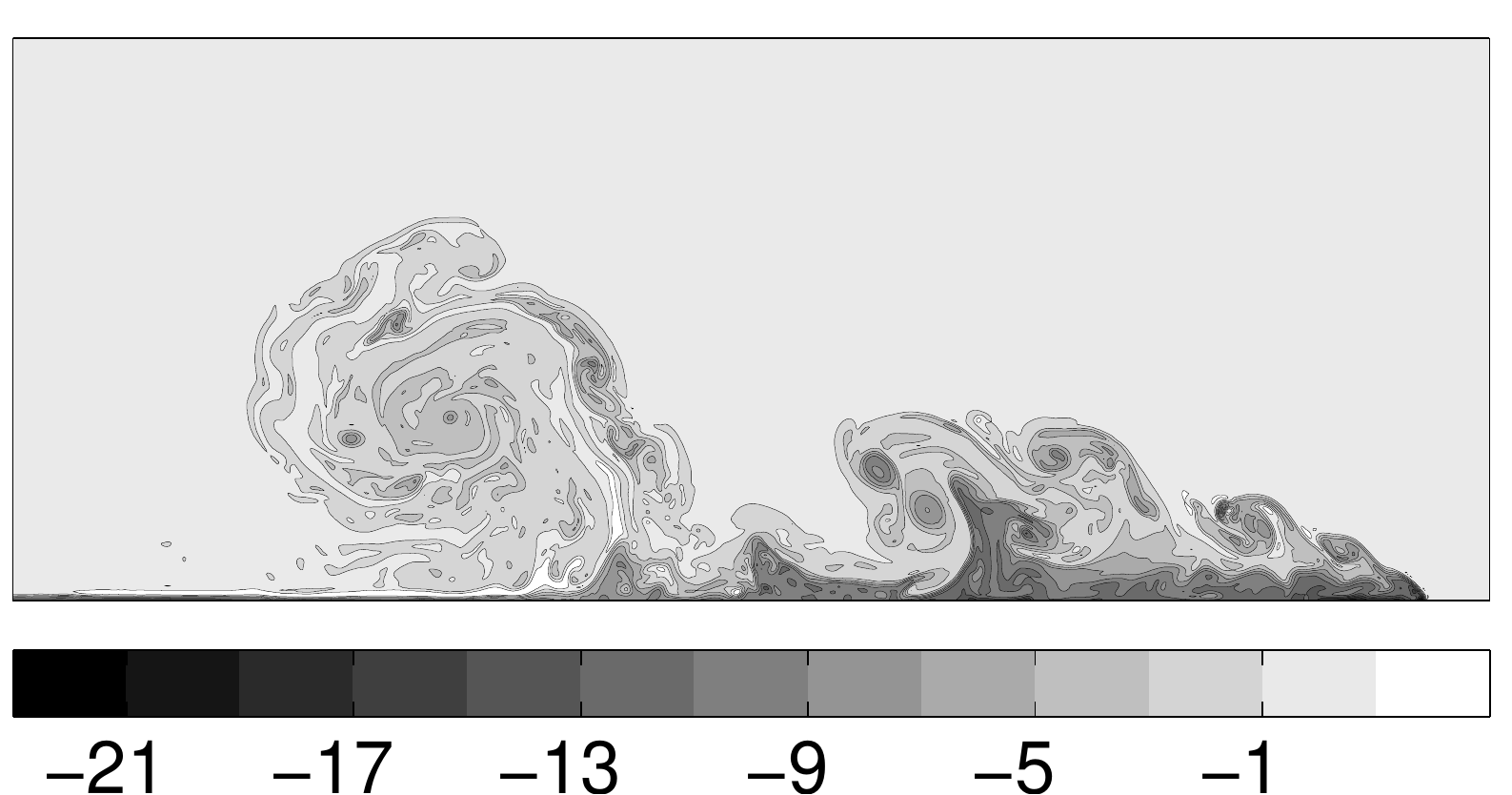}
\end{array}
$$
\caption{Potential temperature perturbation $\theta'$ for the low-viscosity density current \ref{eulerStraka} using 100m, 50m, and 25m nodes at $t=900s$.}
\label{fig:low_mu_straka}
\end{figure}

At such low viscosity, the solution enters the turbulent regime. In such regimes, there is no convergence to any solution as energy cascades to smaller and smaller scales, eventually entering the sub-grid scale domain. Nevertheless, it is interesting to observe whether the model remains stable in this regime. Figure \ref{fig:low_mu_straka} shows the solution at 100m, 50m and 25m resolutions on the three different node layouts. For any given resolution the solution looks completely different depending on the node layout. This is to be expected as changing the node layout in practically the absence of explicit viscosity is equivalent to introducing slight perturbations in the solution. A more robust illustration of this will be given in the test case of a rising thermal bubble, Section \ref{sec:bubble}.


\subsubsection{Translating Density Current, $\mu = 75\, m^2/s$}\label{translating}

\begin{figure}[H]
\centering
\includegraphics[width=0.7\textwidth]{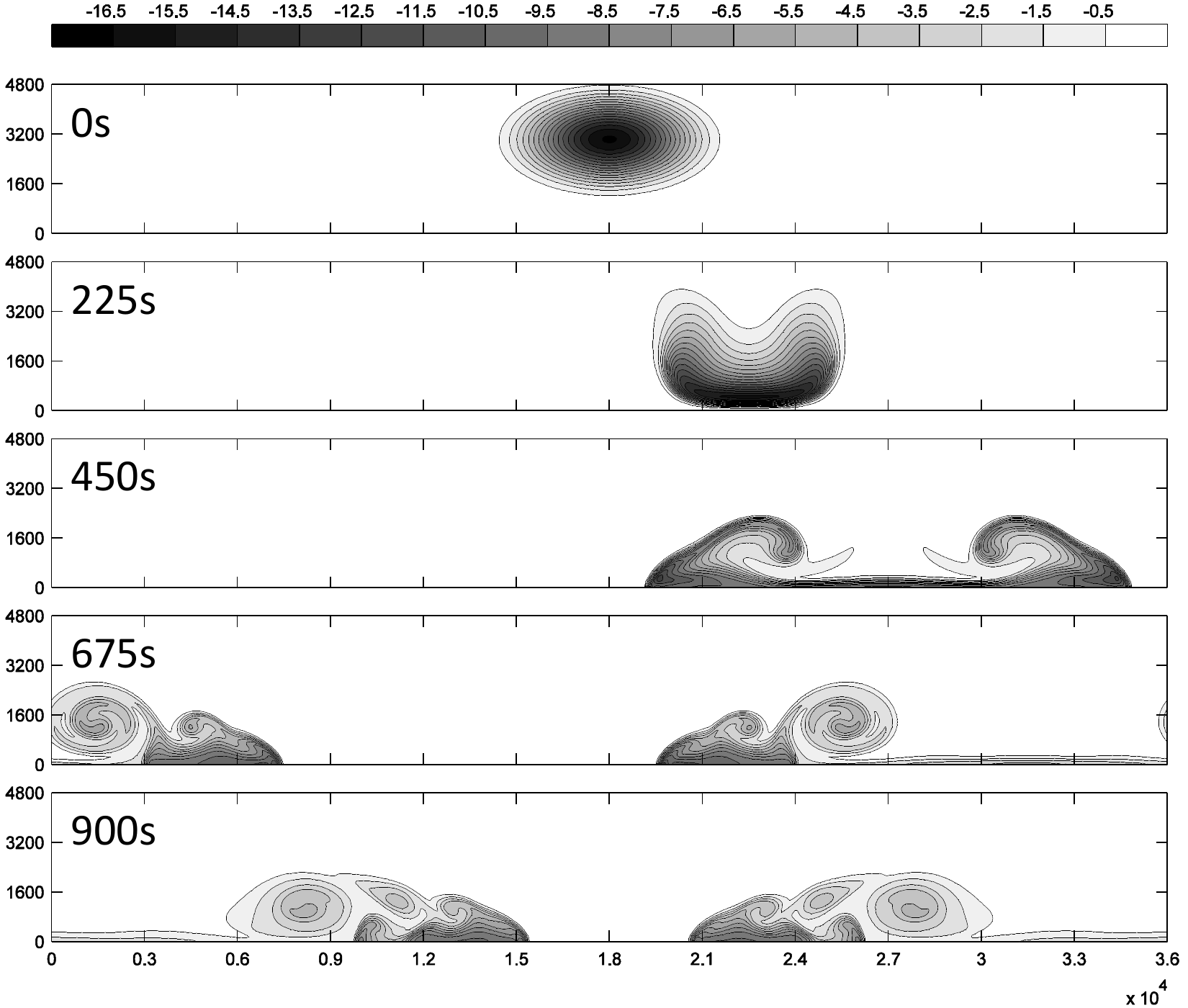}
\caption{Time evolution of the potential temperature $\theta'$ for the translating density current test case.  Snapshots were generated using the 100m RBF-FD solution on hexagonal nodes.}
\label{fig:soln_dnbr}
\end{figure}

This test is the same as in Section \ref{densityCurrent}, except that the domain is now $\left[0,36\right]\times\left[0,6.4\right]\text{ km}^2$ and there is a horizontal background wind of $\overline{u}=20\text{ m}/\text{s}$.  The size of the domain is set up so that at $t=900$s the two ``halves'' of the solution should be symmetric about $x=18$km. The introduction of a background mean flow introduces a large difference in the movement of each half of the solution. The right portion of the outflow has horizontal velocities $\sim 50 \text{ m}/\text{s}$, while the left portion has velocities $\sim 10\text{ m}/\text{s}$.  As a result, it tests the ability of the scheme to translate the features of the solution at the correct speeds and to generate the correct rotor structures that arise from the local shearing instabilities. For this case, only Cartesian and hexagonal nodes are considered, as the case tests the degree to which symmetry is broken in the two halves of the solution at the final time. Figure \ref{fig:soln_dnbr} illustrates the time series of the potential temperature $\theta'$ field, showing how the right part of the solution is advected through the the right side of the domain, with the front locations facing one another at 900s (instead of facing the lateral boundaries as in the previous test case).

\begin{figure}[H]
$$
\begin{array}{c}
\includegraphics[width=0.8\textwidth]{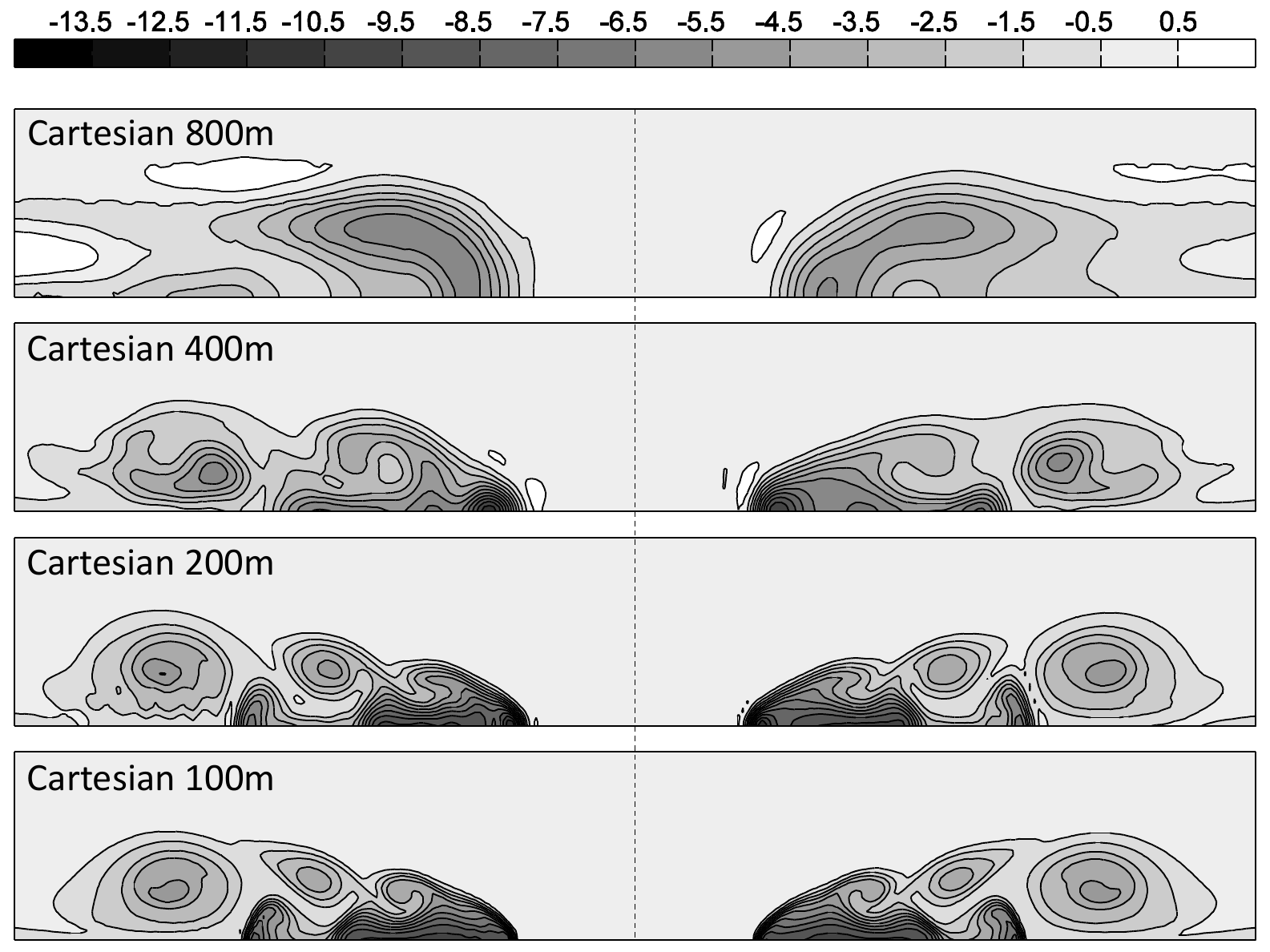} \\
\includegraphics[width=0.8\textwidth]{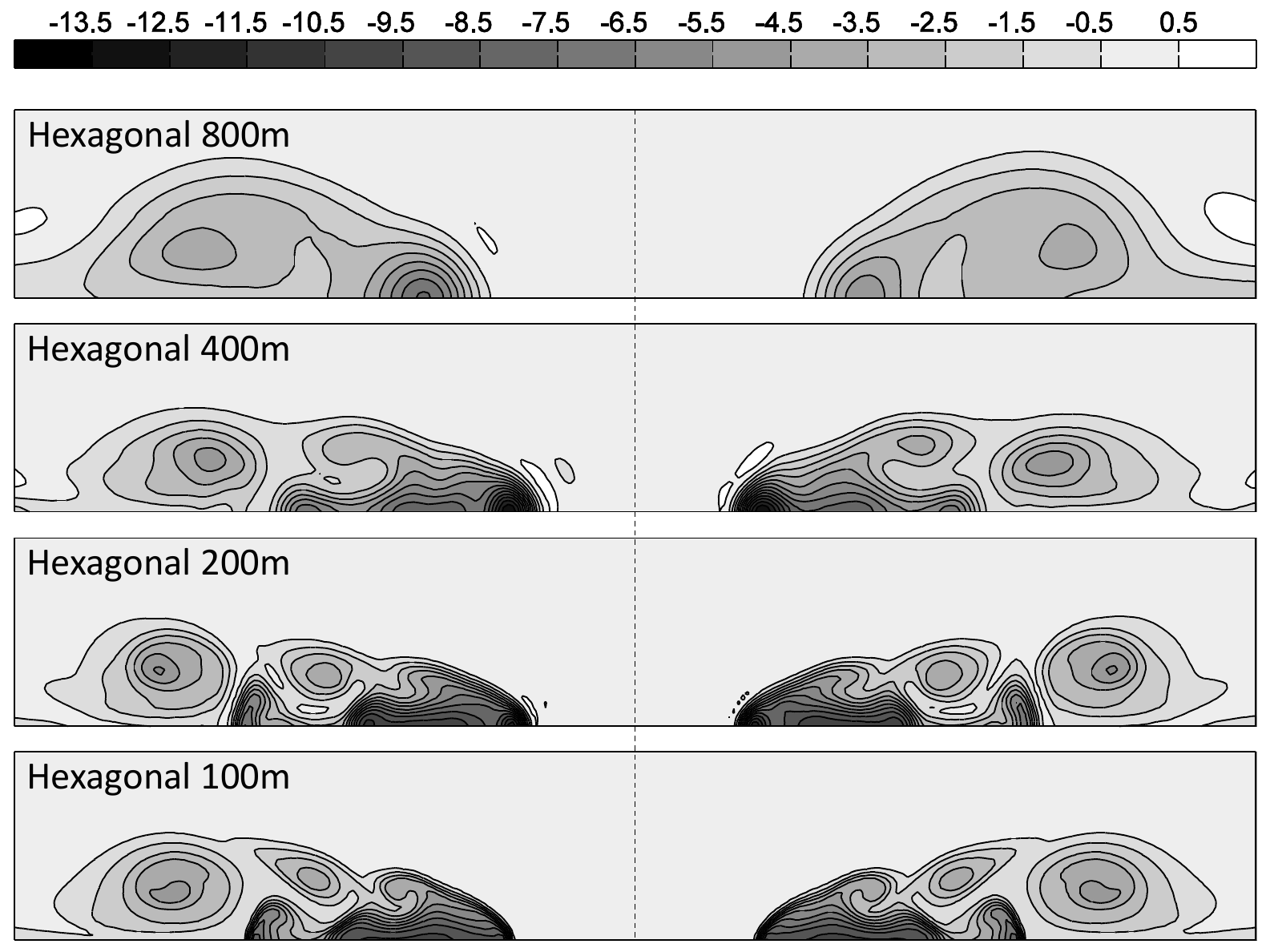}
\end{array}
$$
\caption{Potential temperature perturbation $\theta'$ for the the translating density current at $t=900s$ for Cartesian and hexagonal nodes at given resolutions. The dashed line is the $x = 18$km mark about which the two halves should be symmetric.}
\label{fig:comp_translating}
\end{figure}

In Figure \ref{fig:comp_translating}, the two halves of the solution are compared about the line of symmetry (18km) for Cartesian and hexagonal nodes from 800m to 100m resolutions. The general observations that can be seen are:
\begin{enumerate}
\item The 800m hexagonal node layout performs highly superior to the Cartesian both in terms of symmetry between the two sides, intactness of the large rotor, and its relative location when compared to the 100m case.
\item At both the 400m and 200m Cartesian case, the left half of the solution that has been advected through the right boundary displays a significant amount of Runge phenomena (`wiggles' in the contour lines near the boundary at 200m and severe distortion of the primary and secondary rotor at 400m). This is not the case for hexagonal nodes, which at 400m and 200m, shows relatively nice symmetry between the two sides.
\item At 100m, there is no distinction between the two node sets.
\end{enumerate}
The actual front locations, in terms of there distance from the 18km mark, are given in Table \ref{tbl:translating}. Note that the front on the right is farther from the line of symmetry (18km mark) than the one that has been advected through the boundary for the resolutions 800m to 100m. At 50m, both the Cartesian and hexagonal case is symmetric about the 18km mark. However the distance from the front to the line of symmetry varies between the two cases, 2586m versus 2595m, respectively.

\begin{table}[H]
\centering
\caption{Left and right front locations for the translating density current test case \ref{translating} as given by the distance from the $-0.5$K contour line to the center of the domain, $x=18$km. All values are calculated at the final time, $t=900$s.  Results are for $\phi(r)=r^7$ with up to fourth degree polynomials on a 37-node stencil. A 25m reference solution is given for the hexagonal nodes.}
\begin{tabular}{c|c|c|c}
\\[1ex]
            &   h      &    left front ($m$) &   right front ($m$) \\
\hline\hline
Cartesian   &   800    &    3,065            &   3,215      \\
            &   400    &    1,915            &   2,085      \\
            &   200    &    2,098            &   2,205       \\
            &   100    &    2,487            &   2,512       \\
            &   50     &    2,586            &   2,586       \\
\hline
Hexagonal   &   800    &    3,410            &   3,575      \\
            &   400    &    1,956            &   2,046      \\
            &   200    &    2,165            &   2,260       \\
            &   100    &    2,555            &   2,580        \\
            &   50     &    2,595            &   2,595        \\
\hline
Reference   &   25     &    2,595            &   2,595
\end{tabular}
\label{tbl:translating}
\end{table}


\subsection{Rising Thermal Bubble} \label{sec:bubble}

With this last test case, the paper comes full circle in that the presented RBF-FD method is tested on a problem with sharp gradients and very little boundary interaction as in the advective transport of a scalar variable, but is modeled by the same 2D nonhydrostatic compressible Navier-Stokes equations as in the density current tests.  The only difference is the initial condition is given by a $C^0$ cone-shaped perturbation.  The bubble is warmer than the surrounding atmosphere and thus rises toward the top boundary.  However, the time interval and domain size are chosen so that the bubble never interacts with the boundaries.

There are two variations for this test case:
\begin{enumerate}
\item $\mu=10\text{ m}^2/\text{s}$: By adding a small amount of explicit viscosity, the convergence behavior of the solution can be studied.
\item $\mu=2\times10^{-5}\text{ m}^2/\text{s}$:  At such low viscosity, the bubble is in a turbulent regime, and the behavior of the true solution is unknown.  The purpose of the test is to see if the RBF-FD method can give reasonable results when the initial condition is not even continuously differentiable as well as observe how the instability pattern at the leading edge of thermal changes with the node layout.
\end{enumerate}


\subsubsection{Case $\mu = 10\text{ m}^2/\text{s}$} \label{bubble}

The computational domain is $\left[0,10\right]\times\left[0,10\right]\text{ km}^2$.  The hydrostatic background states are defined by $\overline{\theta}=T_s$ and $\overline{\pi}(z)=1-\frac{g}{c_pT_s}z$, with $T_s=300\text{K}$ being the surface temperature.  The horizontal and vertical velocities and the Exner pressure perturbation $\left(\pi'\right)$ are initially zero, while the potential temperature perturbation is prescribed as a warm cone-shaped ``bubble'' with a jump in the first derivative $\left(C^0\right)$:
\begin{align*}
\left.\theta'\right|_{t=0}=2\max\left\{0,1-r(x,z)/R\right\}.
\end{align*}
Here, $R=1.5\text{ km}$ is the radius of the bubble, and
\begin{align*}
r(x,z)=\sqrt{\left(x-x_c\right)^2+\left(z-z_c\right)^2},\quad\quad\quad\left(x_c,z_c\right)=\left(5\text{ km},3\text{ km}\right).
\end{align*}

The same boundary conditions as in the density current problem are enforced on the top and bottom boundaries
\begin{equation}
w = \frac{\partial^2w}{\partial z^2} = \frac{\partial u}{\partial z} = \frac{\partial\theta'}{\partial z} = 0, \quad\quad \frac{\partial\pi'}{\partial z} = \frac{g\theta'}{c_p\overline{\theta}\left(\overline{\theta}+\theta'\right)}.
\label{eqn:BCbubble}
\end{equation}
The lateral boundary conditions are given by
\begin{align*}
u = \frac{\partial^2u}{\partial x^2} = \frac{\partial w}{\partial x} = \frac{\partial\theta'}{\partial x} = \frac{\partial\pi'}{\partial x} = 0.
\end{align*}

\begin{figure}[H]
$$
\begin{array}{ccccc}
\includegraphics[width=.18\textwidth]{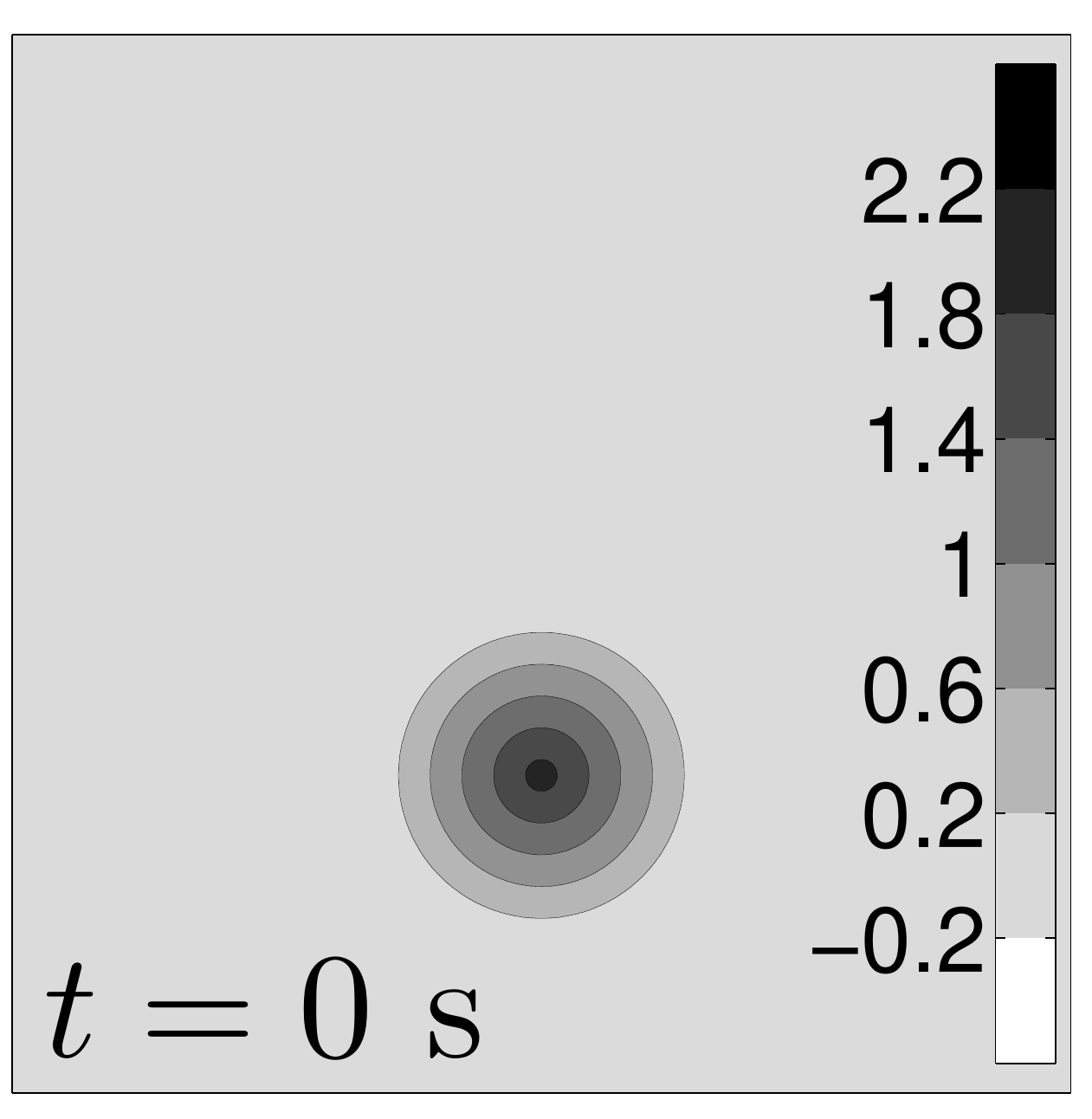}    &
\includegraphics[width=.18\textwidth]{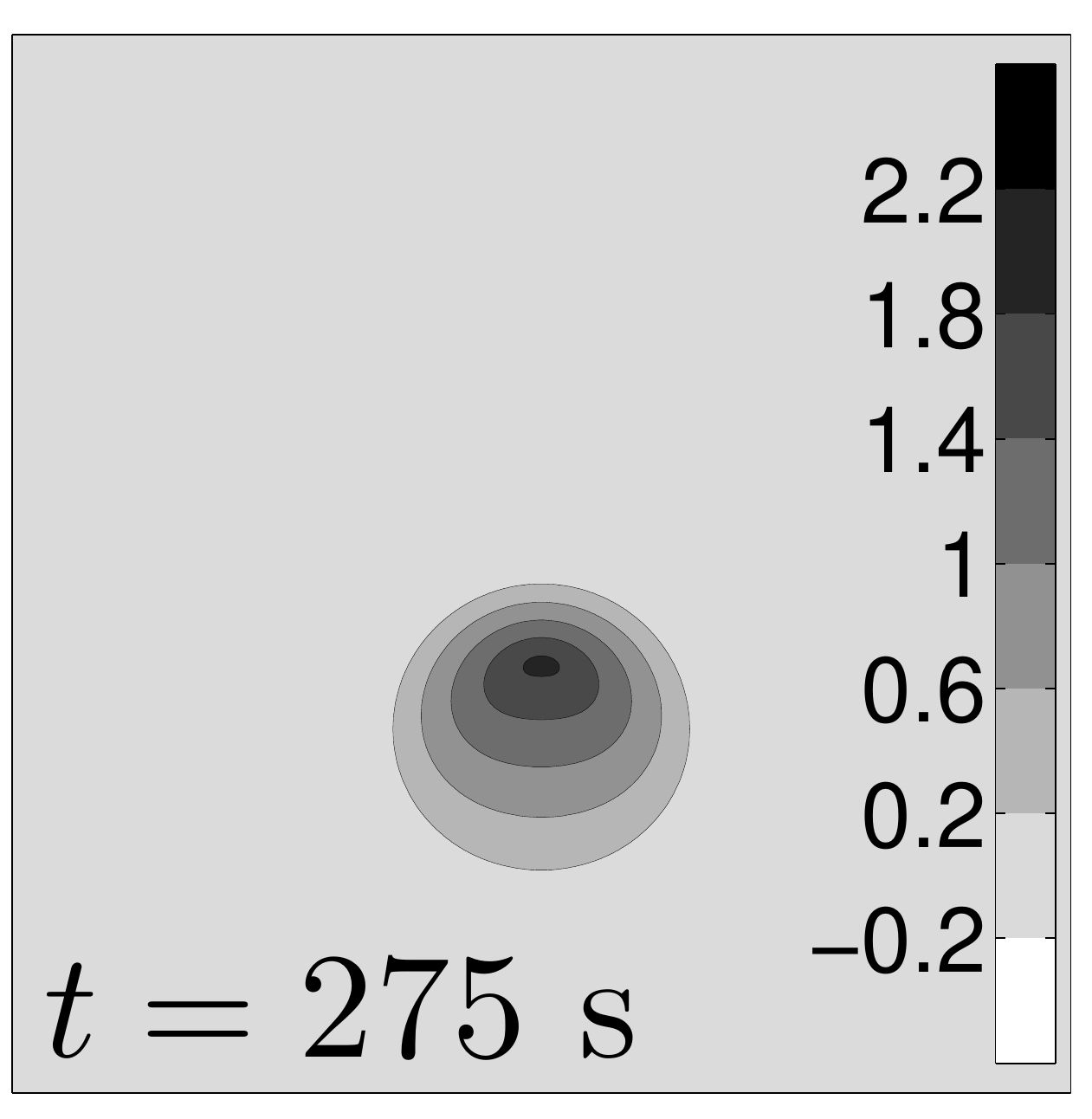}  &
\includegraphics[width=.18\textwidth]{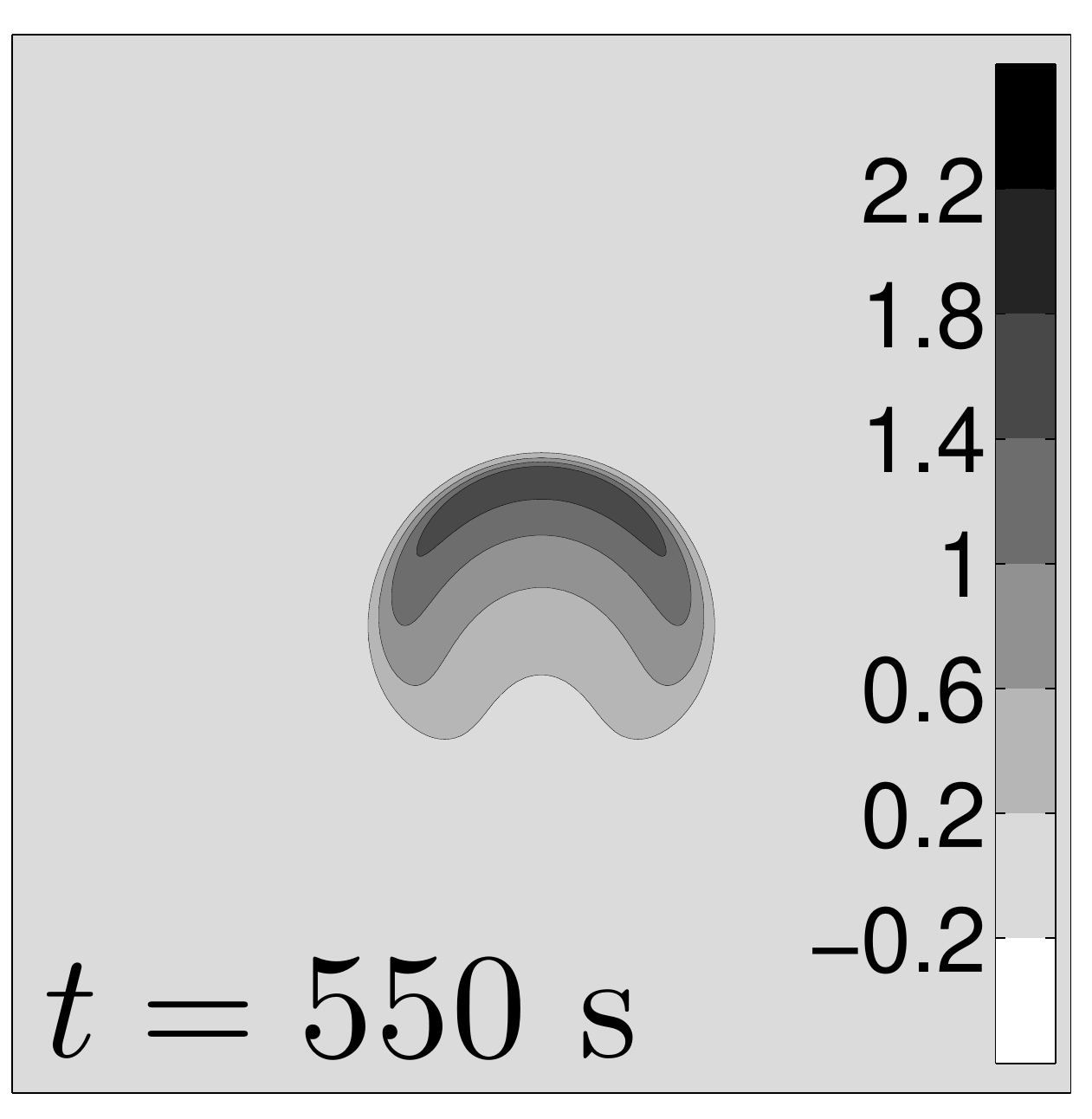}  &
\includegraphics[width=.18\textwidth]{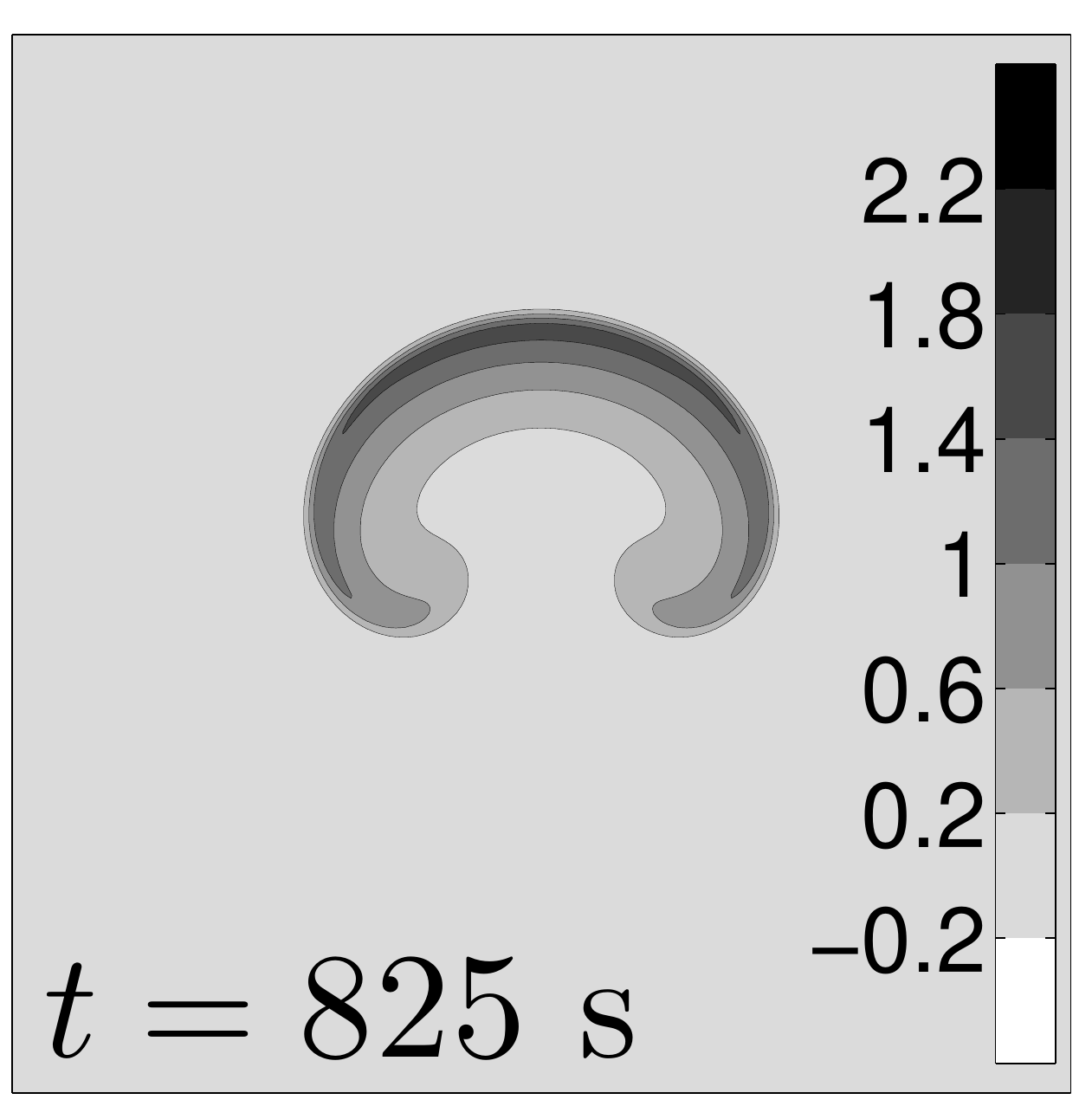}  &
\includegraphics[width=.18\textwidth]{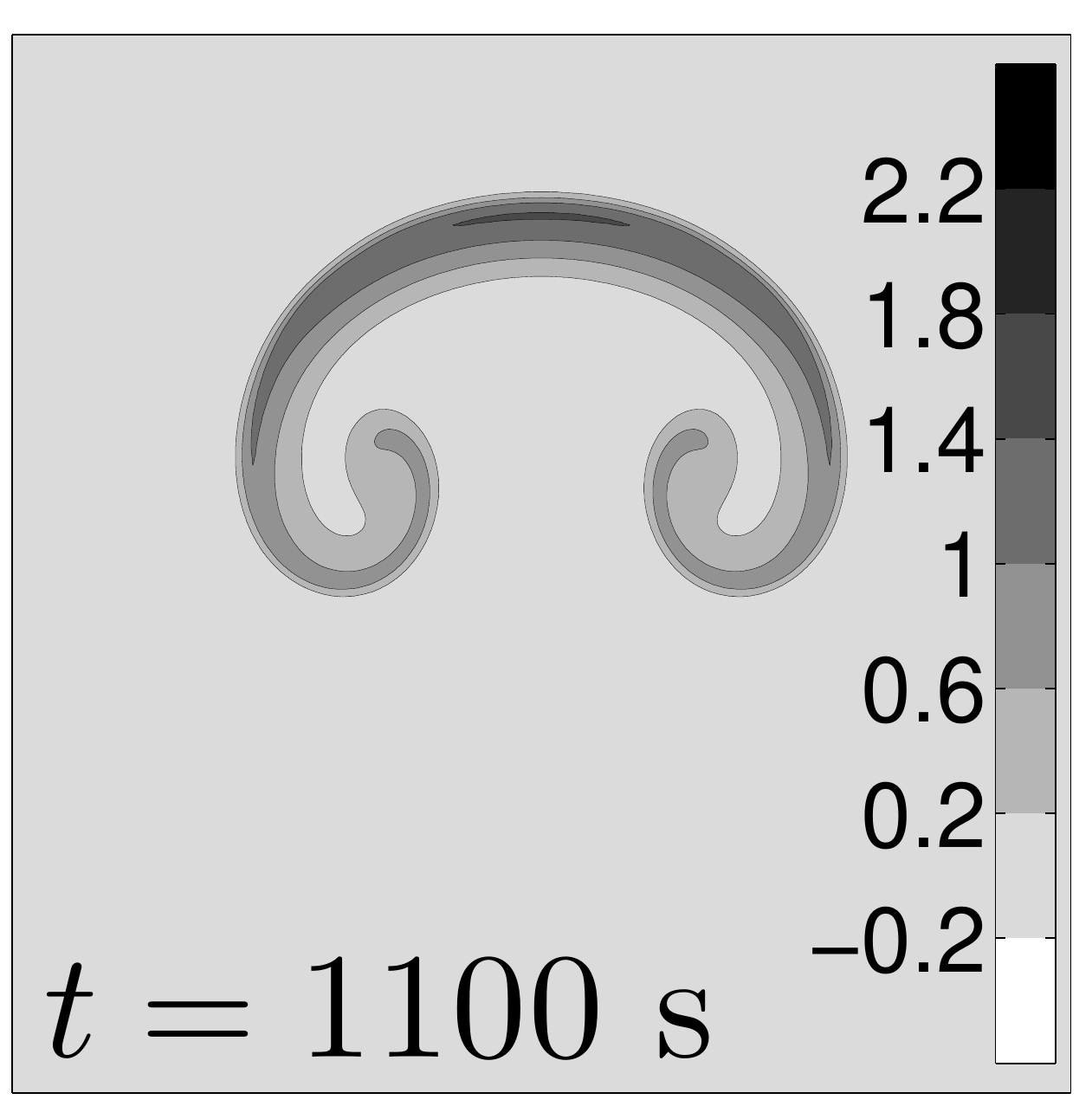}
\end{array}
$$
\caption{Time evolution of the potential temperature $\theta'$ for the $\mu=10\text{ m}^2/\text{s}$ rising thermal bubble.  Snapshots were generated using the 25m RBF-FD solution on hexagonal nodes.}
\label{bubble_mu10_timeSeries}
\end{figure}

\begin{figure}[H]
$$
\begin{array}{cccc}
\text{} & \text{Cartesian} & \text{Hexagonal} & \text{Scattered}       \\
\rotatebox{90}{~~~~~~~~~~~200~m}                                       &
 \includegraphics[width=.25\textwidth]{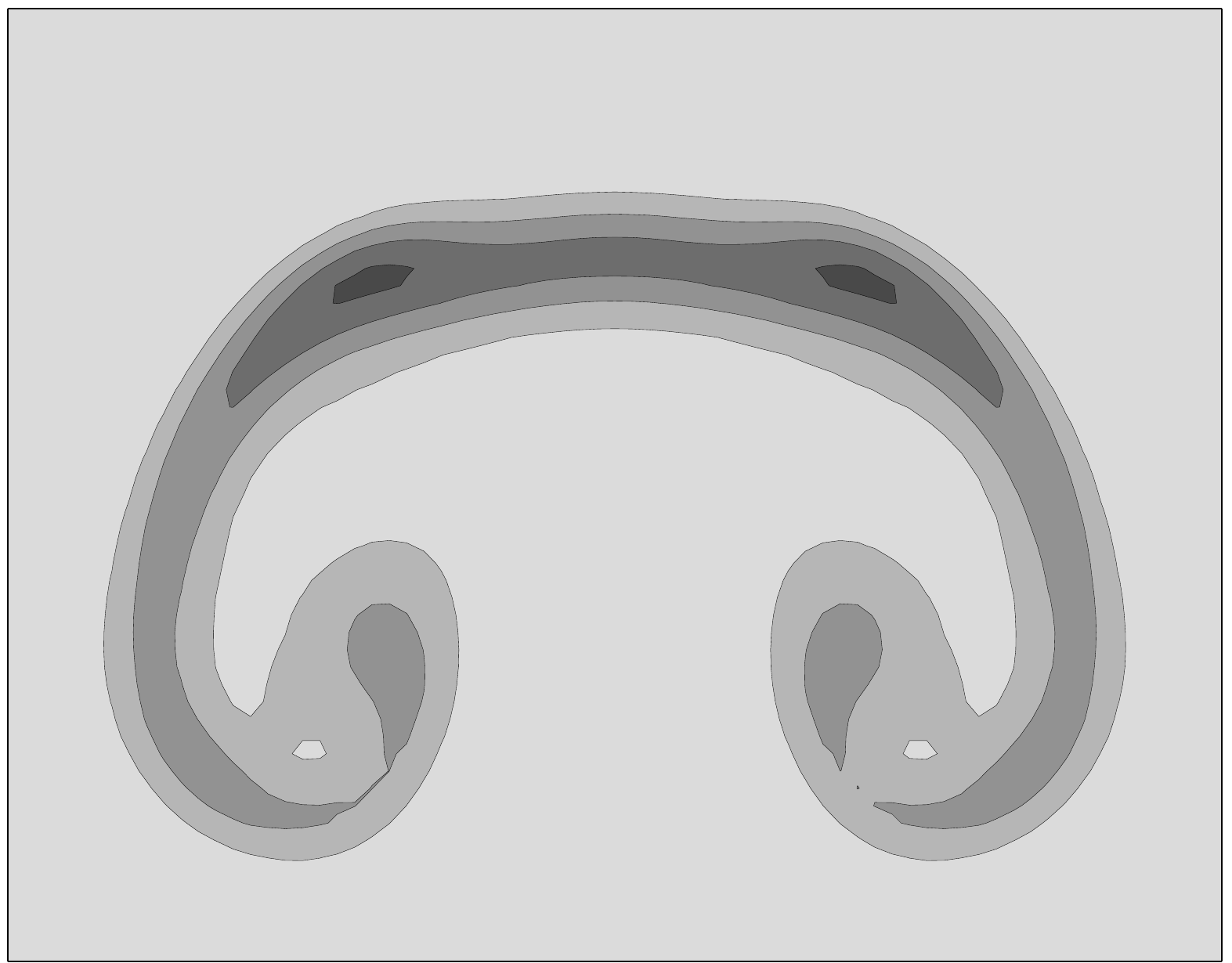} &
\includegraphics[width=.25\textwidth]{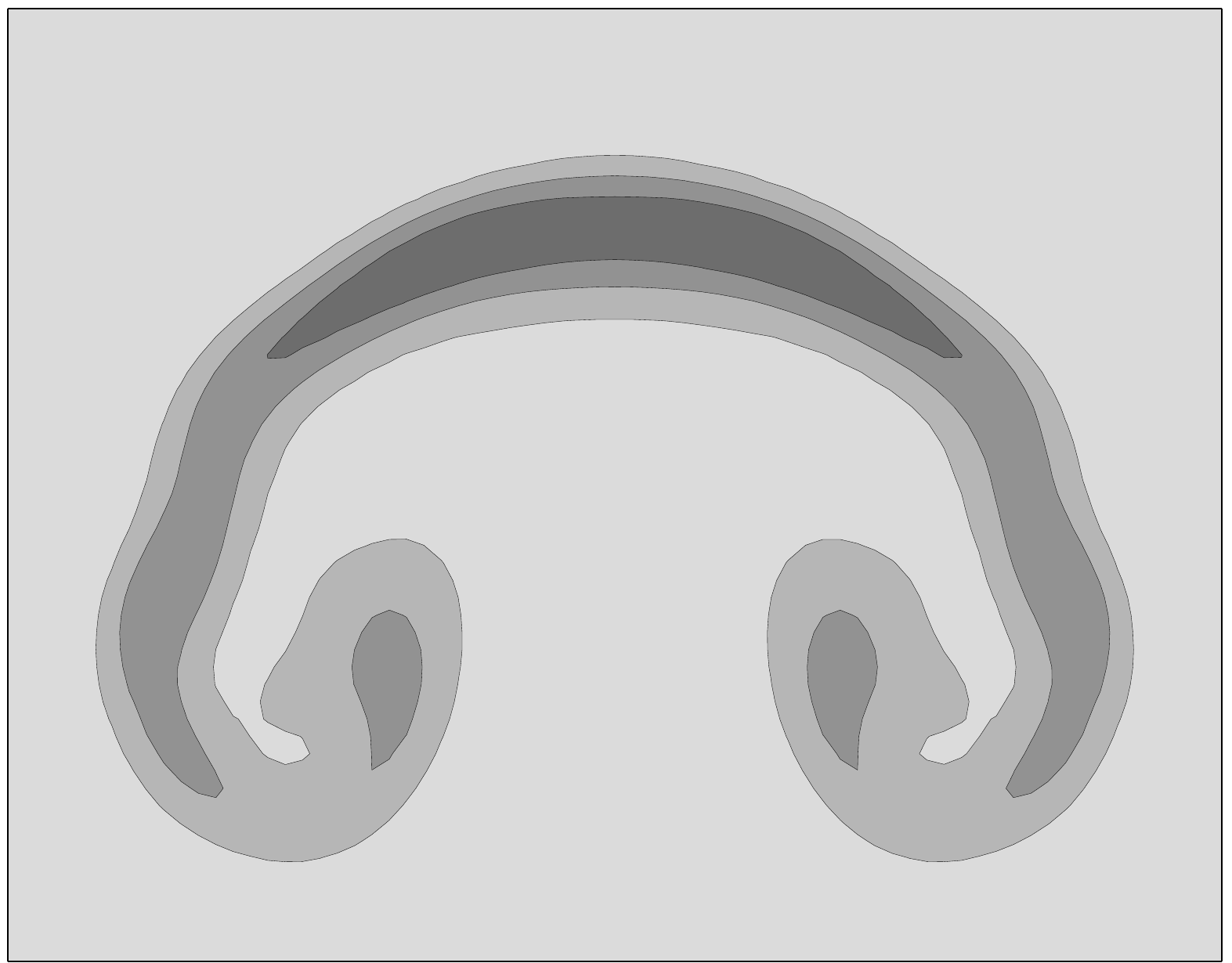}        &
\includegraphics[width=.25\textwidth]{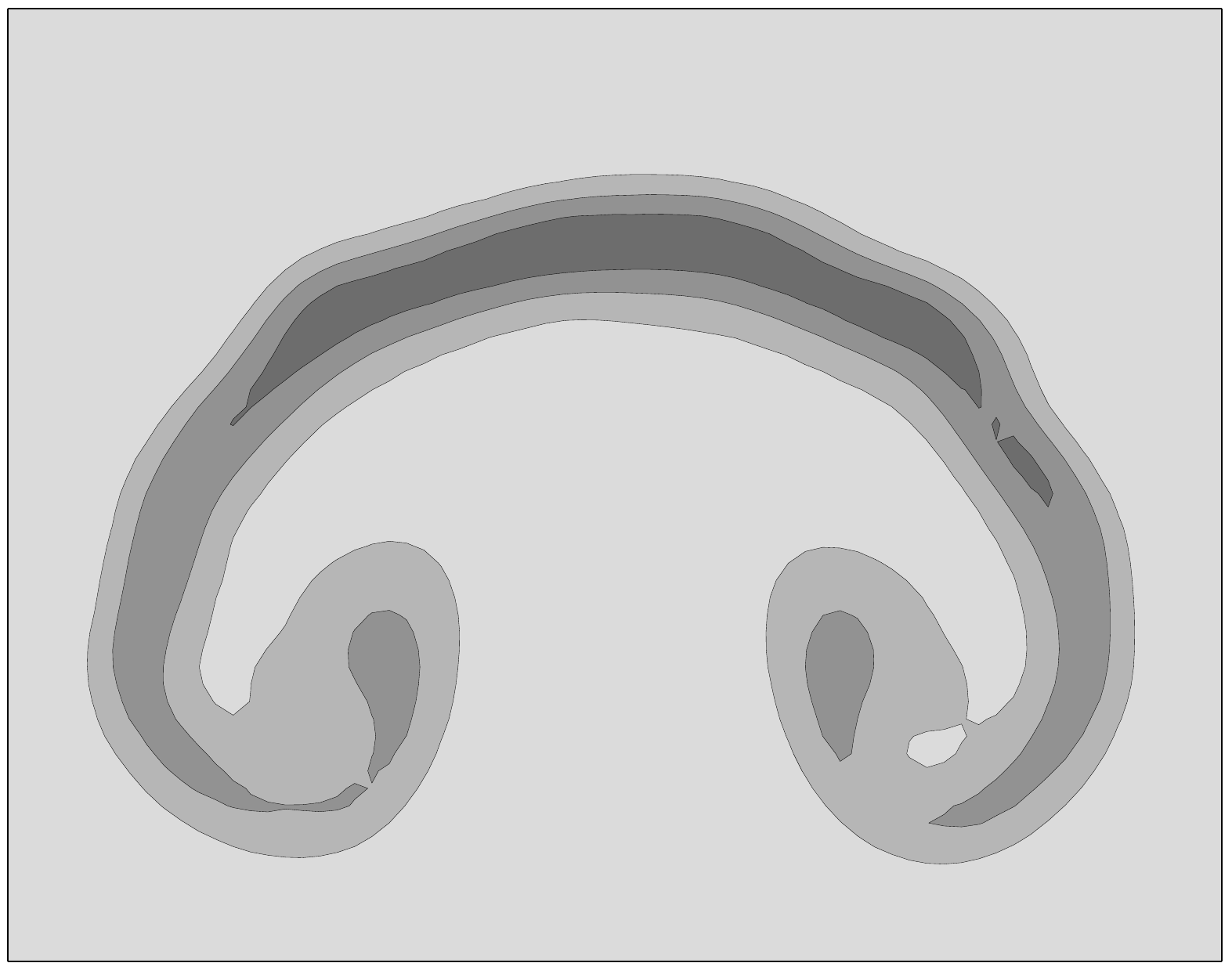}  \\
\rotatebox{90}{~~~~~~~~~~~100~m}                                       &
\includegraphics[width=.25\textwidth]{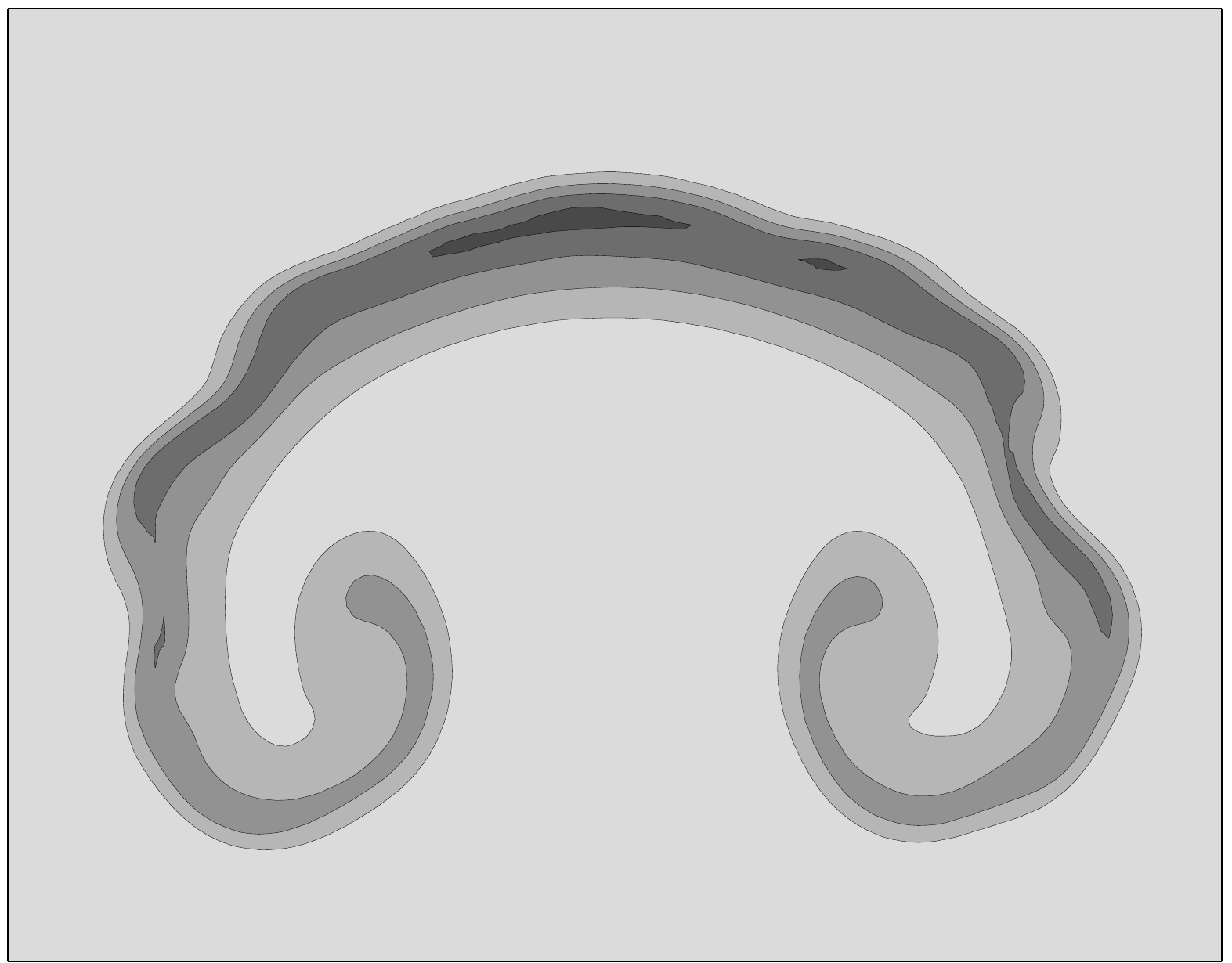}  &
\includegraphics[width=.25\textwidth]{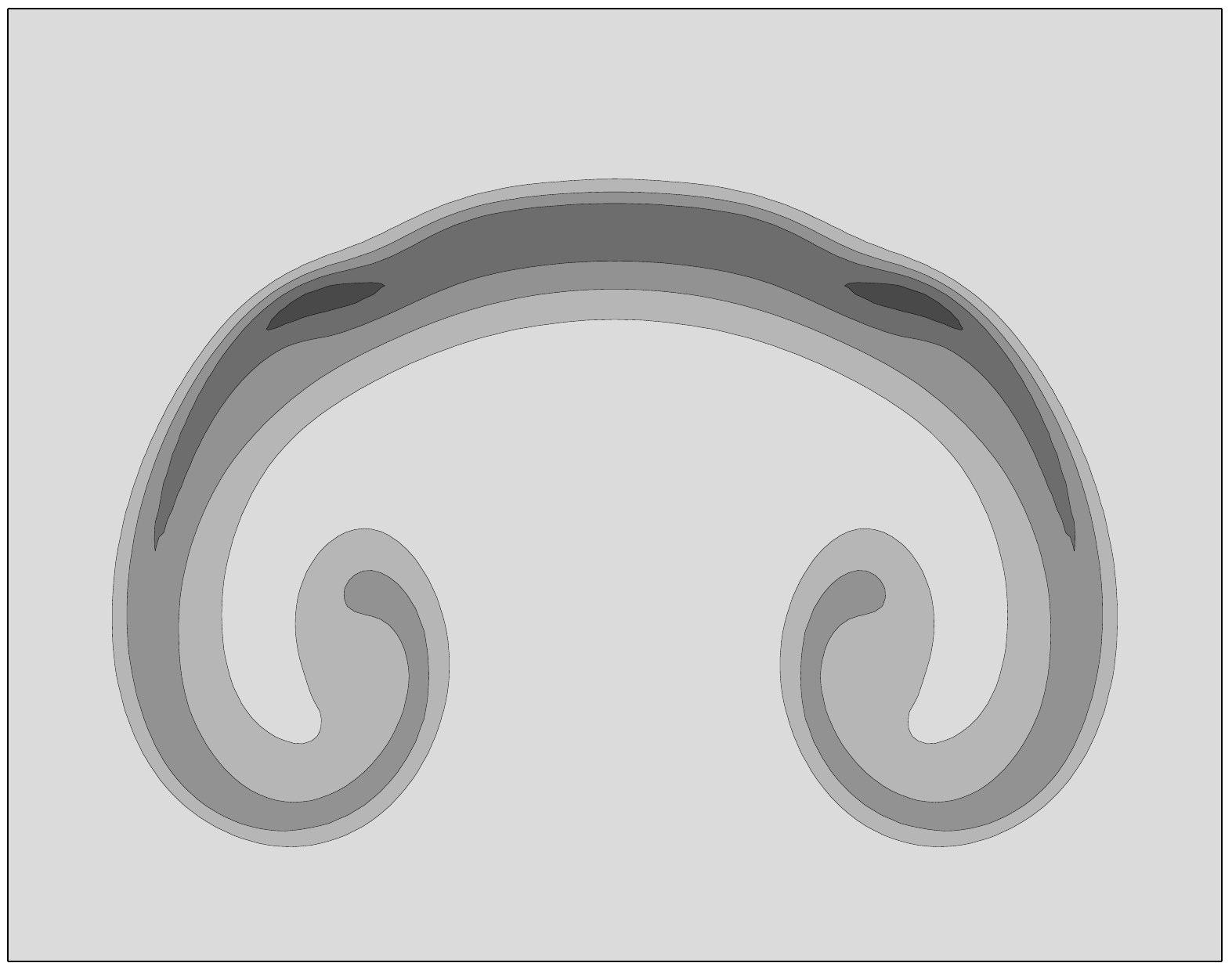}  &
\includegraphics[width=.25\textwidth]{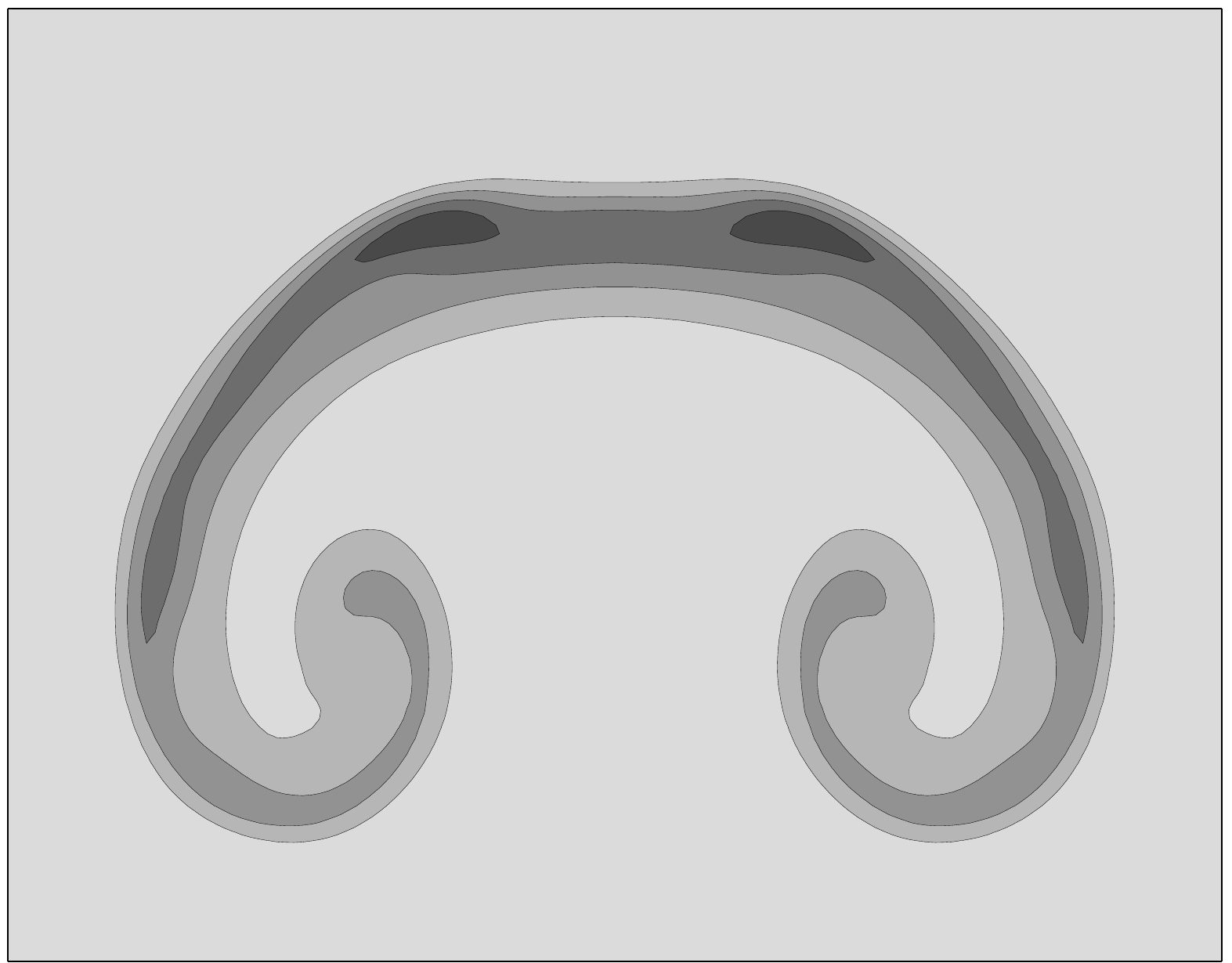}        \\
\rotatebox{90}{~~~~~~~~~~~50~m}                                        &
\includegraphics[width=.25\textwidth]{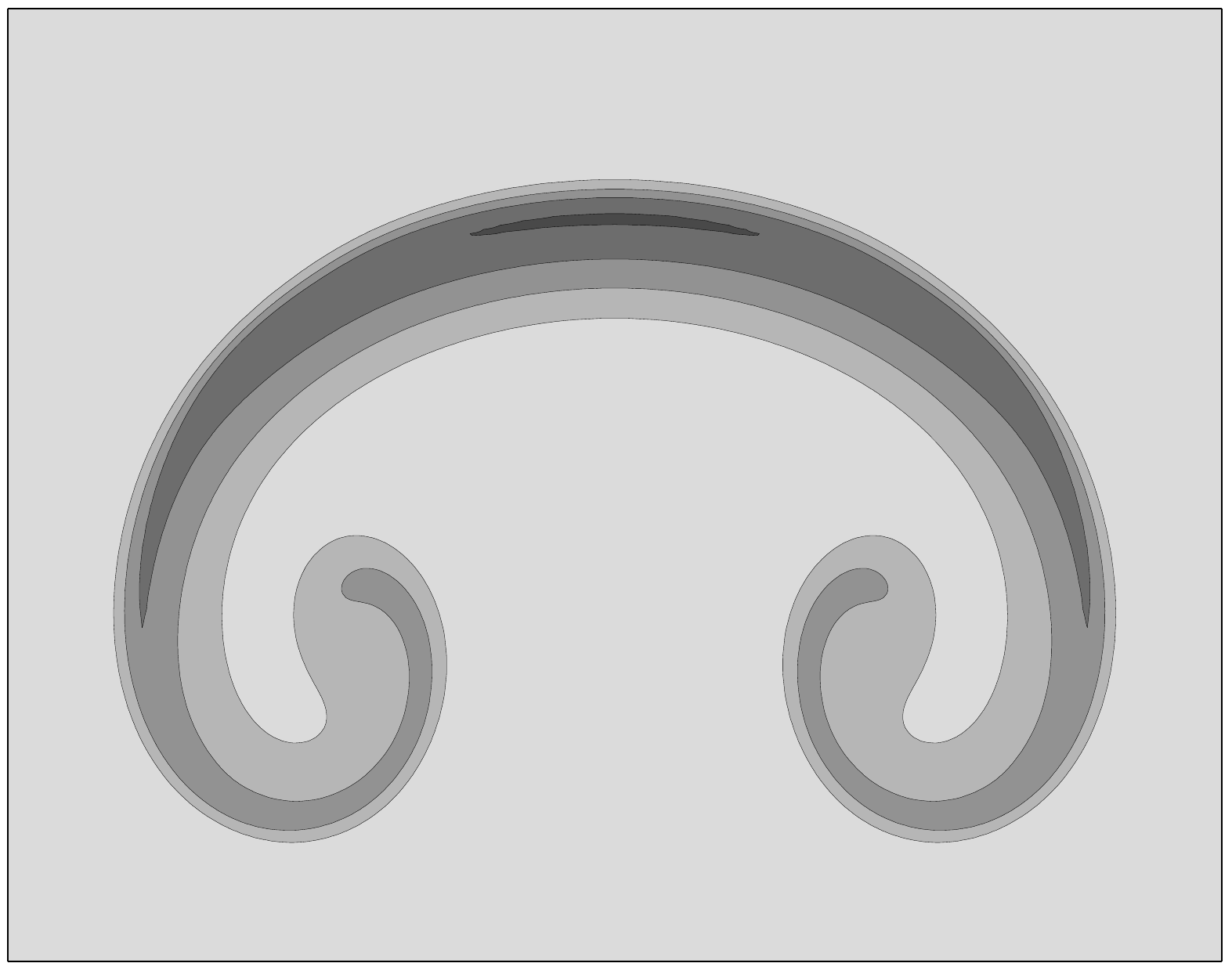}   &
\includegraphics[width=.25\textwidth]{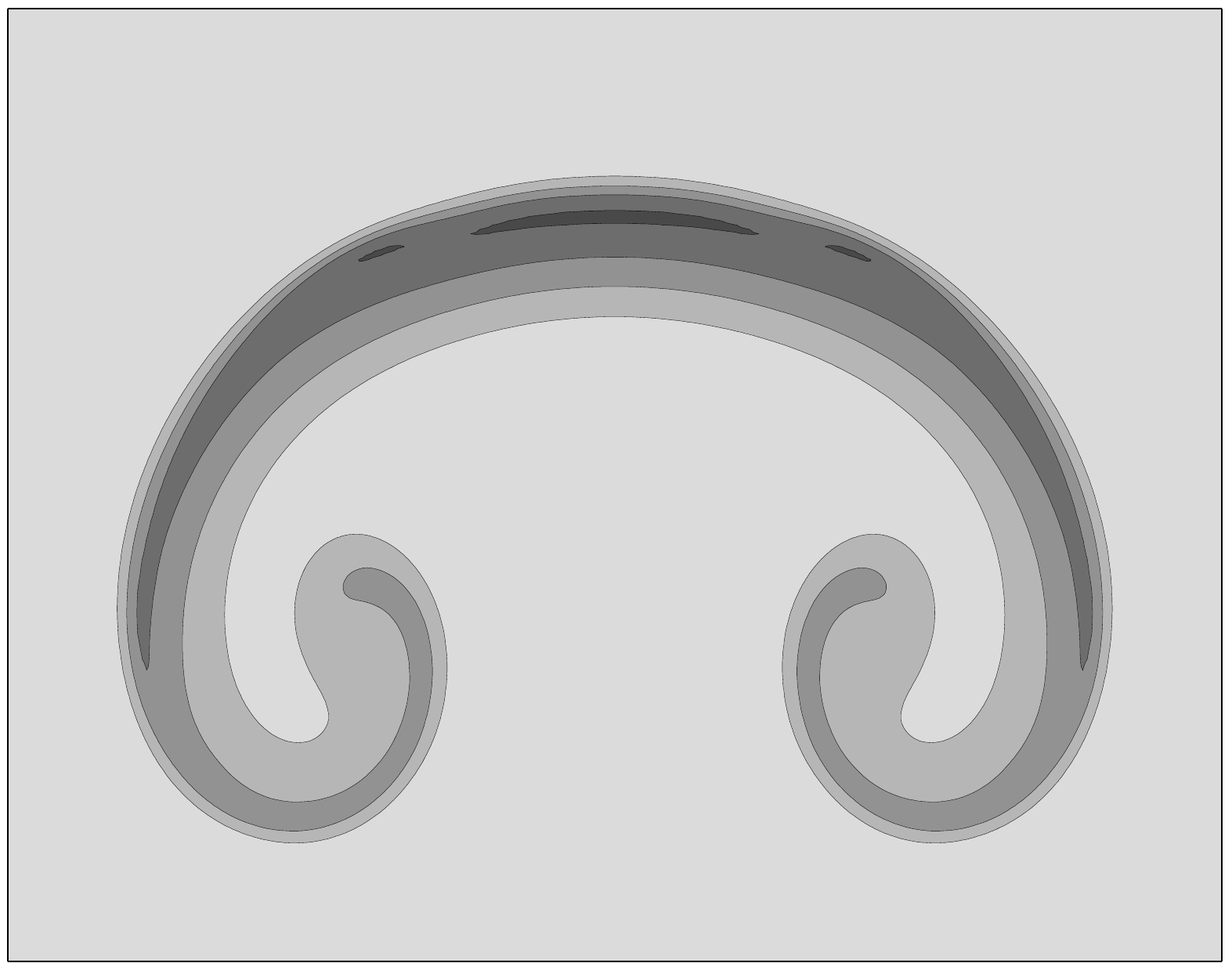}         &
\includegraphics[width=.25\textwidth]{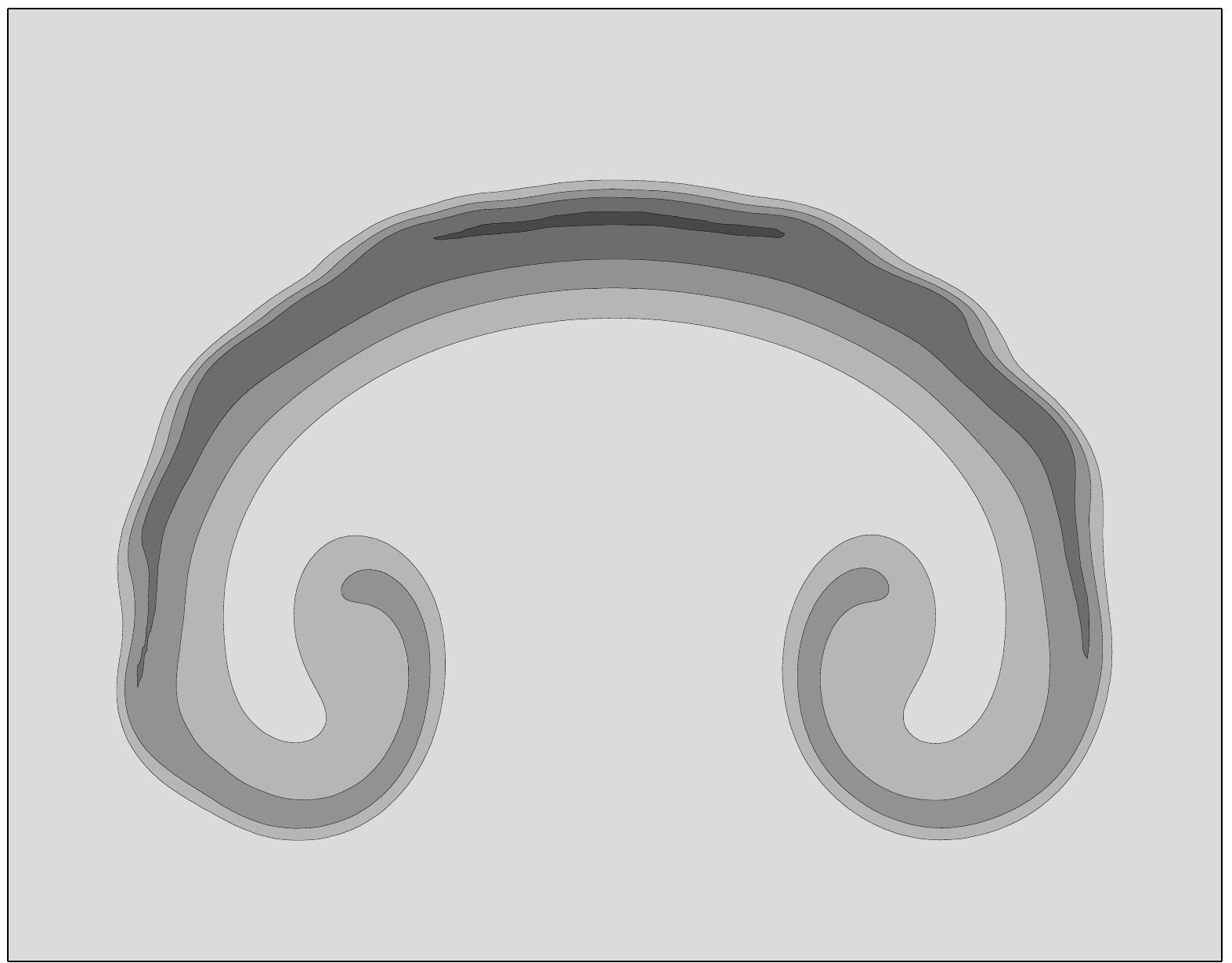}   \\
\rotatebox{90}{~~~~~~~~~~~~~~~~~25~m}                                  &
\includegraphics[width=.25\textwidth]{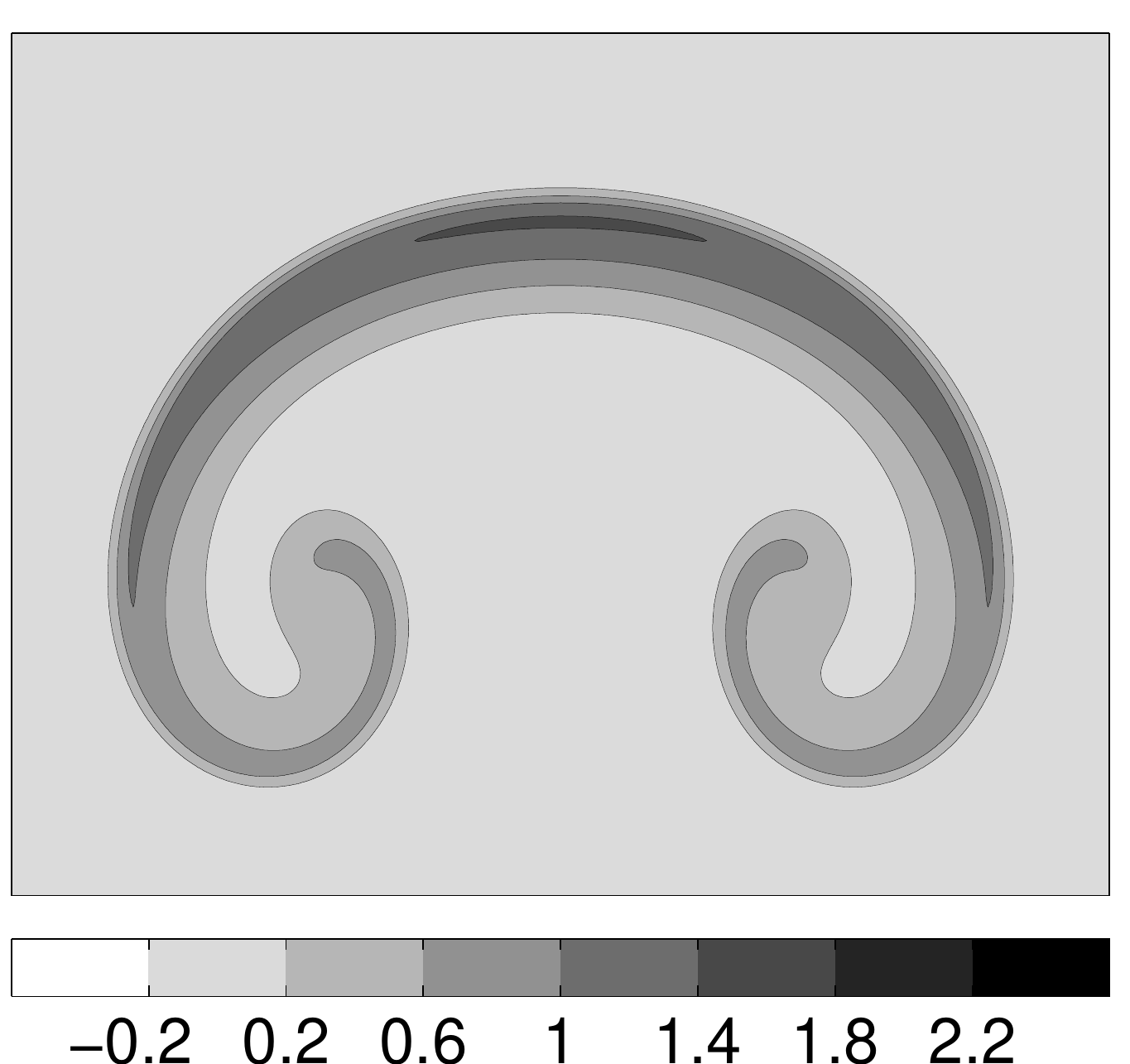}   &
\includegraphics[width=.25\textwidth]{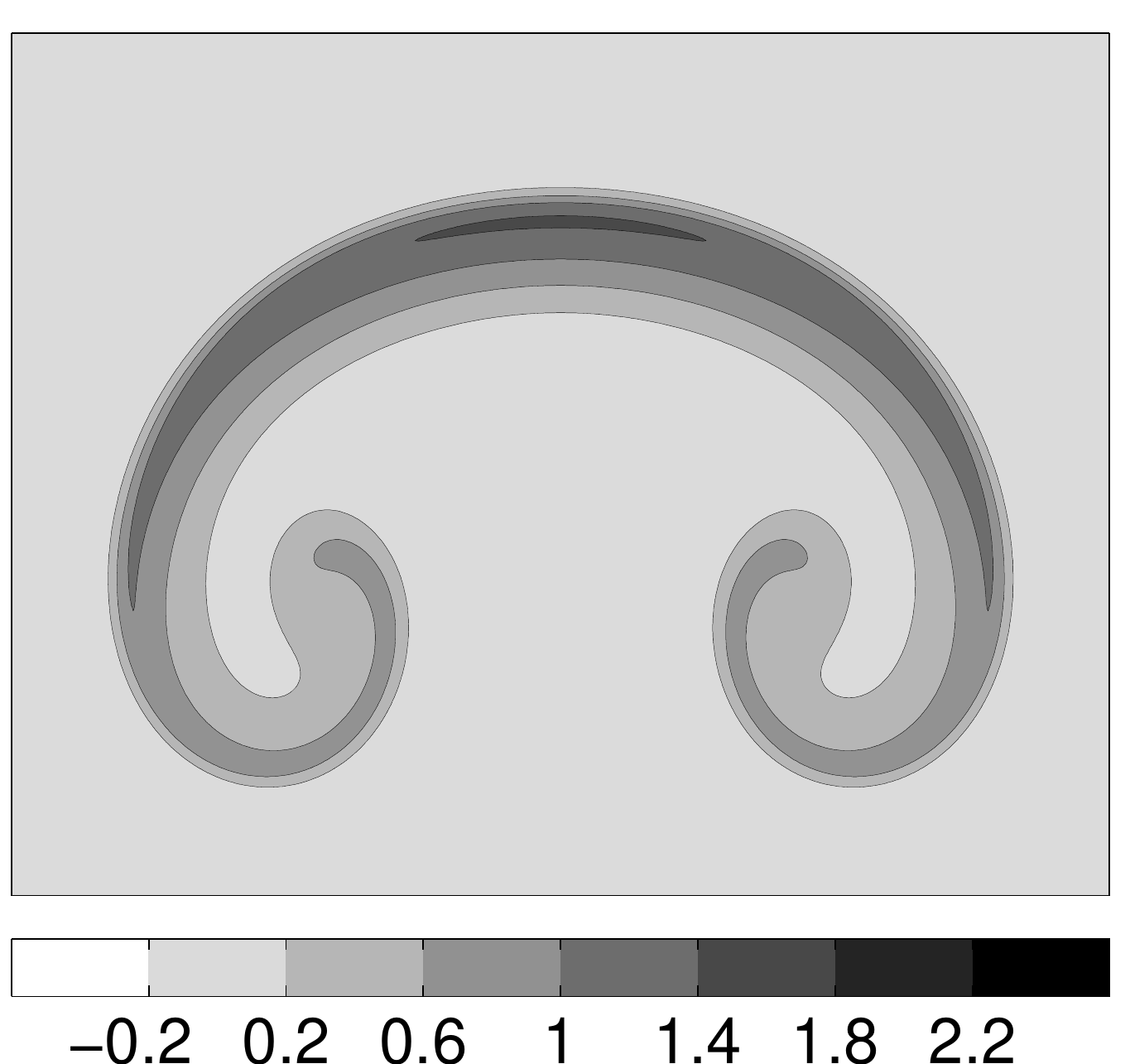}         &
\includegraphics[width=.25\textwidth]{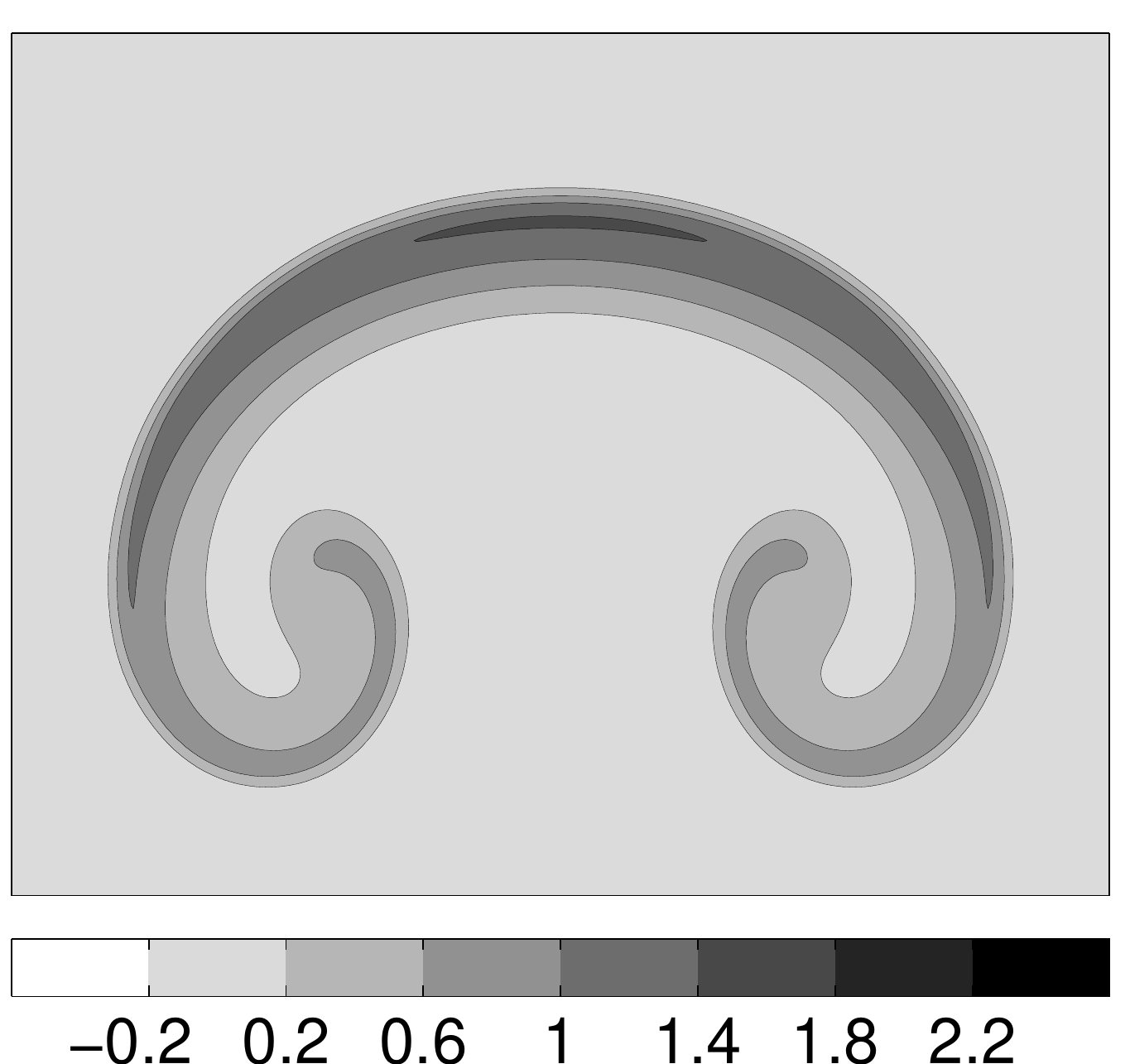}
\end{array}
$$
\caption{Numerical solutions for the rising thermal bubble with $\mu=10\text{ m}^2/\text{s}$ (\ref{bubble}) on the three different types of node distributions  at various resolutions, shown at the final simulation time, $t=1100$s.}
\label{bubble_mu10_res}
\end{figure}

Figure \ref{bubble_mu10_timeSeries} shows the time series of the solution for a 25m resolution ($N = 185,730$) on hexagonal nodes using $r^7$ with up to 4th-order polynomials and a $\Delta^3$-type hyperviscosity. The main purpose of this test is to make sure that, under refinement, all node layouts converge to the same solution (as this will not be the case in the next variation of the test) and to see if the convergence rate follows the predictions of Section \ref{sec:saturation}, even with a $C^0$ initial condition. Figure \ref{bubble_mu10_res} shows the final solution for the three different node layouts from a resolution of 200m to 25m. At 25m resolution, all solutions are visually identical. At coarser resolutions, as 100m, Cartesian and hexagonal nodes are more similar with scattered nodes having more incongruities at the leading edge of the rising bubble. A possible reason for this could be symmetry-breaking associated with scattered node layouts that would affect areas of large shear.

\begin{figure}[H]
\centering
\includegraphics[width=.5\textwidth]{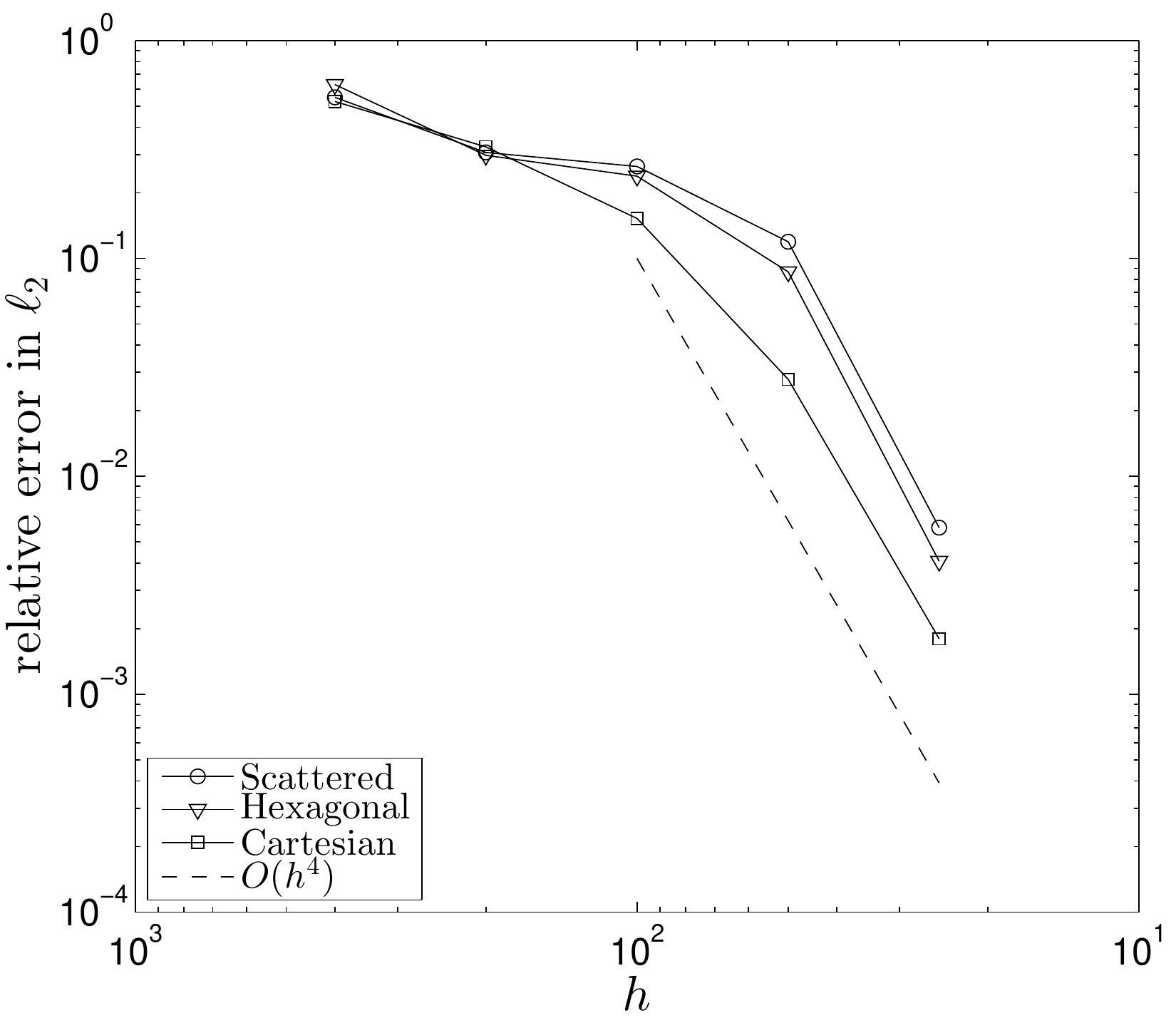}
\caption{Convergence behavior for $\theta'$ in the rising thermal bubble test case \ref{bubble}.   The $h=$400m, 200m, 100m, 50m, and 25m errors were calculated using the 12.5m RBF-FD reference solution.}
\label{convergence_bubble}
\end{figure}

In terms of convergence, Figure \ref{convergence_bubble} shows that even though the initial condition is $C^0$, the method does achieve 4th order convergence as predicted when using up to fourth degree polynomials. All nodes sets converge at fourth order under refinement, with Cartesian giving the best accuracy for $h\leq100$m. Note that at coarser resolutions from 400m too 100m only second order convergence is achieved.

Table \ref{tbl:bubble} shows to what degree the solution has converged in terms of how high the bubble should have risen, $\theta'$,  and $w'$. Note that when comparing against the 12.5m hexagonal node reference solution, all solutions in all variables have converged by 25m. In terms of $\max\{\theta'\}$ and $\max\{w'\}$ all solutions have converged by 50m. The $\min\{\theta'\}$ is almost identical for all node sets while $\min\{w'\}$ varies between node sets for coarser resolutions. In terms of bubble height Cartesian nodes seem to perform the best.

\begin{table}[H]
\centering
\caption{Resolution $\left(h\right)$, minimum and maximum values for $\theta'$ and $w'$, and bubble height at various resolutions for the rising thermal bubble \ref{bubble}.  Results are for $\phi(r)=r^7$ with up to fourth degree polynomials on a 37-node stencil.  The bubble height was determined by the intersection of the $0.1$K contour and the line $x=5$km.}
\label{tbl:bubble}
\begin{tabular}{c|c|c|c|c|c|c}
           &   h (m)   &   $\min\left\{\theta'\right\}$   &   $\max\left\{\theta'\right\}$   &   $\min\{w'\}$   &   $\max\{w'\}$   &   bubble height (m)   \\
\hline\hline
Cartesian  &   200     &   -0.11                          &   1.46                           &   -7.56         &   11.06         &   8,467               \\
           &   100     &   -0.08                          &   1.53                           &   -7.93         &   11.34         &   8,539               \\
           &   50      &   -0.02                          &   1.43                           &   -7.87         &   11.43         &   8,534               \\
           &   25      &    0.00                          &   1.43                           &   -7.74         &   11.43         &   8,535               \\
\hline
Hexagonal  &   200     &   -0.11                          &   1.36                           &   -7.67         &   11.12         &   8,686               \\
           &   100     &   -0.09                          &   1.65                           &   -8.05         &   11.49         &   8,527               \\
           &   50      &   -0.02                          &   1.43                           &   -7.75         &   11.43         &   8,553               \\
           &   25      &    0.00                          &   1.43                           &   -7.74         &   11.43         &   8,535               \\
\hline
Scattered  &   200     &   -0.11                          &   1.25                           &   -7.74         &   10.92         &   8,581               \\
           &   100     &   -0.09                          &   1.48                           &   -8.42         &   11.40         &   8,557               \\
           &   50      &   -0.02                          &   1.43                           &   -8.22         &   11.43         &   8,525               \\
           &   25      &    0.00                          &   1.43                           &   -7.75         &   11.43         &   8,535               \\
\hline
Reference  &   12.5    &    0.00                          &   1.43                           &   -7.74         &   11.43         &   8,535
\end{tabular}
\end{table}


\subsubsection{Case $\mu = 2\times10^{-5}\text{ m}^2/\text{s}$}
\begin{figure}[H]
$$
\begin{array}{cccc}
\text{} & \text{Cartesian} & \text{Hexagonal} & \text{Scattered} \\
\rotatebox{90}{~~~~~~~~~~~200~m}                                 &
\includegraphics[width=.3\textwidth]{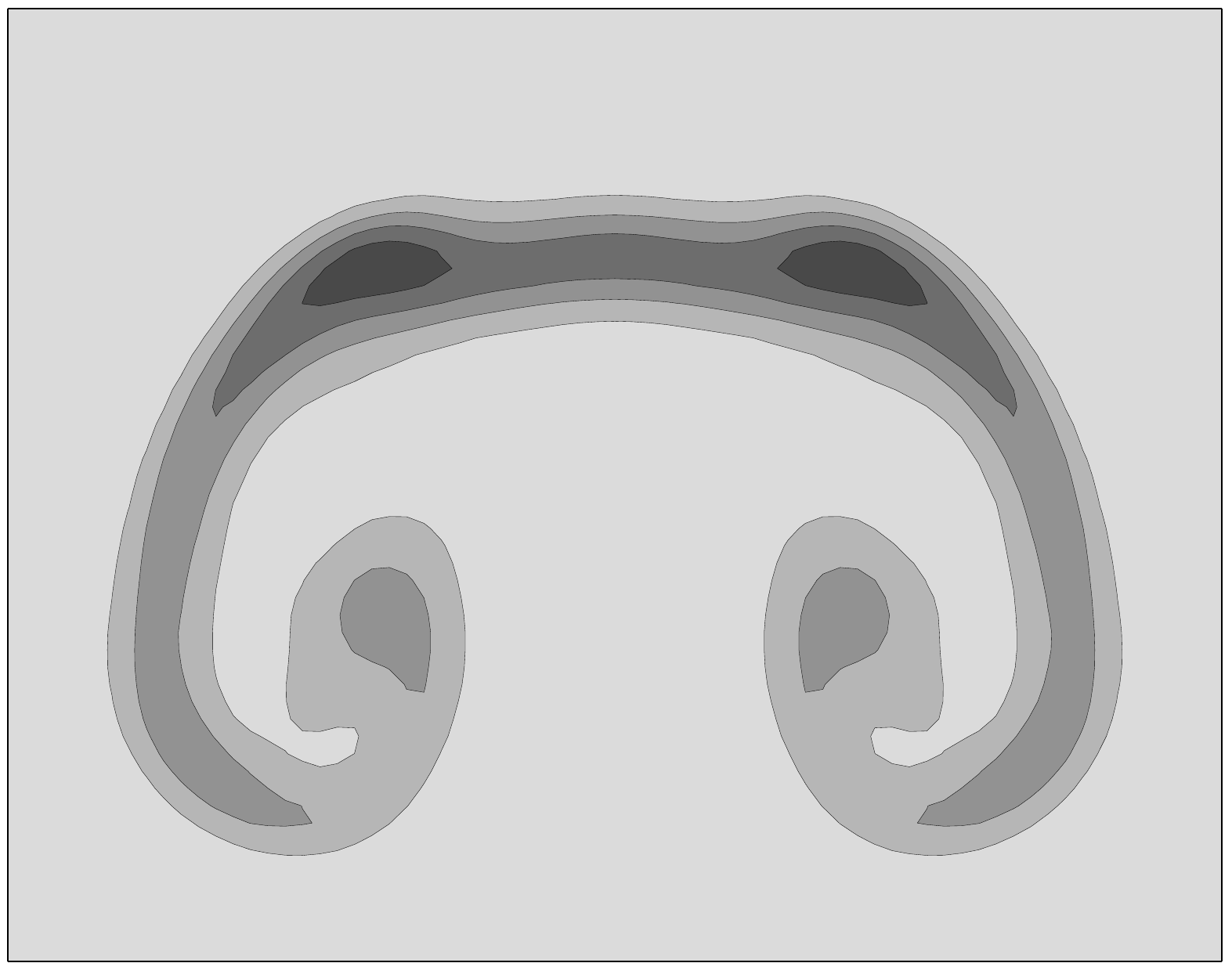}  &
\includegraphics[width=.3\textwidth]{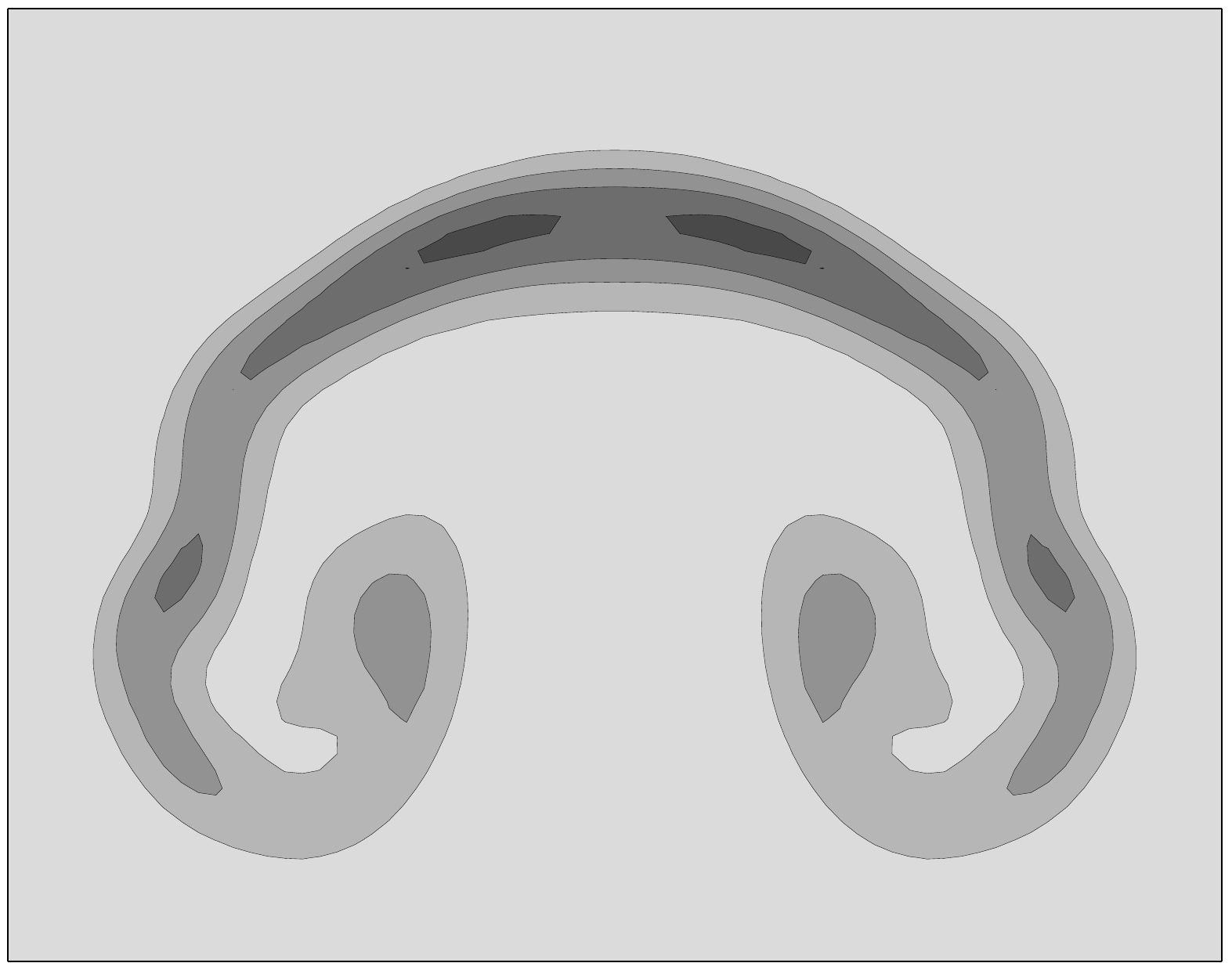}       &
\includegraphics[width=.3\textwidth]{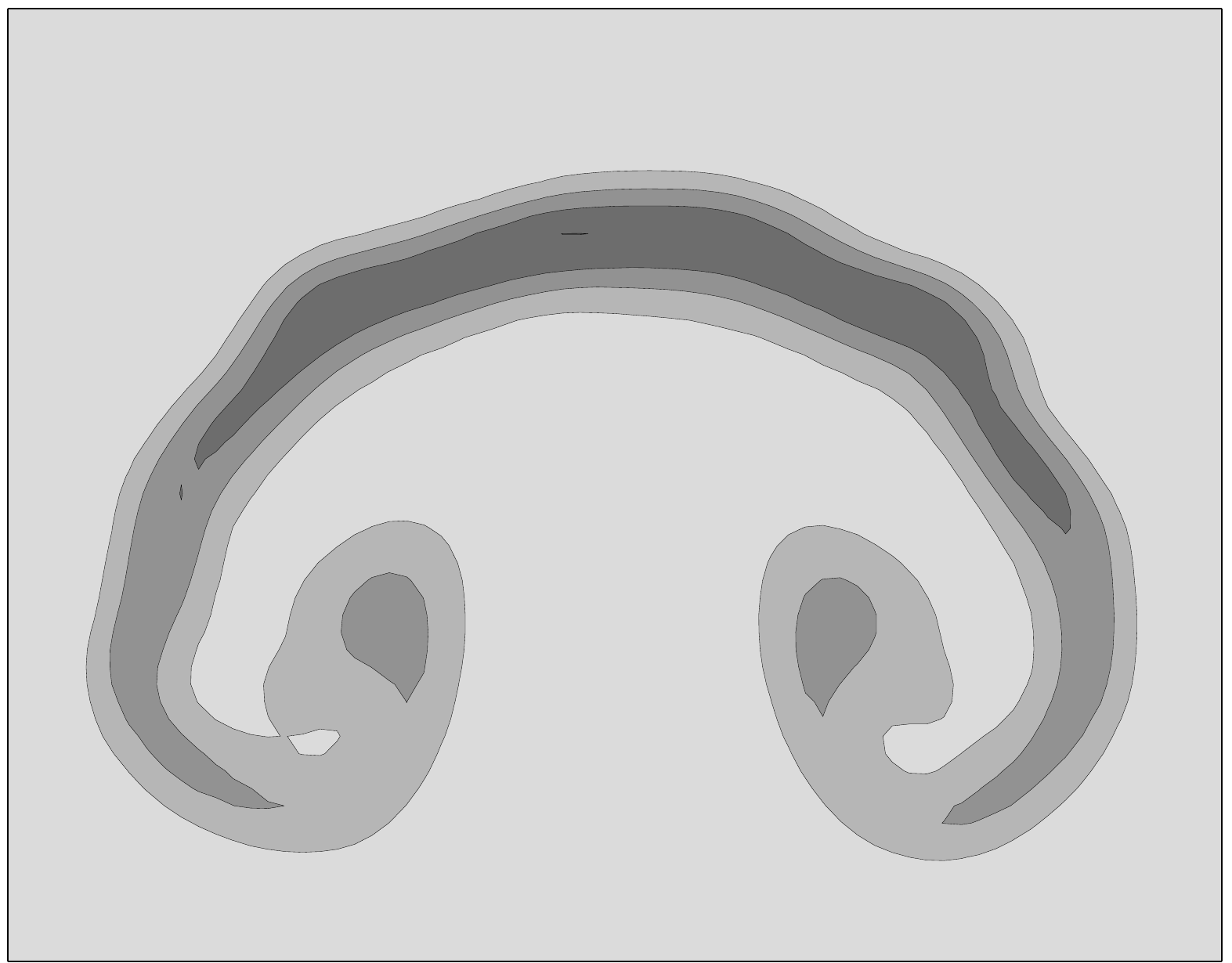}  \\
\rotatebox{90}{~~~~~~~~~~~100~m}                                 &
\includegraphics[width=.3\textwidth]{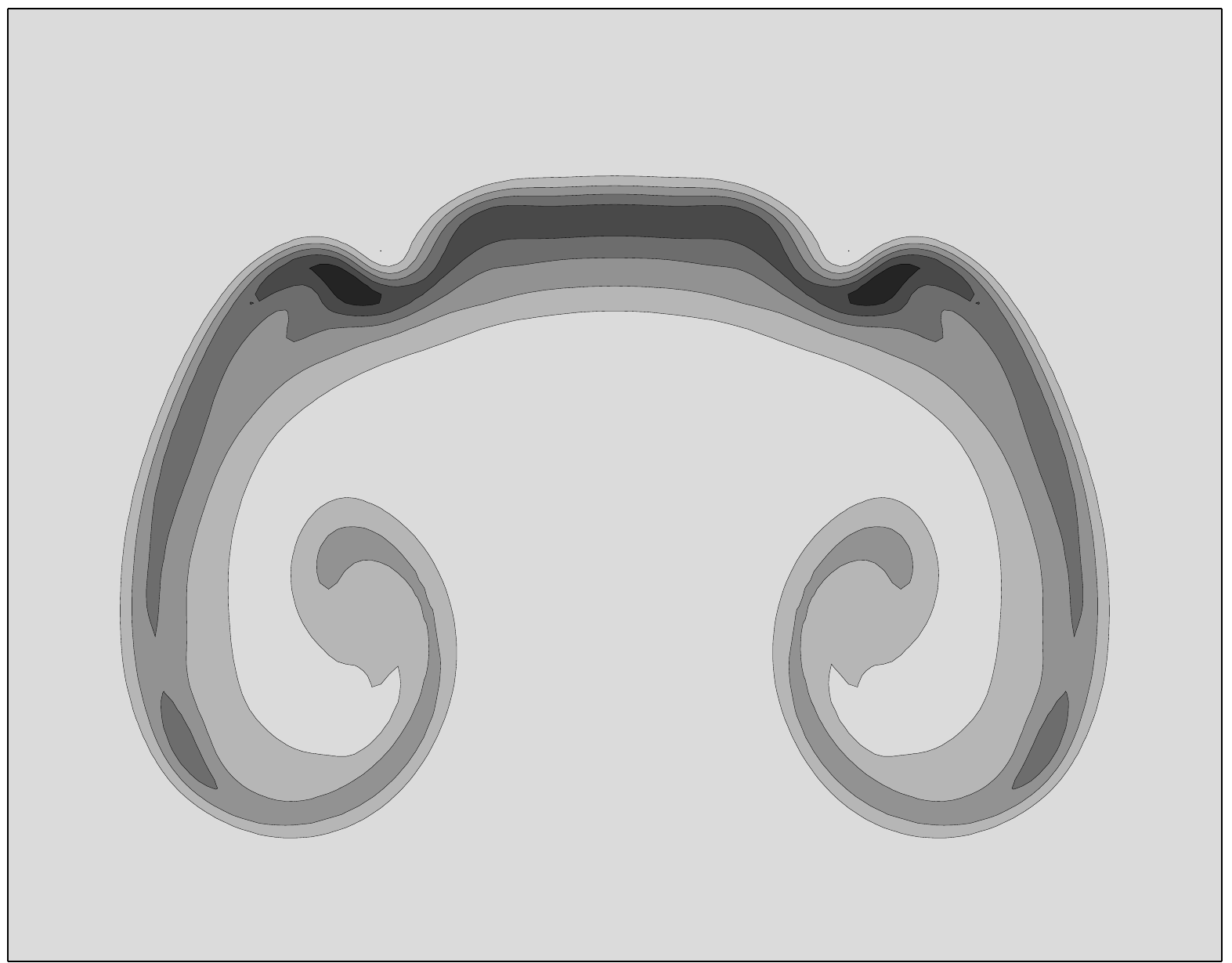}  &
\includegraphics[width=.3\textwidth]{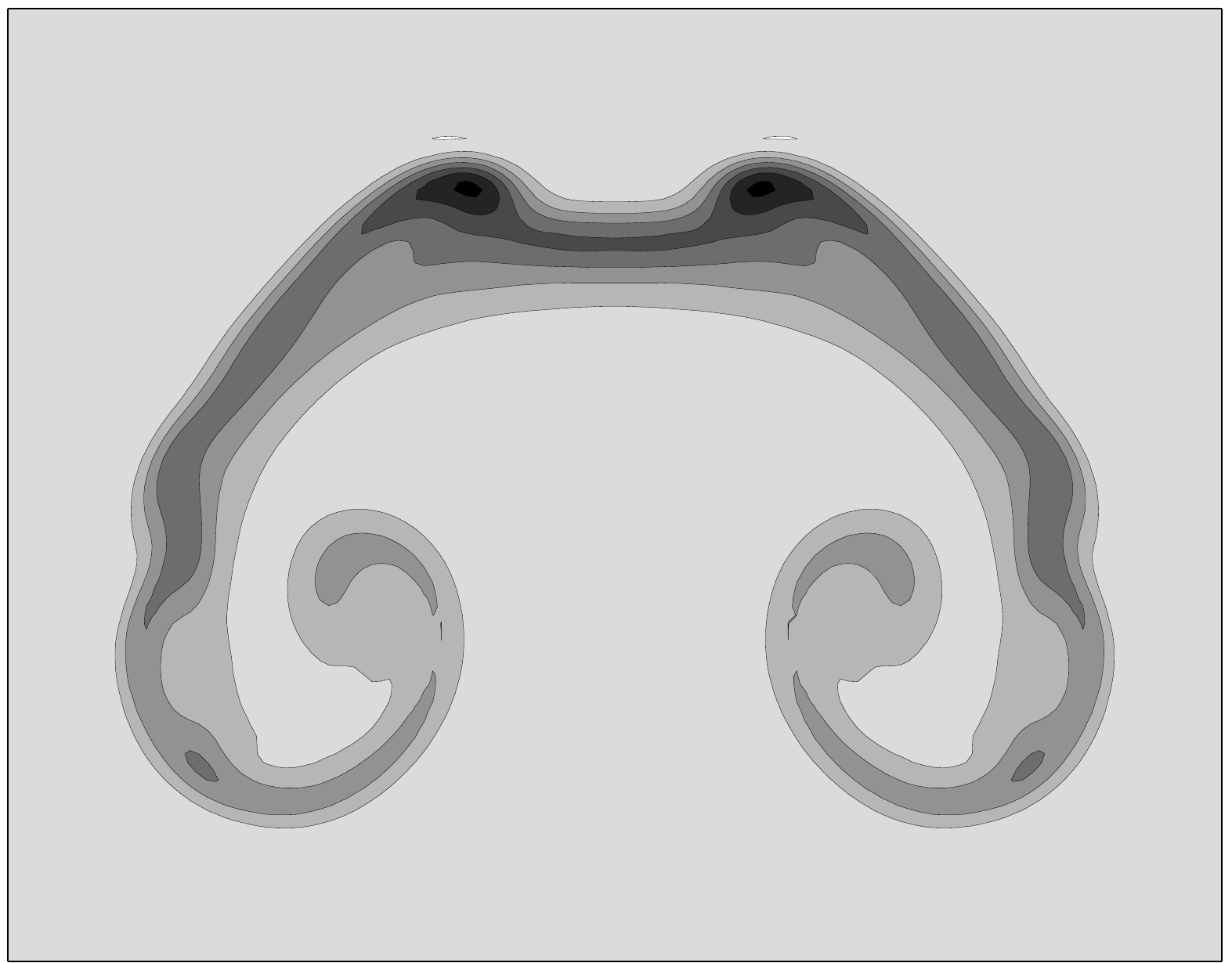}        &
\includegraphics[width=.3\textwidth]{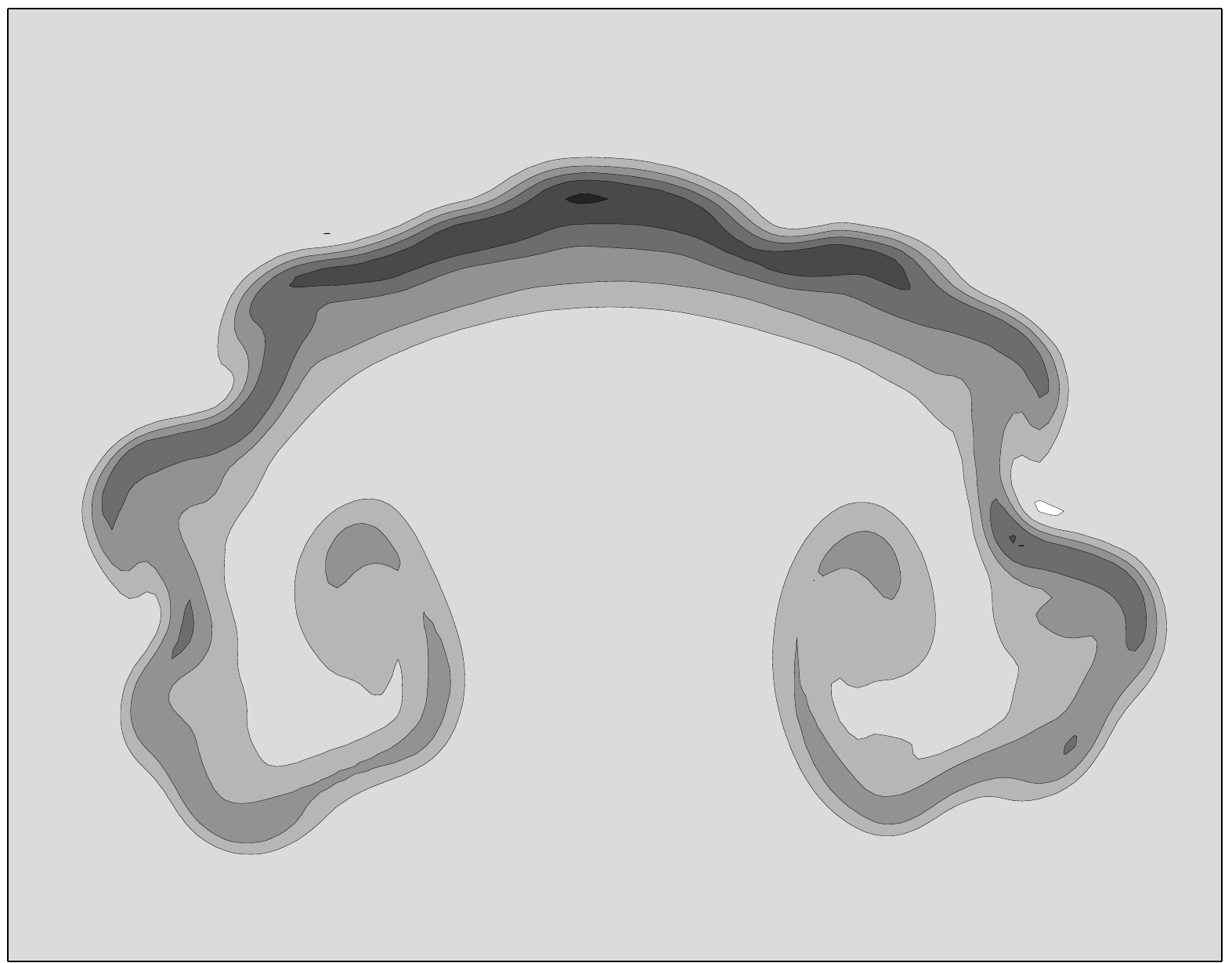} \\
\rotatebox{90}{~~~~~~~~~~~50~m}                                  &
\includegraphics[width=.3\textwidth]{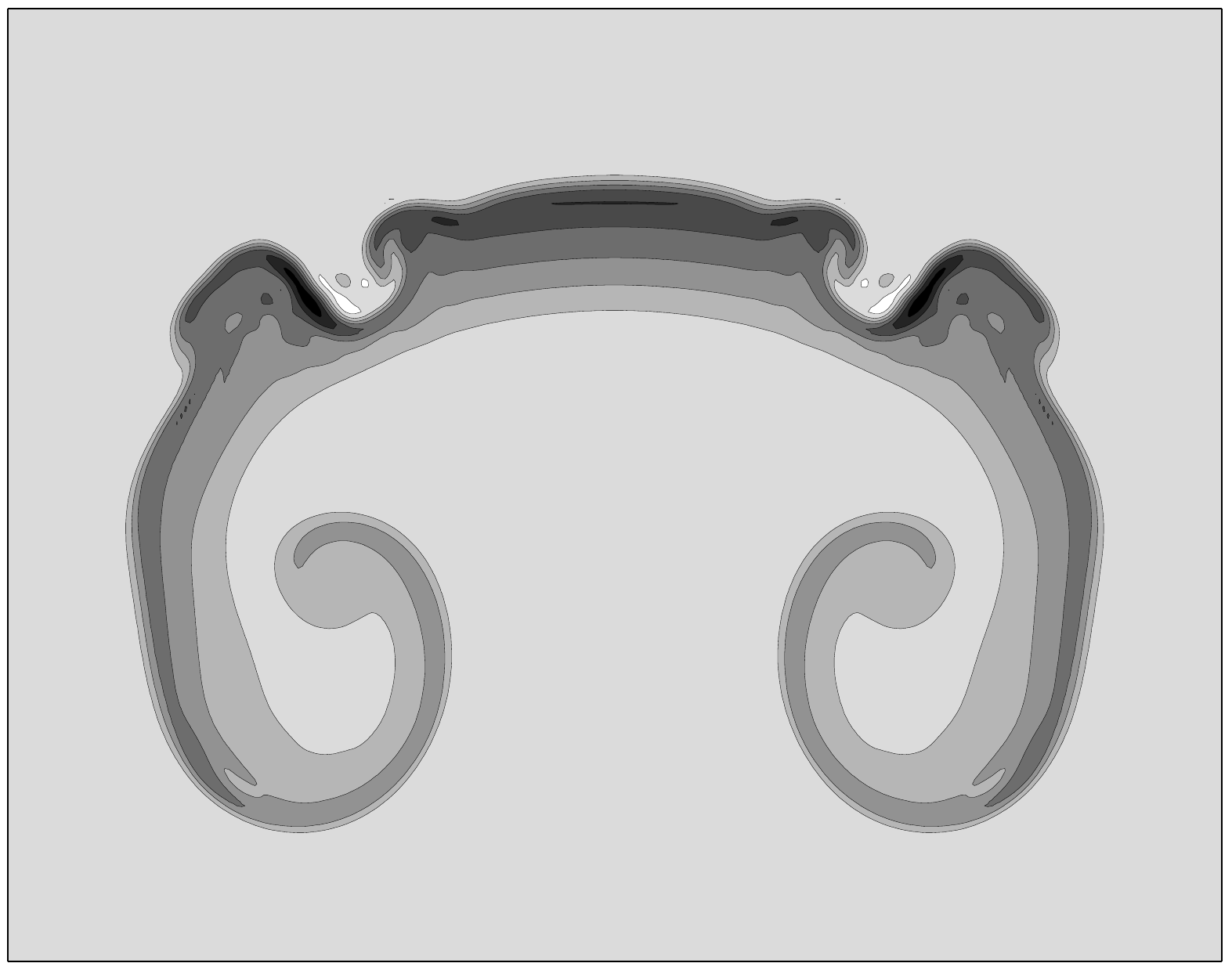}   &
\includegraphics[width=.3\textwidth]{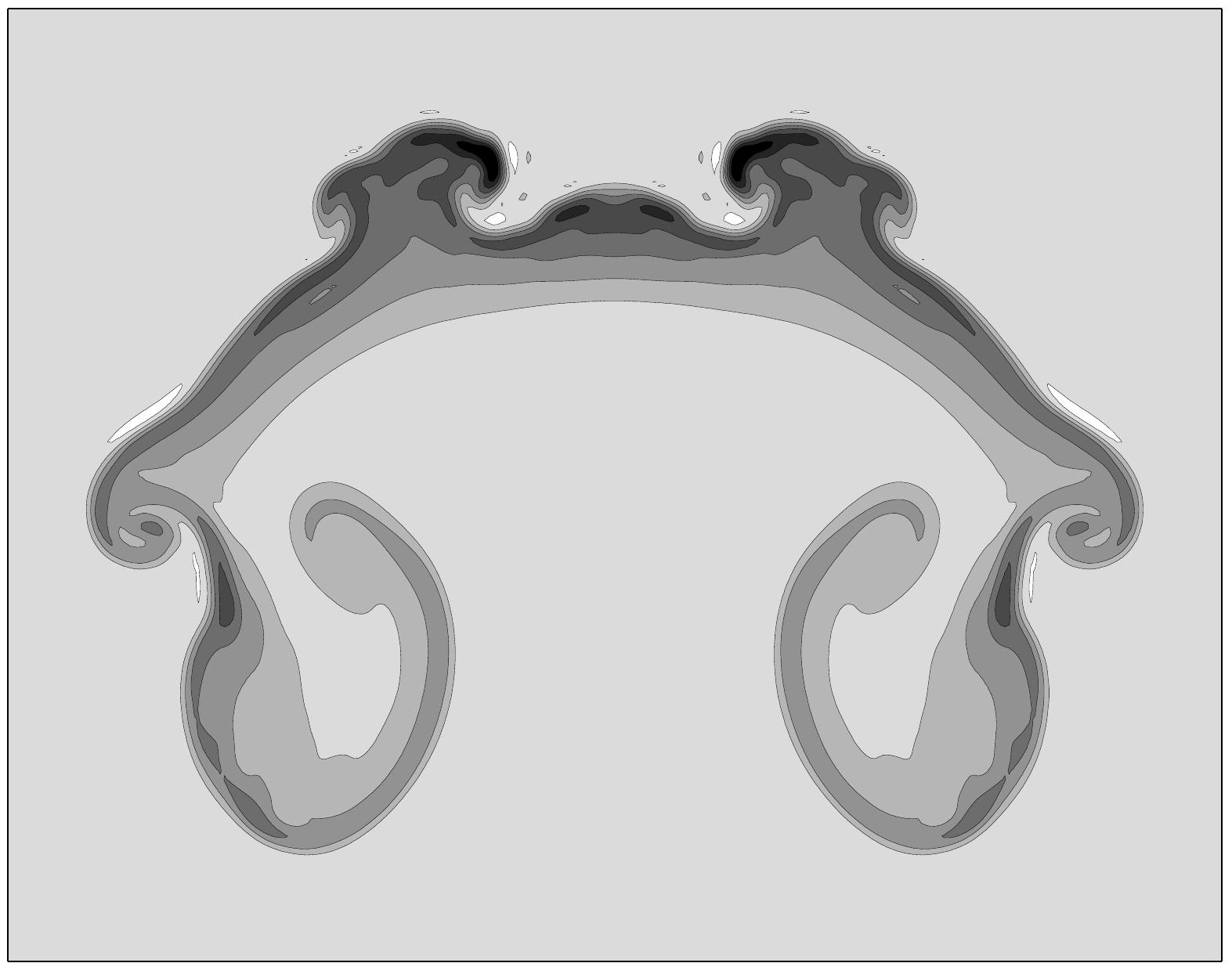}         &
\includegraphics[width=.3\textwidth]{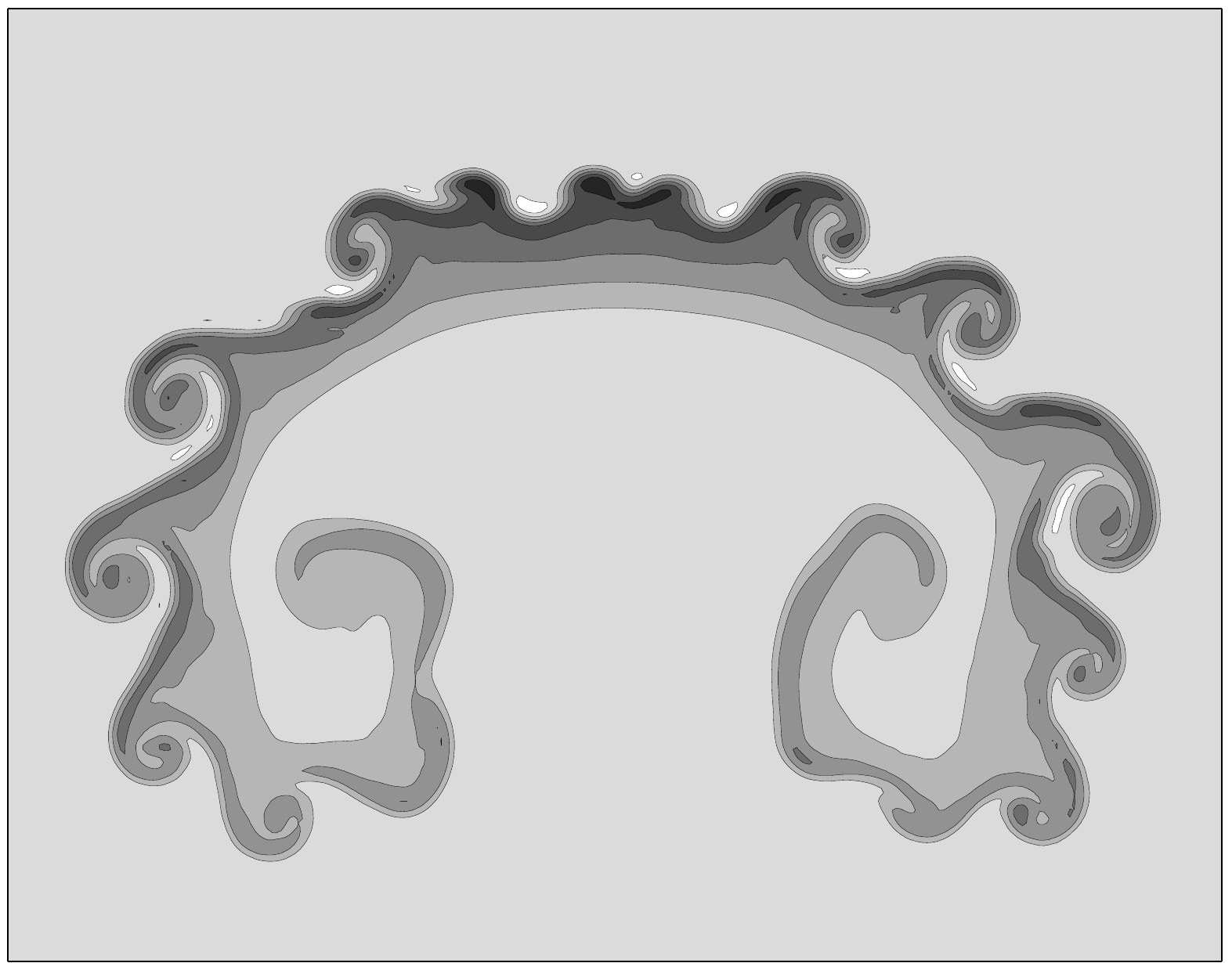}    \\
\rotatebox{90}{~~~~~~~~~~~~~~~~~25~m}                            &
\includegraphics[width=.3\textwidth]{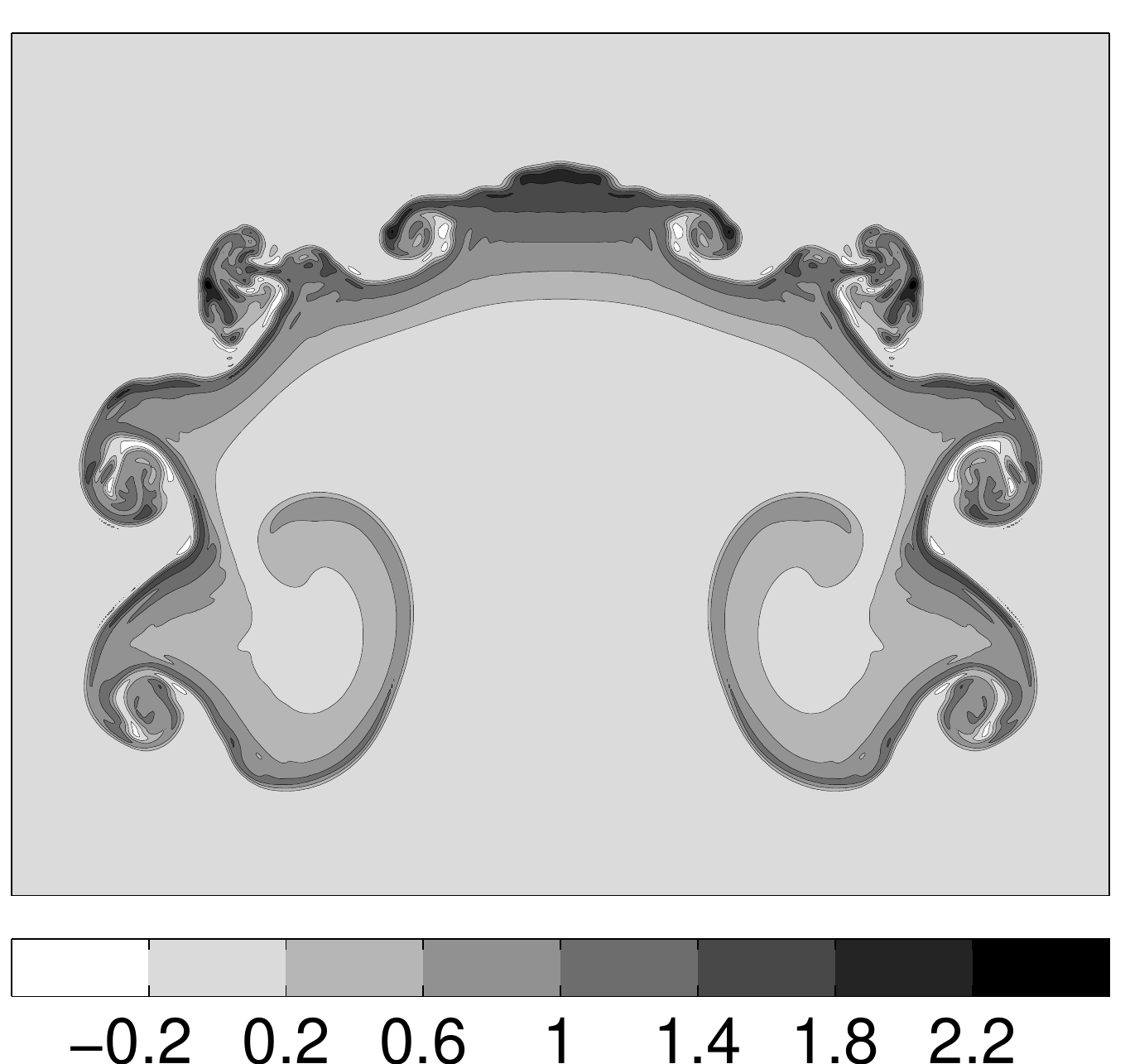}   &
\includegraphics[width=.3\textwidth]{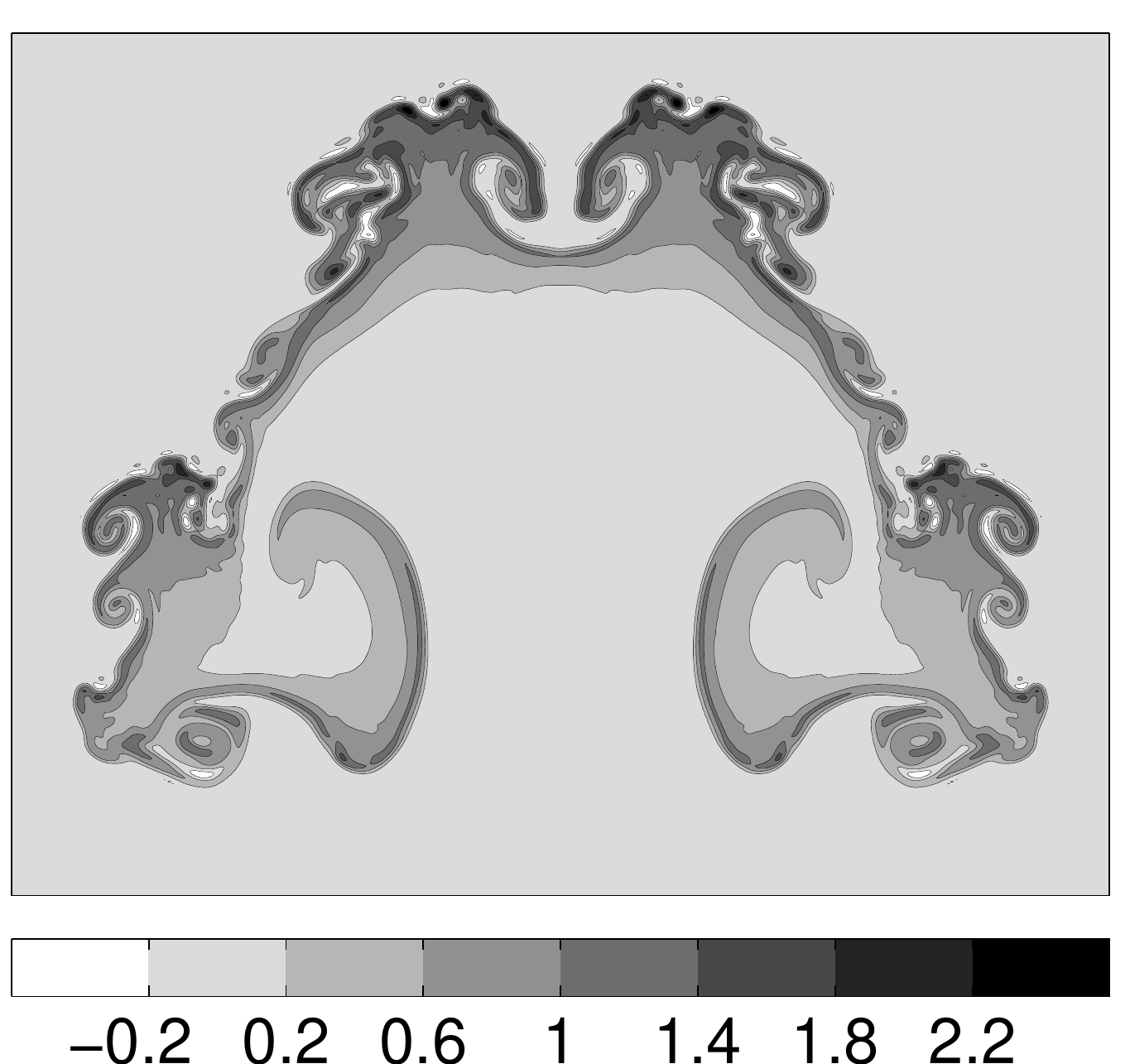}         &
\includegraphics[width=.3\textwidth]{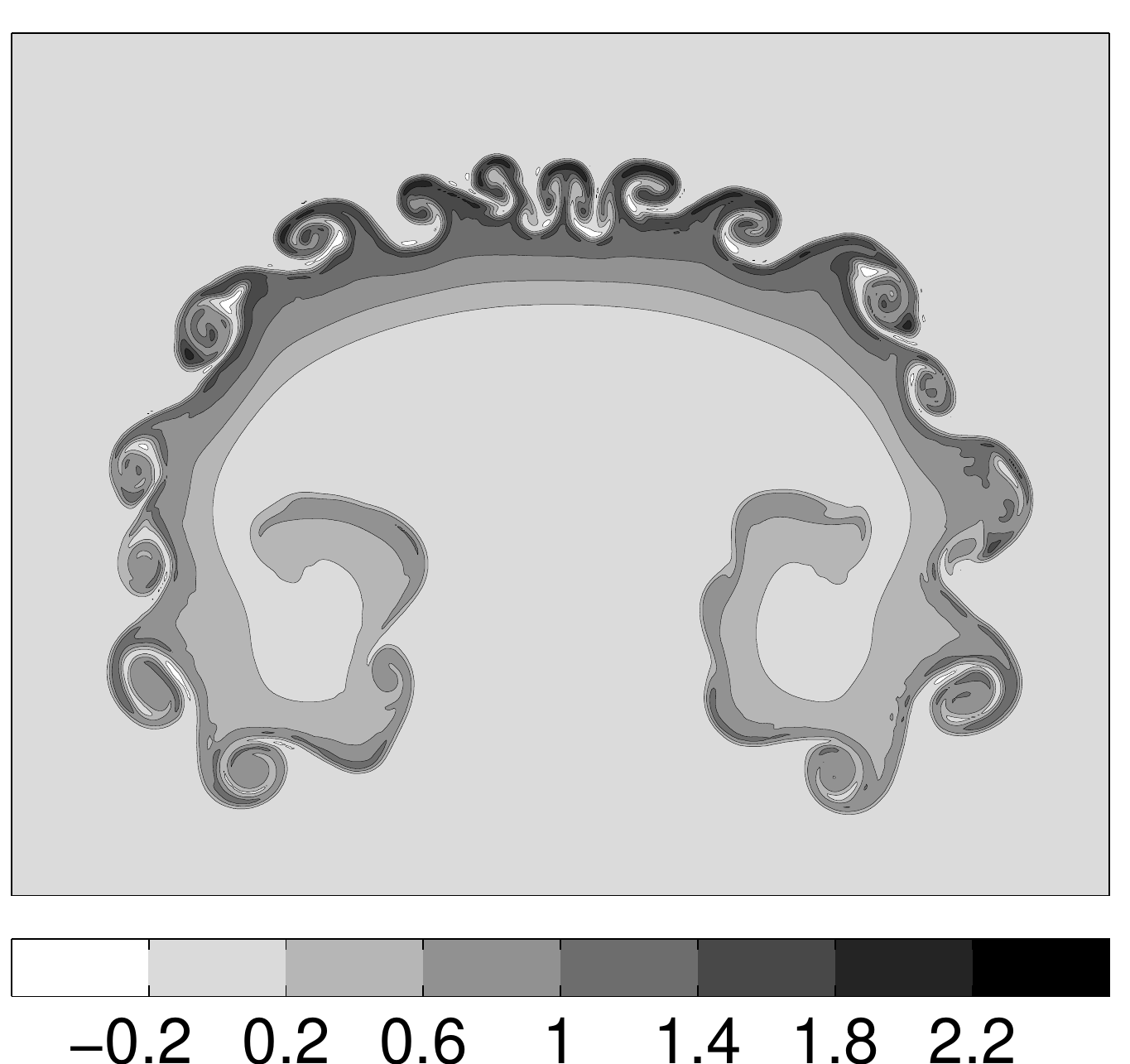}
\end{array}
$$
\caption{Numerical solutions for the rising thermal bubble with $\mu=2\times10^{-5}\text{ m}^2/\text{s}$ on the three different types of node distributions at various resolutions.  All results are shown at the final simulation time, $t=1100$s.}
\label{bubble_mu0_res}
\end{figure}

The rising thermal bubble test case is repeated with the viscosity $\mu$ set to $2\times10^{-5}$. The first purpose of this test case is simply to demonstrate that the proposed RBF-FD method, implemented with such low viscosity and a $C^0$ initial condition, has complete time stability using the same time step and amount of hyperviscosity as is the previous section. Secondly, we are interested in observing how the instability pattern at the leading edge of thermal bubble evolves as the node layout changes. Normally, in numerical testing, to see different evolutions of a solution the initial condition is perturbed. However with RBF-FD, one has the flexibility of leaving the initial condition intact and perturbing the node layout, which in a turbulent regime will lead to different evolutions of the solution. This can be seen in Figure \ref{bubble_mu0_res}. At 200m, there is not much difference between the bubbles. However as can be seen in the 25m results, the shear instability layer at the leading edge of the bubble (darkest contours) develops tight eddies whose structure varies significantly depending on the node layout. In both the Cartesian and hexagonal case, the eddy development is completely symmetric about the midpoint of the bubble due to the symmetry in the node layout, while in the scattered node layout this is not the case (a seemingly more realistic scenario for modeling warm air entrainment in the atmosphere). Furthermore, the scale of the eddies and the degree to which they excite finer scale instabilities varies between the node sets. The Cartesian node layout produces the largest scale eddies as well as the smallest amount of eddies. In contrast, the hexagonal nodes produce a rather strange bubble shape with very fine scale eddy structure.


\section{Conclusions and summary} \label{summary}

In this paper, a modified RBF-FD method is introduced to construct differentiation weights based on a combined RBF-polynomial basis, using RBF polyharmonic splines $\left(\phi(r)=r^m\right)$ with polynomial functions up to degree $l$ in the given dimension of the problem. The method is applied to three standard test cases in the numerical weather prediction community \cite{BlosseyDurran,Straka,Robert93}, with the latter two based on the Navier-Stokes (NS) equations. In addition, the effect of node layout (Cartesian, hexagonal, or scattered) on the error as well the qualitative character of the solution is considered. The following observations are made:

\begin{enumerate}
\item Under refinement, with the inclusion of polynomials, stagnation (saturation) error is evaded.
\item In the absence of boundary effects, the convergence rate is controlled, not by the order of the PHS, but by the highest degree polynomials used.
\item Increasing the order of the PHS marginally increases the accuracy, as the constant that multiples the convergence rate decreases.
\item For stable configurations that require no tuning of the hyperviscosity (e.g. $\gamma = 2^{-6} h^{-2k}$ for all NS tests), the number of nodes in the stencil, $n$, should be approximately twice the number of polynomial basis functions, ($l$+1)($l$+2)$/2$ in 2D. Hence on an $n=37$ node stencil, up to fourth-order polynomials (15 in 2D) are used for the NS equations.
\item In the absence of boundary effects, for a hyperbolic PDE, neither the character of the solution, the error, nor the convergence rate is sensitive to the node layout.
\item In the presence of boundaries, the solution on Cartesian nodes exhibited significant oscillations (Runge phenomena) near the boundary. This is not the case with hexagonal nodes, which was the most effective node layout in correctly capturing the physics, especially at lower resolutions.
\item In all cases, quasi-uniformly scattering the nodes showed no detriment to the quality of the solution, error, or convergence and in the majority of the cases performed better than Cartesian layouts.
\item Decreasing the viscosity by 6 orders of magnitude, (i.e. increasing in the Reynolds number by the same factor), does not require any change to the time step or amount of hyperviscosity added.
\item In the turbulent regime, the type of node layout heavily impacts the location and structure of eddy development on the leading edge of the thermal as well as the degree of excitation of finer scale instabilities.
\end{enumerate}

\noindent \textbf{Acknowledgements}
The authors would like to thank Professor Bengt Fornberg and Dr. Victor Bayona for useful comments and discussions. Dr. Flyer and Mr. Gregory Barnett would like to acknowledge the support of NSF grant DMS-094581. The National Center for Atmospheric Research is sponsored by NSF.

\begin{appendix}

\section{Symmetric Stencils}\label{symmetricStencils}

\begin{figure}[H]
\begin{align*}
\begin{array}{cc}
\includegraphics[width=.5\textwidth]{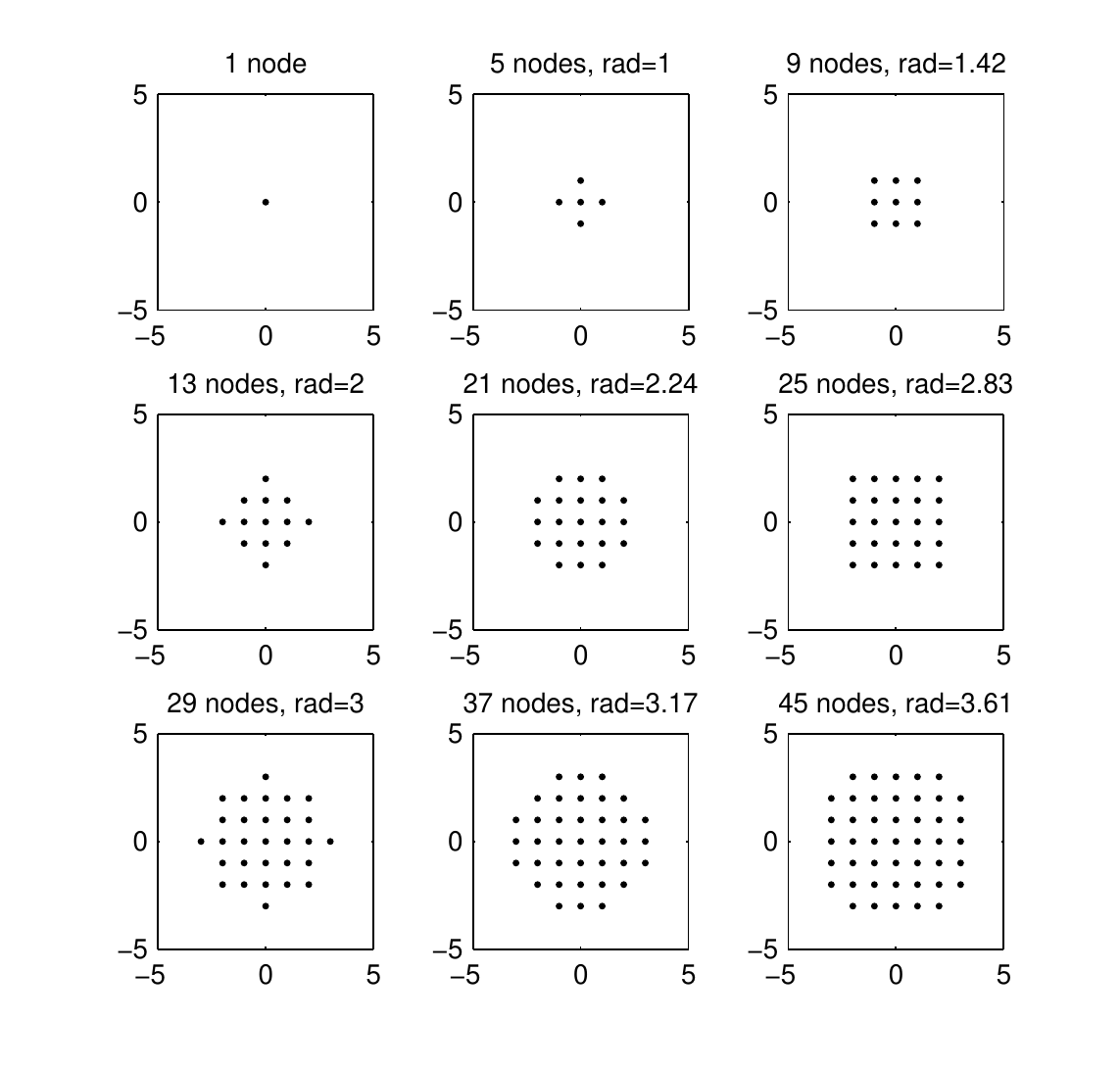} &
\includegraphics[width=.5\textwidth]{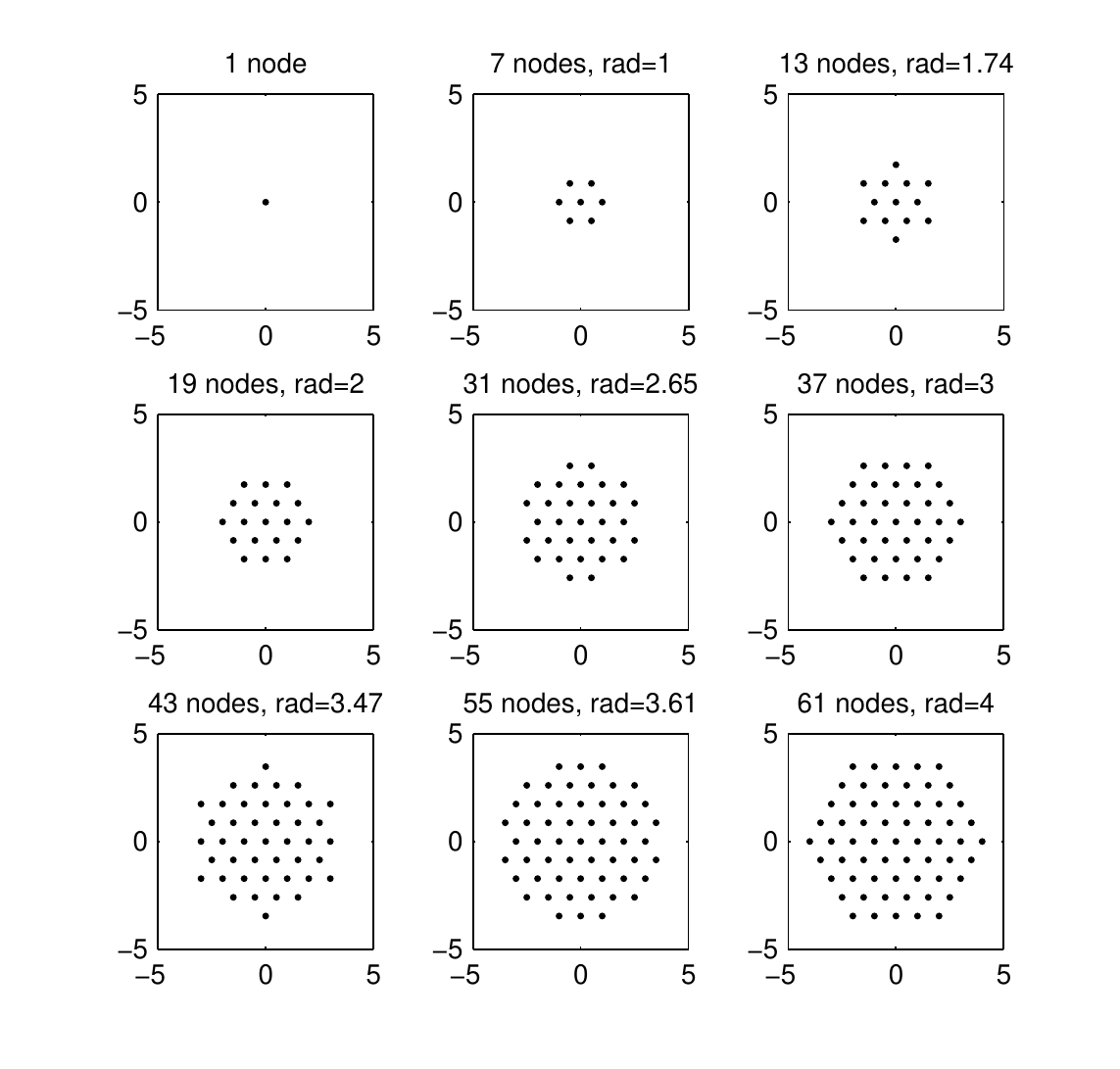}
\end{array}
\end{align*}
\caption{Symmetric stencils for cartesian and hexagonal nodes.  Note that 13 and 37 are the only reasonably small stencil-sizes held in common.}
\end{figure}

\end{appendix}


\end{document}